\documentclass[onecolumn,amsmath,amssymb,nofootinbib,letterpaper,11pt]{article}
\pdfoutput=1
\usepackage{jheppub}
\usepackage{ifpdf}
\usepackage[utf8]{inputenc}
\usepackage[T1]{fontenc}
\usepackage{graphicx}
\input{epsf}
\usepackage{epsfig}
\usepackage{epstopdf}
\usepackage{arcs}
\usepackage{subfigure}
\usepackage{tensor}
\usepackage{braket}
\usepackage{float}
\usepackage[dvipsnames]{xcolor}
\usepackage{amsmath}
\usepackage{morefloats}
\newcommand{\be}{\begin{equation}}
\newcommand{\ee}{\end{equation}}

\renewcommand{\arraystretch}{1.6}
\usepackage{tikz}
\usetikzlibrary{calc}   
\usetikzlibrary{topaths}

\usepackage{hyperref}
\hypersetup{
	colorlinks =WildStrawberry,
	linkcolor =red,
	pdftitle={Leading corrections to Kerr geometry},
	citecolor=NavyBlue,
	filecolor=WildStrawberry,
	urlcolor=WildStrawberry,
}

\usepackage[toc]{appendix}
\usepackage{color,soul} 
\usepackage{datetime}
\newcommand{\bea}{\begin{eqnarray}}
\newcommand{\eea}{\end{eqnarray}}
\newcommand{\ba}{\begin{eqnarray}}
\newcommand{\ea}{\end{eqnarray}}

\newcommand{\beq}{\begin{equation}}
\newcommand{\eeq}{\end{equation}}
\newcommand{\beqa}{\begin{eqnarray}}
\newcommand{\eeqa}{\end{eqnarray}}
\newcommand{\beqar}{\begin{eqnarray*}}
\newcommand{\eeqar}{\end{eqnarray*}}

\newcommand{\req}[1]{(\ref{#1})}

\newcommand{\cL}{\mathcal{L}}
\newcommand{\cN}{\mathcal{N}}

\renewcommand{\href}[2]{#2}

\title{Ringing of rotating black holes in higher-derivative gravity}
\author[a]{Pablo A. Cano,}
\author[a]{Kwinten Fransen}
\author[a]{and Thomas Hertog}

\affiliation[a]{Instituut voor Theoretische Fysica, KU Leuven\\
	Celestijnenlaan 200D, B-3001 Leuven, Belgium\vspace{0.1cm}}
\vspace{0.1cm}
\emailAdd{pabloantonio.cano@kuleuven.be} 
\emailAdd{kwinten.fransen@kuleuven.be}
\emailAdd{thomas.hertog@kuleuven.be}
\abstract{We compute scalar quasinormal mode (QNM) frequencies in rotating black hole solutions of the most general class of higher-derivative gravity theories, to quartic order in the curvature, that reduce to General Relativity for weak fields and are compatible with its symmetries. The wave operator governing the QNMs is not separable, but we show one can extract the QNM frequencies by a projection onto the set of spheroidal harmonics. We have obtained accurate results for the quasinormal frequency corrections relative to Kerr for rotating black holes with dimensionless spins up to $\sim 0.7$. We also discuss to what extent our results carry over to the phenomenologically more relevant case of gravitational QNMs. Finally we provide an ancillary computational package that allows one to generalize our calculations to any effective energy-momentum tensor arising from higher-derivative terms in the effective action.}

\preprint{
}

\begin{document}
\maketitle

\newpage
\section{Introduction}

Gravitational waves (GW) from the inspirals and mergers of black hole binaries \cite{LIGOScientific:2018mvr} encode a wealth of precious information about the gravitational physics of highly warped, dynamical regions of spacetime. Decoding these GW patterns, combined with further advances in GW detection technology, therefore offers a promising route to use GWs as precision probes of General Relativity (GR) \cite{TheLIGOScientific:2016src,Yunes:2016jcc,Berti:2018cxi,Berti:2018vdi,Barack:2018yly,Abbott:2018lct,LIGOScientific:2019fpa} and, in due course, its high-energy completion  \cite{Okounkova:2019zjf,Sennett:2019bpc,Carson:2020cqb,Carson:2020ter,Carson:2020iik,Okounkova:2020rqw}.
The significant theoretical uncertainties on the form of the latter and thus on the scales at which deviations from GR become important, call for a sufficiently general approach.
 
In this spirit we study the final phase of generic mergers of rotating black holes in the most general class of higher-derivative extensions of GR that reduce to GR in the weak field limit and are compatible with its symmetries. Specifically, we consider the theories introduced in Refs.~\cite{Endlich:2017tqa,Cano:2019ore}. In this final `ringdown' phase GWs carry away the multipolar structure of the newly formed remnant, which then relaxes to its final stationary state. The information in GWs generated in this stage is encoded in the complex frequencies of the quasinormal mode (QNM) perturbations of the final state, which describe its response to generic perturbations --- see \cite{Berti:2009kk} and references therein. The final state as well as the QNM frequencies are sensitive to the underlying theory of gravity, and both are known to be highly constrained in GR. Hence this ringdown phase offers great potential to probe GR, and to constrain corrections to it \cite{Berti:2018vdi,Barack:2018yly,Maselli:2019mjd}.

To realize the scientific potential of GWs in this regime, precise and systematic predictions of QNM frequencies will be essential. Calculations of QNM frequencies are rather involved, however, and have often been performed using various approximations. Recently, much progress has been reported on the calculation of QNM frequencies for spherical black holes in a broad range of modifications of GR \cite{Cardoso:2009pk,Blazquez-Salcedo:2016enn,Blazquez-Salcedo:2017txk,Tattersall:2017erk, Tattersall:2018nve,Cardoso:2018ptl,Konoplya:2020bxa} --- see also \cite{Cardoso:2019mqo,McManus:2019ulj} for a parametrized formalism. Coalescences of observational interest, however, typically involve black holes with significant angular momentum \cite{LIGOScientific:2018mvr}. Although some approximations are sometimes implemented to capture the effect of rotation \cite{Carson:2020cqb,Blazquez-Salcedo:2016enn}, the computation of quasinormal frequencies of beyond-Kerr black holes remains an outstanding issue. 
To this end, and in the spirit of the general approach motivated above, we compute QNM frequencies associated with scalar fields in rotating black hole backgrounds in the general class of higher-derivative theories of gravity specified above. 

One remarkable property of Kerr's solution \cite{Kerr:1963ud} is the fact that the master equation for perturbations is separable \cite{PhysRevLett.29.1114,Teukolsky:1973ha}. This allows to reduce the problem to a one-dimensional ``Schr\"odinger-like'' equation with an effective potential, for which many methods are known to compute the QNM frequencies, \textit{e.g.} \cite{PhysRevD.30.295,Schutz:1985km,Leaver:1985ax,doi:10.1063/1.527130,Leaver:1986gd,Berti:2009kk}.
The key technical hurdle that we resolve in this paper is that the wave operator governing the scalar QNM perturbations in rotating black hole backgrounds in higher-derivative gravities is, generically, non-separable. We show that the projection of the wave operator onto the set of spheroidal harmonics yields a consistent second-order ODE for a single variable, enabling us to extract the QNM frequencies in the usual manner.\footnote{See \cite{Zimmerman:2014aha,Mark:2014aja} for a different proposal to resolve non-separability issues.}
The computation of scalar QNM frequencies can be considered a first step towards the calculation of the phenomenologically more interesting gravitational QNMs. In fact, as we discuss in Section \ref{sec:con} below, certain facets of the phenomenology of scalar QNMs carry over to the gravitational QNMs. 

Many of our calculations are analytic yet lengthy, so we provide a number of ancillary files with this submission that allow one to reproduce (and even to extend) most of these computations using Mathematica. The main resource we provide is a package ``CorrectionsKerr.mx'', which includes several functions that enable one to carry out the following computations: 
\begin{enumerate}
\item To solve the modified Einstein's equations for rotating black holes in the above class of theories (cf. Section~\ref{sec:corrKerr}).
\item To find the effective potentials for scalar perturbations (cf. Section \ref{sec:effpotrot}).
\item To find the fundamental complex QNM frequencies through a numerical integration of the wave equation \req{radialeqrot2}.
\end{enumerate} 
The corresponding functions are documented in the example notebook ``QNMKerrCorrections.nb''. The only input needed in this is the effective energy-momentum tensor that appears on the right-hand-side of Einstein's equations --- as in Eq. \req{EinsteinEOM} --- evaluated on the Kerr metric. This is required for step 1 above. We provide these effective energy-momentum tensors for all theories considered in this paper --- see \req{Action} and \req{eq:quarticL} --- in the package ``Tmunu.mx''. However, our methods should also work for any other energy-momentum tensor that arises from higher-derivative terms in the effective action.  
The outline of this paper is as follows:
\begin{itemize}
\item In Sections~\ref{sec:effectiveaction} and \ref{sec:corrKerr} we briefly review the higher-derivative effective actions derived in \cite{Endlich:2017tqa,Cano:2019ore} and the rotating black hole solutions in these theories. 
\item Section~\ref{sec:sphericQNM} is devoted to the study of scalar QNMs of spherically symmetric black holes and serves as a warm-up for the rotating case.
\item In Section \ref{sec:pertrotation} we address the problem of scalar perturbations in the background of a corrected Kerr black hole. We show that the projection of the wave operator onto the spheroidal harmonics yields a second-order ODE for a single variable. From this equation we then derive the effective potential for perturbations. 
\item Section~\ref{sec:QNFRBH} is the core of our paper. We provide a detailed discussion of the corrections to the QNM frequencies of rotating black holes in the theories \req{Action} and \req{eq:quarticL}. 
\item  Assuming a similarity between scalar and gravitational QNMs, in Section~\ref{sec:obsdev} we derive some observational consequences of the deviations of the quasinormal frequencies with respect to the Kerr values and we estimate the bounds on the higher-derivative corrections that can be set with future GW measurements. 
\item We conclude in Section~\ref{sec:con}, where we also comment on the possible analogy between scalar and gravitational QNMs.
\end{itemize} 
Finally, we also include several appendices that contain some technical results.

\section{Higher-derivative effective action}\label{sec:effectiveaction}
We adopt an effective field theory approach and consider the most general higher-derivative corrections to the Einstein-Hilbert action of General Relativity (GR) that are compatible with the symmetries of the theory. These follow from an action of the following form, 
\begin{equation}
S=\frac{1}{16 \pi G}\int d^4x\sqrt{|g|}\left[R+\sum_{n=2}^{\infty}\ell^{2n-2}\mathcal{L}_{(n)}\right]\, ,
\end{equation}
where $\mathcal{L}_{(n)}$ represents the most general diff-invariant Lagrangian containing $2n$ derivatives of the metric, while $\ell$ is a parameter with units of length that controls the overall scale of the corrections. The Lagrangians $\mathcal{L}_{(n)}$ are in general sums over independent curvature invariants weighted by arbitrary couplings $\alpha^{i}_{(n)}$. Inspired by string theory we take the latter to be dynamical scalar fields. 
This family of theories was analyzed in \cite{Cano:2019ore}, where it was shown that the leading corrections to vacuum GR solutions appear at order $\ell^4$ in the metric. They can be studied by means of the following effective action

\begin{equation}\label{Action}
\begin{aligned}
S=&\frac{1}{16\pi G}\int d^4x\sqrt{|g|}\bigg\{R+\alpha_{1} \phi_{1} \ell^2\mathcal{X}_{4}+\alpha_{2}\left(\phi_2 \cos\theta_{m}+\phi_{1}\sin\theta_{m}\right) \ell^2 R_{\mu\nu\rho\sigma} {\tilde R}^{\mu\nu\rho\sigma}\\\
&+\lambda_{\rm ev}\ell^4\tensor{R}{_{\mu\nu }^{\rho\sigma}}\tensor{R}{_{\rho\sigma }^{\delta\gamma }}\tensor{R}{_{\delta\gamma }^{\mu\nu }}+\lambda_{\rm odd}\ell^4\tensor{R}{_{\mu\nu }^{\rho\sigma}}\tensor{R}{_{\rho\sigma }^{\delta\gamma }} \tensor{\tilde R}{_{\delta\gamma }^{\mu\nu }}-\frac{1}{2}(\partial\phi_{1})^2-\frac{1}{2}(\partial\phi_{2})^2\bigg\}\, ,
\end{aligned}
\end{equation}
where  $\mathcal{X}_{4}=R_{\mu\nu\rho\sigma} R^{\mu\nu\rho\sigma}-4R_{\mu\nu}R^{\mu\nu}+R^2$ is the Gauss-Bonnet density, while ${\tilde R}^{\mu\nu\rho\sigma}=\frac{1}{2}\epsilon^{\mu\nu\alpha\beta}\tensor{R}{_{\alpha\beta}^{\rho\sigma}}$ is the dual Riemann tensor. Here we have assumed that the scalar fields  $\phi_{1}$, $\phi_{2}$ are massless. The inclusion of a mass term for the scalars greatly complicates the analysis of the solutions of this theory. In addition, one might consider more general coupling functions $f_i(\phi_1,\phi_2)$ instead of the shift-symmetric case we are assuming. However, at weak coupling the scalars will be of order $\ell^2$ with respect to their vacuum values, and hence it is enough to take the linear expansion of $f_i(\phi_1,\phi_2)$ in order to study the leading corrections to GR solutions. Ignoring the constant terms (which are topological) and performing a rotation of the scalars, it is always possible to express the action as in \req{Action} --- see \cite{Cano:2019ore}. 
A priori, one might wish to add other terms to the effective theory \eqref{Action}, such as $RR_{\mu\nu\rho\sigma} R^{\mu\nu\rho\sigma}$. However, it turns out that all of these can be removed by means of field redefinitions. The physical properties of black holes, such as the quasinormal frequencies of interest here, are invariant under changes of frame. Thus \eqref{Action} preserves generality.

Conversely one might wish to further constrain the theory \req{Action}. To preserve parity, for example, one should set $\lambda_{\rm odd}=\theta_m=0$. To eliminate non-minimally coupled scalars one can set $\alpha_1=\alpha_2=0$, which in practice is equivalent to truncating the scalars. One might also want to set the cubic terms to zero, since it has been argued that these give rise to causality violations unless the action is supplemented with an infinite tower of higher-spin modes \cite{Camanho:2014apa}. 

Finally, if all corrections in \req{Action} are discarded or ruled out on physical grounds, one would need to consider the $\mathcal{O}(\ell^6)$ corrections. In Ref.~\cite{Endlich:2017tqa} it was shown that these corrections can be written as the following combination of quartic invariants,\footnote{Note that one could consider as well dynamical couplings in this case, nevertheless their effect is subleading in this case. } 
\begin{equation}\label{eq:quarticL}
S_{(4)}=\frac{\ell^6}{16\pi G}\int d^4x\sqrt{|g|}\left\{\epsilon_1 \mathcal{C}^2+\epsilon_2\tilde{\mathcal{C}}^2+\epsilon_3\mathcal{C}\tilde{\mathcal{C}}\right\}\, ,
\end{equation}
where 
\begin{equation}
\mathcal{C}=R_{\mu\nu\rho\sigma} R^{\mu\nu\rho\sigma}\, ,\quad \tilde{\mathcal{C}}=R_{\mu\nu\rho\sigma} \tilde{R}^{\mu\nu\rho\sigma}\, .
\end{equation}
This theory was proposed in \cite{Endlich:2017tqa} as an effective field theory extension of GR that is particularly suitable to be tested using gravitational waves since it involves no new degrees of freedom. Black hole solutions of the theory \req{eq:quarticL} were studied in \cite{Cardoso:2018ptl}. The dynamics of inspiraling black hole binaries and the LIGO/Virgo observational constraints on corrections of the form \req{eq:quarticL} were studied in \cite{Sennett:2019bpc}.

For both theoretical and phenomenological reasons it is clearly of interest to carry out the analysis of the quasinormal modes of static and rotating black holes for both sets of corrections ($\mathcal{O}(\ell^4)$ and $\mathcal{O}(\ell^6)$). Due to limitations of space and time, in this paper we focus primarily on the corrections included in \req{Action}, since they dominate if present. However, for the sake of completeness and in light of further developments along these lines, we also provide a (less detailed) analysis of the corrections to the quasinormal modes induced by the quartic terms in \req{eq:quarticL}. Furthermore, the ancillary files that we include with this paper allow one to perform the analysis of the corrections to the Kerr metric and to the quasinormal frequencies for general higher-derivative theories. 

\section{Rotating black holes in higher-derivative gravity}\label{sec:corrKerr}
In this section we review the equations of motion of the theories \req{Action} and \req{eq:quarticL} and how to find rotating black hole solutions as perturbative corrections to the Kerr metric. These solutions were studied in Refs.~\cite{Cano:2019ore} and \cite{Cardoso:2018ptl}, respectively, although spinning black holes in particular theories captured by \req{Action} (\textit{e.g.}, Einstein-scalar-Gauss-Bonnet gravity \cite{Kanti:1995vq,Alexeev:1996vs,Torii:1996yi} or dynamical Chern-Simons gravity \cite{Alexander:2009tp}) have been largely studied in the literature using different approaches \cite{Yunes:2009hc,Pani:2011gy,Kleihaus:2011tg,Yagi:2012ya,Ayzenberg:2014aka,Maselli:2015tta,Kleihaus:2015aje,Delsate:2018ome,Reall:2019sah,Burger:2019wkq,Adair:2020vso}. 

The equations of motion arising from \req{Action} can be written as 
\begin{eqnarray}
\label{EinsteinEOM}
G_{\mu\nu}&=&T^{\text{scalars}}_{\mu\nu}+T^{\text{cubic}}_{\mu\nu}\ ,\\
\label{scalarEOM1}
\nabla^2 \phi_1&=&-\alpha_1 \ell^2 R_{\mu\nu\rho\sigma}R^{\mu\nu\rho\sigma} -\alpha_2 \ell^2\sin\theta_m  R_{\mu\nu\rho\sigma}\tilde R^{\mu\nu\rho\sigma}\ , \\
\label{scalarEOM2}
\nabla^2 \phi_2&=&-\alpha_2 \ell^2  \cos \theta_m  R_{\mu\nu\rho\sigma}\tilde R^{\mu\nu\rho\sigma} \ ,
\end{eqnarray}
where the effective energy-momentum tensors appearing in the left-hand-side of Einstein's equation are given by

\begin{equation}
\begin{aligned}
T^{\text{scalars}}_{\mu\nu}=&-\alpha_1 \ell^2g_{\nu\lambda}\delta^{\lambda \sigma \alpha\beta}_{\mu\rho\gamma\delta}  R^{\gamma\delta}{}_{\alpha\beta} \nabla^\rho\nabla_\sigma \phi_1+4\alpha_2 \ell^2  \nabla^\rho\nabla^\sigma\left[\tilde R_{\rho(\mu\nu)\sigma}\,\left(\cos \theta_m \phi_2+\sin \theta_m \phi_1\right)\right] \\
&+\frac{1}{2}\left[\partial_\mu \phi_1 \partial_\nu \phi_1-\frac{1}{2}g_{\mu\nu} \left(\partial \phi_1\right)^2\right]+\frac{1}{2}\left[\partial_\mu \phi_2 \partial_\nu \phi_2-\frac{1}{2}g_{\mu\nu} \left(\partial \phi_2\right)^2\right] \ ,
\end{aligned}
\end{equation}
and 

\begin{equation}
\begin{aligned}
T^{\text{cubic}}_{\mu\nu}=&\lambda_{\text{ev}}\ell^4\left[3 R_\mu{}^{\sigma \alpha \beta} R_{\alpha\beta}{}^{\rho\lambda} R_{\rho\lambda \sigma \nu}+\frac{1}{2}g_{\mu\nu}\tensor{R}{_{\alpha\beta }^{\rho\sigma}}\tensor{R}{_{\rho\sigma }^{\delta\gamma }}\tensor{R}{_{\delta\gamma }^{\alpha\beta }}-6 \nabla^\alpha\nabla^\beta\left(R_{\mu\alpha\rho \lambda}R_{\nu\beta}{}^{\rho\lambda}\right)\right]\\
&+\lambda_{\text{odd}}\ell^4\bigg[-\frac{3}{2}\tensor{R}{_{\mu}^{\rho\alpha\beta}}\tensor{R}{_{\alpha\beta\sigma\lambda}}\tensor{\tilde R}{_{\nu\rho}^{\sigma\lambda}}-\frac{3}{2}\tensor{R}{_{\mu}^{\rho\alpha\beta}}\tensor{R}{_{\nu\rho\sigma\lambda}}\tensor{\tilde R}{_{\alpha\beta}^{\sigma\lambda}}+\frac{1}{2}g_{\mu\nu}\tensor{R}{_{\mu\nu }^{\rho\sigma}}\tensor{R}{_{\rho\sigma }^{\delta\gamma }} \tensor{\tilde R}{_{\delta\gamma }^{\mu\nu }}\\
&+3\nabla^{\alpha}\nabla^{\beta}\left(\tensor{R}{_{\mu\alpha\sigma\lambda}}\tensor{\tilde R}{_{\nu\beta}^{\sigma\lambda}}+\tensor{R}{_{\nu\beta\sigma\lambda}}\tensor{\tilde R}{_{\mu\alpha}^{\sigma\lambda}}\right)\bigg] \, .
\end{aligned}
\end{equation}
We note that  the scalars acquire a non-trivial value of order $\ell^2$ on account of their non-minimal coupling to the curvature. On the other hand, both $T^{\text{scalars}}_{\mu\nu}$ and $T^{\text{cubic}}_{\mu\nu}$ are of order $\ell^4$, and hence the metric can expanded as $g_{\mu\nu}=g_{\mu\nu}^{(0)}+\ell^4 g_{\mu\nu}^{(4)}$, where $g_{\mu\nu}^{(0)}$ is a solution of the zeroth-order Einstein's equations and $ g_{\mu\nu}^{(4)}$ is a perturbative correction. Regarding the quartic terms \req{eq:quarticL}, their contribution to the Einstein's equations can be recasted as an effective energy-momentum tensor that reads
\begin{align}
&T^{\rm quartic}_{\mu\nu}=-\ell^6R_{\nu\alpha\beta\sigma}\left[\tensor{R}{_{\mu}^{\alpha\beta\sigma}}\left(2\epsilon_1\mathcal{C}+\epsilon_3\tilde{\mathcal{C}}\right)-\tensor{\tilde{R}}{_{\mu}^{\alpha\beta\sigma}}\left(2\epsilon_2\tilde{\mathcal{C}}+\epsilon_3\mathcal{C}\right)\right]\\ \notag
&+\frac{\ell^6}{2}g_{\mu\nu}\left(\epsilon_1 \mathcal{C}^2+\epsilon_2\tilde{\mathcal{C}}^2+\epsilon_3\mathcal{C}\tilde{\mathcal{C}}\right)-2\ell^6\nabla^{\alpha}\nabla^{\beta}\left[\tensor{R}{_{\mu\alpha\nu\beta}}\left(2\epsilon_1\mathcal{C}+\epsilon_3\tilde{\mathcal{C}}\right)-\tensor{\tilde{R}}{_{\mu\alpha\nu\beta}}\left(2\epsilon_2\tilde{\mathcal{C}}+\epsilon_3\mathcal{C}\right)\right]\, .
\end{align}
In this case, the correction to the metric is of order $\ell^6$, so that  $g_{\mu\nu}=g_{\mu\nu}^{(0)}+\ell^6 g_{\mu\nu}^{(6)}$.
Now we are going to assume that $g_{\mu\nu}^{(0)}$ is the Kerr metric and our goal is to compute the corresponding corrections to this solution. 

First, it is necessary to write down an appropriate metric ansatz in order to describe the deviations to the Kerr geometry.  Following \cite{Cano:2019ore} we write
\begin{equation}\label{rotatingmetric}
\begin{aligned}
ds^2=&-\left(1-\frac{2 M \rho}{\Sigma}-H_1\right)dt^2-\left(1+H_2\right)\frac{4 M a \rho (1-x^2)}{\Sigma}dtd\phi+\left(1+H_3\right)\Sigma\left(\frac{d\rho^2}{\Delta}+\frac{dx^2}{1-x^2}\right)\\
&+\left(1+H_4\right)\left(\rho^2+a^2+\frac{2 M  \rho a^2(1-x^2)}{\Sigma}\right)(1-x^2)d\phi^2\, ,
\end{aligned}
\end{equation}
where $\Sigma$ and $\Delta$ are given by
\begin{equation}
\Sigma=\rho^2+a^2x^2\, ,\qquad \Delta=\rho^2-2M\rho+a^2\, .
\end{equation}
and where $H_i=H_i(\rho,x)$ are four functions characterizing the corrections. For $H_i=0$ the metric above reduces to Kerr, and since we are only interested in perturbative corrections, the functions $H_i$ are determined from the linearized Einstein equations,
\begin{equation}
G_{\mu\nu}^{L}\Big|_{H_i}=T_{\mu\nu}^{\rm eff}\Big|_{g_{\mu\nu}^{(0)}}\, ,
\end{equation}
where $G_{\mu\nu}^{L}$ is the linearized Einstein tensor on the Kerr background, and the right-hand side is one of the effective energy-momentum tensors shown above, evaluated on the zeroth-order metric. 

The equations of motion do not allow for a fully analytic solution for arbitrary angular momentum. It is nevertheless possible to expand the functions $\phi_{1,2}$, $H_i$ in a series in the spin parameter $\chi=a/M$. This yields
\begin{equation}\label{chiexp}
\phi_{1,2}=\sum_{n=0}^{\infty}\chi^n\sum_{p=0}^{n}\sum_{k=0}^{k_{\rm max}}\phi_{1,2}^{(n,p,k)}x^p\rho^{-k}\, ,\quad  H_{i}=\sum_{n=0}^{\infty}\chi^n\sum_{p=0}^{n}\sum_{k=0}^{k_{\rm max}}H_{i}^{(n,p,k)}x^p\rho^{-k}\, ,\quad i=1,2,3,4\, .
\end{equation}
That is, every term in the $\chi$-expansion is a polynomial in $x$ and in $1/\rho$, 
where $\phi_{1,2}^{(n,p,k)}$, $H_{i}^{(n,p,k)}$ are constant coefficients and the value of $k_{\rm max}$ depends on $n$ and $p$. These coefficients are determined by the equations of motion, except for a few ones that have to be fixed by the boundary conditions. For the scalars we can set
\begin{equation}
\hskip2cm \phi_{1}^{(n,0,0)}=\phi_{2}^{(n,0,0)}=0\, ,\quad n=0,1,2,\ldots\, ,
\end{equation}
so that $\phi_1$ and $\phi_2$ vanish asymptotically (any other value is equivalent because the theory \req{Action} is shift-symmetric).
For the functions $H_i$ we must impose
\begin{equation}\label{bcond2}
H_1^{(n,0,0)}=0\, ,\quad H_3^{(n,0,0)}=H_4^{(n,0,0)}=-\frac{H_3^{(n, 0,1)}}{M}\, ,\quad H_2^{(n,0,0)}=-\frac{H_3^{(n,0,0)}}{2}\, .
\end{equation}
These conditions guarantee that the solution is asymptotically flat and that the total mass and angular momentum are not ``renormalized'' by the corrections, \textit{i.e.} $M$ and $J=aM$ are still the physical mass and angular momentum. 

For the theories with $\mathcal{O}(\ell^4)$ corrections, the solution up to order $\mathcal{O}(\chi^3)$ is given in the Appendix of \cite{Cano:2019ore}, and a higher-order $\chi$-expansion is available in the ancillary files associated to that paper. For the theories with quartic corrections in \req{eq:quarticL}, the few first terms in the $\chi$-expansion were found in \cite{Cardoso:2018ptl} (possibly with a different ansatz though). With this paper we provide an automatized Mathematica package that computes the functions $H_i$ up to the desired order in $\chi$ for both sets of corrections, and that can easily be extended to other types of higher-derivative terms. 

Ref.~\cite{Cano:2019ore} has studied several properties of the rotating black hole geometries \req{rotatingmetric}. Here we limit ourselves to two observations that are relevant for our analysis. First, the form of the ansatz means that the position of the horizon in terms of the radial coordinate is not modified,\footnote{This is correct as long as  $|\chi|<1$. The extremal case should be considered separately.}
\begin{equation}
\rho_{+}=M\left(1+\sqrt{1-\chi^2}\right)\, .
\end{equation}
Moreover, the horizon is regular, which serves, of course, as important input in the study of the behaviour of fields on this geometry.
Second, the light ring frequencies and radii of this back hole were also computed. This will allow us to compare our results for the QNMs in the eikonal limit with the prediction coming from the geometric optics estimate.

\section{Scalar perturbations of spherically symmetric black holes}\label{sec:sphericQNM}
Our goal is to find the QNM frequencies of a free scalar field perturbation in the black hole backgrounds we have just reviewed. Before considering these scalar perturbations on fully rotating black hole geometries, it is convenient to study first the case of spherically symmetric black holes, whose treatment is much simpler. Besides, the results that we obtain here will serve as a useful test for the rotating case, since we must recover the same quasinormal frequencies in the non-spinning limit.\footnote{Useful since our method in the rotating case differs from the one we employ in the static case.}

When we set $\chi=0$, the black hole metric  \req{rotatingmetric}  becomes
\begin{equation}
ds^2=-f(\rho)dt^2+g(\rho) d\rho^2+r(\rho)^2d\Omega_{(2)}^2\, ,
\end{equation}
where
\begin{align}
f(\rho)&=1-\frac{2 M}{\rho}-H_1\, , \qquad g(\rho)=\frac{1+H_3}{\left(1-\frac{2 M}{\rho}\right)}\, ,\qquad r(\rho)^2=\rho^2\left(1+H_3\right)\, .
\end{align}
Since the purpose of this section is mainly illustrative, we consider the $\mathcal{O}(\ell^4)$ corrections only --- those in Eq.~\req{Action} --- 
in which case the functions $H_1$ and $H_3$ read

\begin{align}
\notag
H_1&=\frac{\alpha_1^2 \ell^4}{M^4}\left(1-\frac{2 M}{\rho}\right)\left(-\frac{208 M^6}{11 \rho ^6}-\frac{1616 M^5}{165 \rho ^5}-\frac{1174 M^4}{231 \rho ^4}+\frac{1868 M^3}{1155 \rho ^3}+\frac{1117 M^2}{1155 \rho^2}+\frac{1117 M}{1155 \rho }\right)\\
& +\frac{\lambda_{\rm ev}\,\ell^4}{M^4}\left(1-\frac{2 M}{\rho}\right)\left(\frac{24 M^6}{11 \rho ^6}+\frac{40 M^5}{33 \rho ^5}+\frac{160 M^4}{231 \rho ^4}+\frac{32 M^3}{77 \rho ^3}+\frac{64 M^2}{231 \rho ^2}+\frac{64 M}{231
	\rho }\right)\\
\notag
H_3&=\frac{\alpha_1^2 \ell^4}{M^4}\left(-\frac{368 M^6}{33 \rho ^6}-\frac{1168 M^5}{165 \rho ^5}-\frac{1102M^4}{231 \rho ^4}-\frac{404M^3}{1155 \rho ^3}-\frac{19M^2}{1155 \rho ^2}+\frac{1117 M}{1155  \rho }-\frac{1117}{1155}\right)\\
&+\frac{\lambda_{\rm ev}\,\ell^4}{M^4}\left(-\frac{392M^6}{11 \rho ^6}+\frac{8M^5}{33 \rho ^5}+\frac{40M^4}{231 \rho ^4}+\frac{32M^3}{231 \rho ^3}+\frac{32M^2}{231 \rho ^2}+\frac{64M}{231 \rho }-\frac{64}{231}\right)
\end{align}
We note that the horizon is exactly placed at $\rho=2M$, but the price to pay is that $\rho$ is not the standard radial coordinate measuring the area of spheres, which instead is $r(\rho)$. Observe that these coordinates do not coincide even asymptotically. Indeed, we have
\begin{equation}
r(\rho)=\rho\left(1-\frac{1117 \alpha_1^2 \ell^4}{1155M^4}-\frac{64\lambda_{\rm ev}\,\ell^4}{231 M^4}\right)+\frac{1117 \alpha_1^2 \ell^4}{1155M^4}+\frac{64\lambda_{\rm ev}\,\ell^4}{231 M^4}+\mathcal{O}(\rho^{-1})\, .
\end{equation}
Thus, it is important to bear in mind the distinction between $\rho$ and $r$.  

\subsection{Effective potential}\label{sec:effectiverot}
Now, let us consider a test scalar field satisfying the wave equation in this background,\footnote{In principle this could be one of the scalar fields already appearing in the action \req{Action}. However, since those fields have a non-vanishing background value, one cannot neglect the coupling between the scalar and gravitational perturbations, and hence one would need to solve a coupled system of equations. Thus, one should imagine that \req{scalarWave} corresponds to the approximation in which the geometry is kept fixed, or either, that $\psi$ is a different field from those appearing in \req{Action}.} 
\begin{equation}\label{scalarWave}
\nabla^2\psi=0 \,.
\end{equation}
Writing
\begin{equation}\label{eqexp2}
\psi=\int_{-\infty}^{\infty}d\omega e^{-i\omega t}\sum_{l=0}^{\infty}\sum_{m=-l}^{l}Y_{l}^{m}(\theta,\varphi)\psi_{l,m}(r,\omega)\, ,
\end{equation}
where $Y_{l}^{m}$ are the spherical harmonics, we obtain a radial equation for each component $\psi_{l,m}$,
\begin{equation}\label{psieq1spheric}
\frac{1}{r^2\sqrt{fg}}\frac{d}{d \rho}\left(\sqrt{\frac{f}{g}}r^2\frac{d\psi_{l,m}}{d \rho}\right)+\left(\frac{\omega^2}{f}-\frac{l(l+1)}{r^2}\right)\psi_{l,m}=0\, .
\end{equation}
We can now massage this expression in order to bring it into a more standard form. First, since the amplitude of $\psi_{l,m}$ decays as $\sim 1/r$ at infinity, it is convenient to work with the variable
\begin{equation}
\hat\psi_{l,m}=r \psi_{l,m}\, ,
\end{equation}
Next, switching the tortoise coordinate $\rho_*$, defined as
\begin{equation}\label{drho*drho}
\frac{d\rho_*}{d\rho}=\sqrt{\frac{g}{f}}\, ,
\end{equation}
Eq.~\req{psieq1spheric} takes the canonical form
\begin{equation}\label{eq:perturbeq1}
\frac{d^2 \hat\psi_{l,m}}{d\rho_{*}^2}+\left(\omega^2-V_{l,m}\right)\hat\psi_{l,m}=0\, ,
\end{equation}
where the effective potential $V_{l,m}$ is given by
\begin{equation}
V_{l,m}=\frac{fl(l+1)}{r^2}+\frac{1}{r}\sqrt{\frac{f}{g}}\left(\sqrt{\frac{f}{g}}r'\right)'\, ,
\end{equation}
and where each ``prime'' denotes a derivative with respect to $\rho$. Evaluating this expression and expanding up to first order in $\ell^4$, we get
\begin{align}
\notag
V_{l,m}&=\left(1-\frac{2 M}{\rho}\right)\left(\frac{l (1+l)}{\rho ^2}+\frac{2 M}{\rho ^3}\right)+\frac{\alpha_1^2 \ell^4}{M^6}\left(1-\frac{2 M}{\rho}\right)\bigg\{l (1+l) \bigg(\frac{992 M^8}{33 \rho ^8}+\frac{928 M^7}{55 \rho ^7}\\ \notag
&+\frac{2276 M^6}{231 \rho ^6}-\frac{488 M^5}{385 \rho ^5}-\frac{366 M^4}{385 \rho ^4}-\frac{2234 M^3}{1155
	\rho ^3}+\frac{1117 M^2}{1155 \rho ^2}\bigg)+\frac{21184 M^9}{33 \rho ^9}+\frac{6208 M^8}{165 \rho ^8}\\ \notag&+\frac{472 M^7}{21 \rho ^7}-\frac{59504 M^6}{1155 \rho
	^6}-\frac{476 M^5}{165 \rho ^5}-\frac{1348 M^4}{231 \rho ^4}+\frac{1117 M^3}{385 \rho ^3}\bigg\}+\frac{\lambda_{\rm ev}\,\ell^4}{M^6}\left(1-\frac{2 M}{\rho}\right)\\ \notag
&\bigg\{l (1+l) \bigg(\frac{368 M^8}{11 \rho ^8}-\frac{16 M^7}{11 \rho ^7}-\frac{200 M^6}{231 \rho ^6}-\frac{128 M^5}{231 \rho ^5}-\frac{32 M^4}{77 \rho ^4}-\frac{128 M^3}{231 \rho
	^3}+\frac{64 M^2}{231 \rho ^2}\bigg)\\
&+\frac{17056 M^9}{11 \rho ^9}-\frac{21488 M^8}{33 \rho ^8}-\frac{40 M^7}{21 \rho ^7}-\frac{96 M^6}{77 \rho ^6}-\frac{32 M^5}{33 \rho
	^5}-\frac{320 M^4}{231 \rho ^4}+\frac{64 M^3}{77 \rho ^3}\bigg\}\, ,
\label{eq:sphericpotential}
\end{align}
while integrating \req{drho*drho} we get the tortoise coordinate 
\begin{align}
\notag
\rho_{*}&=\rho+2M\log\left(\frac{\rho}{2M}-1\right)\\ \notag
&-\frac{\alpha_1^2 \ell^4}{M^3}\left[\frac{1117 \rho}{2310 M}+\frac{73}{60} \log \left(1-\frac{2 M}{\rho}\right)+\frac{6719 M}{2310 \rho}+\frac{7451 M^2}{2310
	\rho^2}+\frac{1316 M^3}{495 \rho^3}+\frac{62 M^4}{33 \rho^4}\right]\\
&-\frac{\lambda_{\rm ev}\,\ell^4}{M^3}\left[\frac{32 \rho}{231 M}+\frac{1}{4}\log \left(1-\frac{2M}{\rho}\right)+\frac{109M}{154\rho}+\frac{391M^2}{462\rho^2}+\frac{14M^3}{11 \rho^3}+\frac{23 M^4}{11 \rho^4}\right]\, .
\end{align}

\begin{figure}[t]
	\begin{center}
		\includegraphics[width=0.48\textwidth]{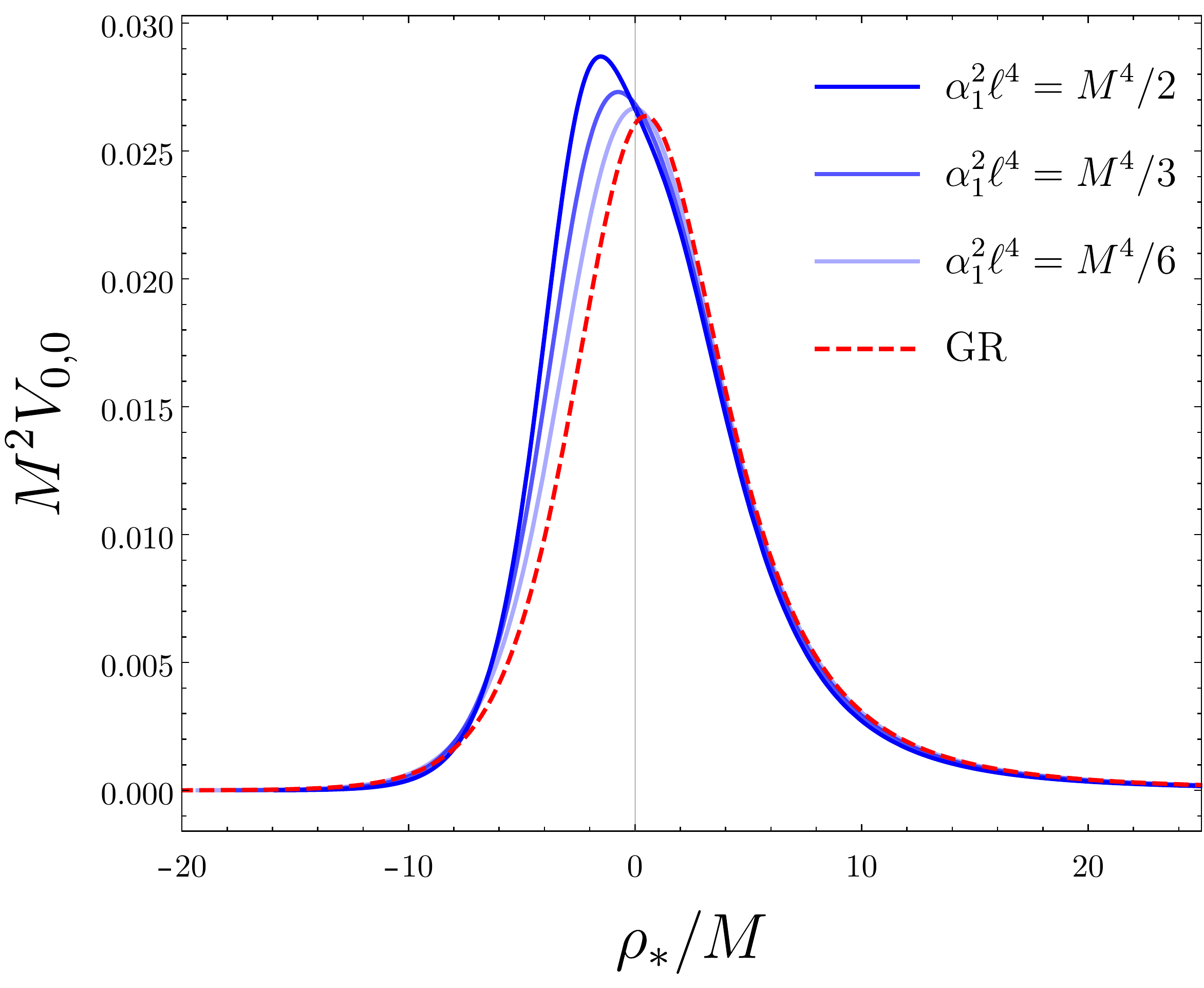} \hskip0.2cm
		\includegraphics[width=0.48\textwidth]{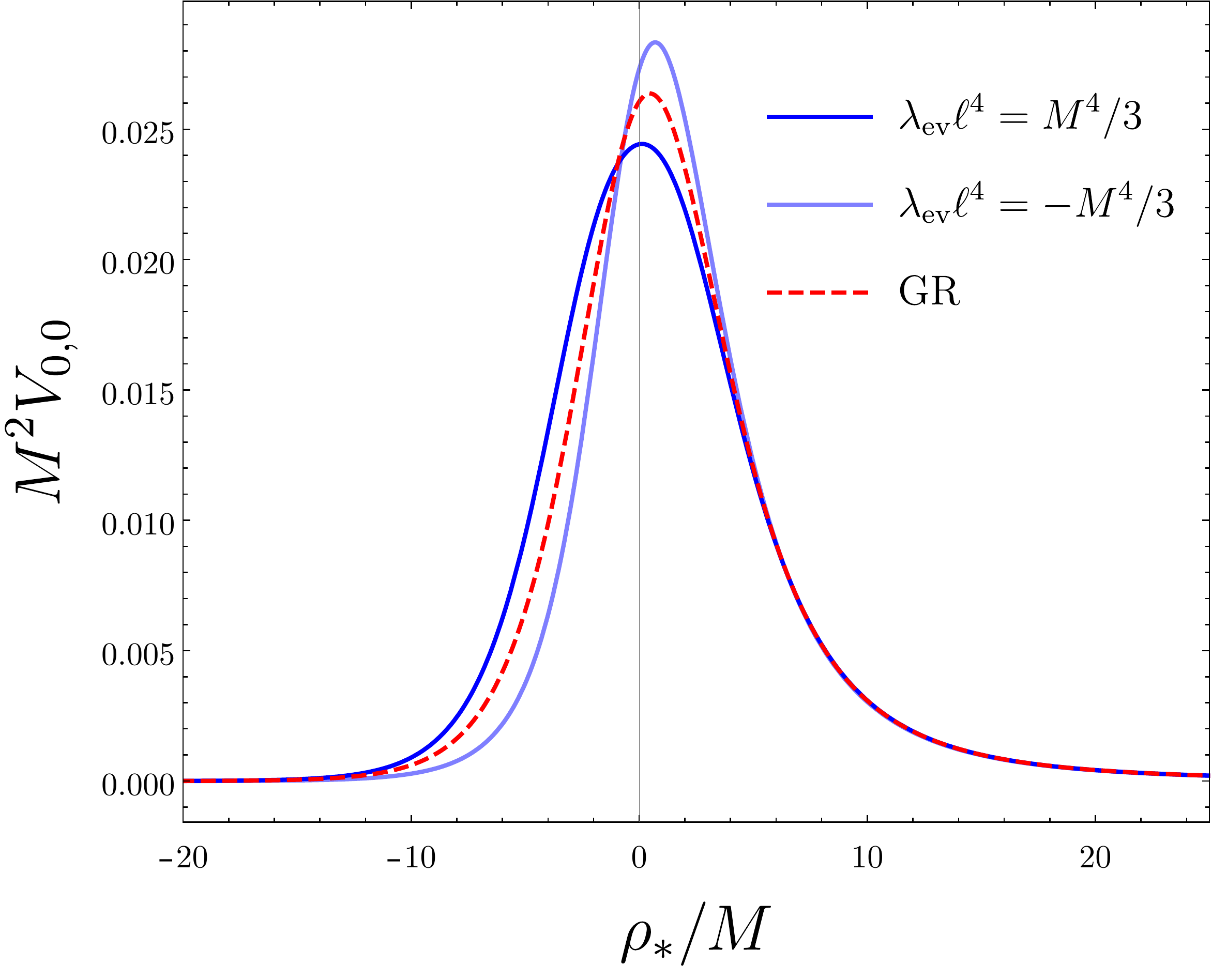}
		\caption{Effective potentials for scalar perturbations with $l=0$ in spherically symmetric black holes. Left: we take various values of $\alpha_1$ while keeping $\lambda_{\rm ev}=0$. Right: we vary $\lambda_{\rm ev}$ while keeping $\alpha_1=0$. In both cases the red dashed line corresponds to the Schwarzschild potential. }
		\label{fig:potentialschi0}
	\end{center}
\end{figure}

It is interesting to plot some of these potentials to see the effect of the corrections. In Fig.~\ref{fig:potentialschi0} we show the potential corresponding to $l=0$ for several values of $\alpha_1$ and $\lambda_{\rm ev}$, and we compare these to the predictions of GR. Thus, we can see that the effect of $\alpha_1$ is to increase the peak of the potential, while for $\lambda_{\rm ev}$ the effect depends on the sign: $\lambda_{\rm ev}<0$ makes the peak higher and sharper and $\lambda_{\rm ev}>0$ has the exact opposite effect.

\subsection{Quasinormal frequencies}
To compute the quasinormal frequencies we must solve Eq.~\req{eq:perturbeq1} with the boundary conditions
\begin{equation}
\hat\psi_{l,m}\propto\begin{cases}
e^{i\omega \rho_{*}}\quad &\text{when} \quad \rho_{*}\rightarrow\infty\\
e^{-i\omega \rho_{*}}\quad &\text{when} \quad \rho_{*}\rightarrow-\infty\, ,
\end{cases} 
\end{equation}
which represent the absence of modes coming from infinity and from the horizon.
Since we are only interested in the linear perturbative corrections to the quasinormal frequencies, in general we can write
\begin{equation}
M\omega_{l,n}= M\omega^{(0)}_{l,n}+\frac{\ell^4}{M^4}\left(\alpha_{1}^{2}\Delta\omega^{(1)}_{l,n}+\lambda_{\rm ev}\Delta\omega^{(\rm ev)}_{l,n}\right)\, ,
\end{equation}
where $\omega^{(0)}_{l,n}$ is the uncorrected value of the frequency and $\Delta\omega^{(1)}_{l,n}$ and $\Delta\omega^{(\rm ev)}_{l,n}$ are dimensionless constants. The index $n$ labels the overtone (starting with $n=0$) and we are already taking into account that the frequencies do not depend on $m$. To find the coefficients $\Delta\omega^{(1)}_{l,n}$ and $\Delta\omega^{(\rm ev)}_{l,n}$ we first obtain the QNM frequencies of the full potential \req{eq:sphericpotential} for several (small) values of $\alpha_{1}$ and $\lambda_{\rm ev}$, and then perform a linear fit which yields these coefficients, along with $\omega^{(0)}_{l,n}$.

We have used several methods to compute the QNM frequencies. For all values of $l$, the fundamental mode can be obtained with excellent precision by performing a numerical integration of the wave equation --- a sketch of this method is explained in  appendix~\ref{app:methods}. The frequencies for $l=0,\ldots,5$ computed in this way are shown in Table~\ref{table:sphericn0}.

\bgroup
\def\arraystretch{1.2}
\begin{table*}[t]
	\centering
	\begin{tabular}{|c|c|c|c|}
		\hline
		$l$ & $M\omega^{(0)}_{l,0}$& $\Delta\omega^{(1)}_{l,0}$&$\Delta\omega^{(\rm ev)}_{l,0}$\\
		\hline
		0&$0.110453- 0.104897\, i$ &$0.0512+ 0.00131\, i$ &$0.0260+ 0.0152\, i $\\
		\hline
			1&$0.292936- 0.0976602\, i$ &$0.0708+ 0.00682\, i$ &$0.0106+ 0.0133\, i $\\
		\hline
			2&$0.483644- 0.0967584\, i$ &$0.1045+ 0.00763\, i$ &$ 0.0128+ 0.00996\, i $\\
		\hline
			3&$0.675366- 0.0964993\, i$ &$0.1411+  0.00781\, i$ &$0.0177+0.00850\, i $\\
		\hline
			4&$0.867415- 0.0963914\, i$ &$0.1786+ 0.00787\, i$ &$0.0231+ 0.00783\, i $\\
		\hline
			5&$1.05961- 0.0963365\, i$ &$0.2166+ 0.00790\, i$ &$0.0284+  0.00749\, i $\\
		\hline
	\end{tabular}
	\caption{Fundamental quasinormal frequencies $(n=0)$ for several values of $l$.}
	\label{table:sphericn0}
\end{table*}
\egroup

However, the numeric integration fails to produce the overtones and we must resort to approximate methods for those, such as the WKB method or the P\"oschl-Teller approximation. 
Unfortunately, we have checked that these methods are unreliable for low values of $l$; they do not even produce the correct values of the coefficients $\Delta\omega^{(1)}_{l,0}$ and $\Delta\omega^{(\rm ev)}_{l,0}$ shown in Table~\ref{table:sphericn0}. For $l\ge 3$ the WKB method starts yielding a more consistent result for these coefficients, however the precision for the overtones is probably smaller.
 Thus, these approximate methods will only work for sufficiently large $l$. In the eikonal limit the WKB method or the P\"oschl-Teller approximation should actually give the exact result for the quasinormal frequencies, and they allow us to obtain an analytic result.  For instance, the P\"oschl-Teller method gives the following approximation to the quasinormal frequencies \cite{Berti:2009kk}:
\begin{equation}
\omega^{\rm PT}_n=\sqrt{V_0-\alpha^2/4}-i\alpha\left(n+\frac{1}{2}\right)\, ,\quad n=0,1,\ldots
\label{eqn:PT}
\end{equation}
where $V_0$ corresponds to the maximum of the potential, $V_0\equiv V(\rho_{*}^{\rm max})$, while \begin{equation}
\alpha=\sqrt{\frac{-V''(\rho_{*}^{\rm max})}{2 V_0}}\, .
\end{equation}
Evaluating these quantities for the potential \req{eq:sphericpotential} and taking $l\rightarrow\infty$, we get 
\begin{equation}
\omega_{l,n}=\frac{(2l+1)}{6 \sqrt{3}M}\left(1+\frac{4397 \alpha_1^2 \ell^4 }{21870 M^4}+\frac{20 \lambda_{\rm ev}\ell^4 }{729 M^4}\right) -i\frac{(2n+1)}{6 \sqrt{3}M}\left(1-\frac{1843 \alpha_1^2 \ell^4 }{21870 M^4}-\frac{52 \lambda_{\rm ev}\ell^4}{729 M^4}\right)\, .
\end{equation}
One can see, for instance, that for $l=5$, $n=0$ this formula yields
\begin{equation}\label{PTstatic55}
\omega_{5,0}=1.0585-0.0962\, i+\frac{\alpha_1^2 \ell^4 }{M^4}\left(0.2128+0.0081\, i\right)+\frac{\lambda_{\rm ev}\ell^4 }{M^4}\left(0.0290+0.0069\, i\right)\, ,
\end{equation}
rather close to the value shown in Table~\ref{table:sphericn0}.
Besides being analytical, this result is interesting because the behaviour of the quasinormal frequencies in the eikonal limit is supposed to be universal, regardless of the spin of the perturbations. Thus, we expect that the formula above also predicts the corrections to the gravitational quasinormal frequencies in the eikonal limit.

\section{Scalar perturbations of rotating black holes}\label{sec:pertrotation}
We now move on to test fields $\psi$ in the rotating black hole backgrounds \req{rotatingmetric}, satisfying
\begin{equation}
\nabla^2\psi=0\,.
\end{equation}
Since we have the Killing vectors $\partial_t$ and $\partial_{\phi}$, we can separate those variables right away by writing 
\begin{equation}\label{eqexp2}
\psi=\int_{-\infty}^{\infty}d\omega \sum_{m=-\infty}^{\infty}e^{i(m\phi-\omega t)}\psi_{m,\omega}(\rho,x)\, ,
\end{equation}
so that we have 
\begin{equation}
\nabla^2\psi= \int_{-\infty}^{\infty}d\omega\sum_{m=-\infty}^{\infty}e^{i(m\phi-\omega t)}\mathcal{D}^2_{m,\omega}\psi_{m,\omega}\,,
\end{equation}
and every component satisfies  an equation of the form
\begin{equation}\label{perteqmw}
\mathcal{D}^2_{m,\omega}\psi_{m,\omega}=0\, .
\end{equation}
The explicit form of the operator $\mathcal{D}^2_{m,\omega}$ in terms of the $H_i$ functions appearing in the metric \req{rotatingmetric} is shown in the appendix \ref{app:laplacian}. 
When we insert the values of these functions given by the solution described in Section~\ref{sec:corrKerr}, we realize that the operator $\mathcal{D}^2_{m,\omega}$ is not separable, \textit{i.e.}, it does not allow for solutions of the form $\psi_{m,\omega}(\rho,x)=R(\rho)X(x)$.
 
\subsection{How to separate a non-separable equation}
Needless to say, the difficulty of solving a non-separable equation is several orders of magnitude higher than the separable spherical case above. In this section we present a resolution of this problem. We start by expanding the perturbation $\psi_{m,\omega}(\rho,x)$ in a basis of angular functions. A natural possibility is to expand $\psi_{m,\omega}$ using the associated Legendre polynomials $P^m_l(x)$,\begin{equation}\label{exppsiP}
\psi_{m,\omega}=\sum_{l=|m|}^{\infty}P^m_l(x) R_{l,m}(\rho)\, .
\end{equation}
Note that when the sum in $m$ in \req{eqexp2} is taken into account, we are summing over the spherical harmonics, $Y_{l}^{m}(\theta,\phi)=e^{i m\phi}P^m_l(\cos\theta)$, which are a basis of functions on the sphere, and hence this is a valid expansion of the solution. Now if we were to project $\mathcal{D}^2_{m,\omega}\psi_{m,\omega}$ with $P^m_{l'}(x)$ then we would obtain an infinite-dimensional system of ODEs for the variables  $R_{l,m}(\rho)$. Due to the smoothness, a numerical solution of a truncated set of these equations would have an exponential convergence. A more efficient decomposition of the scalar perturbation takes into account that $\mathcal{D}^2_{m,\omega}\psi_{m,\omega}$ is \emph{almost} separable, in the sense that its zeroth order part it separable, and the non-separability comes only from the perturbative corrections. Thus, let us write
\begin{equation}
\mathcal{D}^2_{m,\omega}=\mathcal{D}^2_{(0) m,\omega}+\lambda \mathcal{D}^2_{(1) m,\omega}\, ,
\end{equation}
where $\mathcal{D}^2_{(0) m,\omega}$ is the operator corresponding to Kerr and where $\lambda$ is a parameter that controls the corrections (\textit{e.g.}, $\lambda=\ell^4$). Now, it is a well-known fact that the operator $\mathcal{D}^2_{(0) m,\omega}$ is separable, and its angular eigenfunctions are the spheroidal harmonics, $S_{l,m}(x; c)$, where $c=a\omega$. Thus, we might use, instead of the $P^m_l(x)$, the $S_{l,m}(x; c)$ in order to expand the perturbation: 
\begin{equation}\label{exppsiS}
\psi_{m,\omega}=\sum_{l=|m|}^{\infty}S_{l,m}(x; c) R_{l,m}(\rho)\, .
\end{equation}
An important remark here is that, for a fixed $m$, the spheroidal functions $S_{l,m}$ are spanned by the associated Legendre polynomials $P^m_l$. In fact, as we review in appendix \ref{app:spheroidal}, the $S_{l,m}$ can be expanded in a series of the form \cite{meixner1954}
\begin{equation}
S_{l,m}(x; c)=\sum_{n=|m|}^{\infty}c^{2(n-|m|)}a_{n,m}P^m_n(x)\, ,
\label{eqn:spheroidalexpand}
\end{equation}
which converges exponentially fast. Therefore, Eq. \req{exppsiS} is perturbatively equivalent to a rewriting of \req{exppsiP}, and hence a valid expansion, since we are spanning the whole set of functions on the sphere. 

The zeroth-order operator $\mathcal{D}^2_{(0) m,\omega}$ is given by
\begin{align}
\mathcal{D}^2_{(0) m,\omega}\psi&=\frac{1}{\Sigma}\partial_{\rho}\left(\Delta\partial_{\rho}\psi\right)+\frac{1}{\Sigma}\partial_{x}\left((1-x^2)\partial_{x}\psi\right)\\
&-\frac{\psi}{\Delta\Sigma}\bigg[\frac{(\Sigma-2M\rho)m^2}{(1-x^2)}+4Ma\rho m\omega-\left(2M\rho a^2+\frac{\Sigma(\rho^2+a^2)}{(1-x^2)}\right)\omega^2\bigg]\, ,
\end{align}
Acting on the function \req{exppsiS} yields
\begin{equation}\label{D0mw}
\mathcal{D}^2_{(0) m,\omega}\psi_{m,\omega}=\frac{1}{\Sigma}\sum_{l}S_{l,m}(x; c)\left( \mathcal{D}^2_{(0) m,\omega,\rho}-A_{l,m}(c)\right)R_{l,m}(\rho)\, ,
\end{equation}
where we have taken into account the equation satisfied by the spheroidal harmonics,
\begin{equation}
\left(1-x^2\right)S''_{l,m}-2 x S'_{l,m}+ \left(A_{l,m}(c)+c^2 x^2-\frac{m^2}{1-x^2}\right)S_{l,m}=0\, .
\label{eqn:spheroidal}
\end{equation}
where $c\equiv a \omega$ and $A_{l,m}(c)$ are the angular separation constants, corresponding to the eigenvalues of the previous equation. On the other hand, $\mathcal{D}^2_{(0) m,\omega,\rho}$ is the radial operator given by 
\begin{equation}
\mathcal{D}^2_{(0) m,\omega,\rho}R=\frac{d}{d\rho}\left(\Delta\frac{d R}{d\rho}\right)+\frac{R}{\Delta} \left(a^2 m^2-4 a m M \rho  \omega + \omega ^2(\rho ^4+a^2\rho^2+ 2a^2\rho M )\right)\, .
\end{equation}
Then, let us make the following observation: at zeroth-order, the quasinormal modes of Eq.~\req{perteqmw} are given by the functions with a definite value of $m$ and $l$, \textit{i.e.}, each quasinormal mode contains  a single term in the expansion \req{exppsiS} above. Once the corrections are included, we expect that the quasinormal modes will contain all the possible terms in the series due to the non-separability of the equation. However, since they will be given by the zeroth-order ones plus a perturbative correction, the possible ``off-diagonal'' terms will all be of order $\lambda$. Thus, we can single out one term in the $l$ expansion and write 
\begin{equation}
\psi_{m,\omega}=S_{l_0,m}(x; c) R_{l_0,m}(\rho)+\lambda\sum_{l\neq l_0}S_{l,m}(x; c) R_{l,m}(\rho)\, ,
\end{equation}
where we are making explicit that the terms with $l\neq l_0$ are of order $\lambda$. Therefore, inserting this into the equation \req{perteqmw}, using \req{D0mw}, and neglecting quadratic terms in $\lambda$, we have
\begin{align}
(\rho^2+a^2 x^2)\mathcal{D}^2_{m,\omega}\psi_{m,\omega}&=S_{l_0,m}(x; c)\left( \mathcal{D}^2_{(0) m,\omega,\rho}-A_{l_0,m}(c)\right)R_{l_0,m}(\rho)\\
&+\lambda(\rho^2+a^2 x^2) \mathcal{D}^2_{(1) m,\omega}\left(S_{l_0,m}(x; c) R_{l_0,m}(\rho)\right)\\
&+\lambda\sum_{l\neq l_0}S_{l,m}(x; c)\left( \mathcal{D}^2_{(0) m,\omega,\rho}-A_{l,m}(c)\right)R_{l,m}(\rho)+\mathcal{O}(\lambda^2)\, .
\end{align}
Finally, taking in account that the spheroidal harmonics with the same $m$ are orthogonal,\footnote{There are other possible conventions for the spheroidal harmonics but we will use this normalization.} $\langle S_{l,m}|S_{l',m}\rangle=\delta_{ll'}$,
projecting onto $S_{l_0,m}(x; c)$ yields one radial equation that only involves $R_{l_0,m}(\rho)$:
\begin{align}
&\left( \mathcal{D}^2_{(0) m,\omega,\rho}-A_{l_0,m}(c)\right)R_{l_0,m}(\rho)\\
&+\lambda\int_{-1}^{1}dx S_{l,m}(x; c)(\rho^2+a^2 x^2) \mathcal{D}^2_{(1) m,\omega}\left(S_{l_0,m}(x; c) R_{l_0,m}(\rho)\right)=0\, .
\end{align}
This can also be written, without the need to split the operator $\mathcal{D}^2_{m,\omega}$, as
\begin{equation}\label{eq:projectedop}
\int_{-1}^{1}dx S_{l,m}(x; c)(\rho^2+a^2 x^2) \mathcal{D}^2_{m,\omega}\left(S_{l,m}(x; c) R_{l,m}(r)\right)=0\, .
\end{equation}
Expressed in an even more compact way, we have essentially proven that the radial operator obtained from the projection $\mathcal{D}^2_{l,m,\omega}=\langle S_{l,m}|(\rho^2+a^2 x^2) \mathcal{D}^2_{m,\omega} |S_{l,m}\rangle$ actually gives a consistent equation. Thus, the solutions of this equation will provide us with the quasinormal frequencies at first order in $\lambda$. 

Before passing to the next subsection, let us clarify that in this analysis we are not making any assumption yet on $R_{l,m}$. That is, we are not assuming that it can be decomposed as $R_{l,m}^{(0)}+\lambda R_{l,m}^{(1)}$ --- such decomposition is assumed only on the coefficients of the angular expansion. However, another way to derive the corrections to the quasinormal modes consists precisely in decomposing the scalar field as $\psi=\psi^{(0)}+\lambda \psi^{(1)}$, and similarly with the frequency $\omega=\omega^{(0)}+\lambda\omega^{(1)}$, and to perform a perturbative treatment in $\lambda$. In that case, one can derive an explicit formula for $\omega^{(1)}$, analogous to the one obtained in perturbation theory of quantum mechanics: 
\begin{equation}
\omega^{(1)}\sim -\frac{\langle\psi^{(0)}|\mathcal{D}^2_{(1)}|\psi^{(0)}\rangle}{\langle\psi^{(0)}|\partial_{\omega}\mathcal{D}^2_{(0)}|\psi^{(0)}\rangle}
\end{equation}
This result was derived in \cite{Zimmerman:2014aha} and later used in \cite{Mark:2014aja} in order to obtain the quasinormal frequencies of weakly charged Kerr-Newman black holes. However, one difficulty with this method is that the quasinormal modes $\psi^{(0)}$ are not normalizable, and in order to define the inner products one has to extend the integral to the complex plane. 
It would be interesting to explore this method elsewhere in order to compute higher-derivative corrections to the Kerr QNFs. 
Here we will base our analysis on the method that we have explained before, from where we can extract effective potentials for the perturbations, as we now show. 

\subsection{Effective potentials}\label{sec:effpotrot}
To proceed, we must perform the integral in \req{eq:projectedop} in order to get an effective radial equation for the QNMs. 
To do so, we expand the spheroidal functions $S_{l,m}(x;c)$ in a power series in $c$ --- see Appendix \ref{app:spheroidal} --- and then  expand the integrand simultaneously in $c$ and in $\chi$. This is doubly convenient since our solution for the $H_i$ functions in the metric \req{rotatingmetric} is also expressed as a series in $\chi$. Let us also note that, although $c=\chi M\omega$, it is convenient to keep an independent expansion for both variables, since, in principle, $c$ could be large even if $\chi$ is small (for instance, this happens in the eikonal limit). When the integrand is expanded in this way, we are able to perform the integral analytically order by order. The result in all cases is a second-order equation of the form
\begin{equation}\label{radialeqrot1}
A\frac{d^2 R}{d \rho^2}+B\frac{d R}{d \rho}+C R=0\, ,
\end{equation}
where $A$, $B$ and $C$ are functions of $\rho$, and we are omitting the $l,m$ subindices in order to reduce the clutter. Now our task is to bring this equation to the more familiar form of the stationary, one-dimensional Schr\"odinger equation, for which we need to remove the friction term in the equation above. There are two transformations that we can consider: a change of variable for the radial coordinate $y=y(\rho)$, and a rescaling of $R$, so that
\begin{equation}
\frac{dy}{d\rho}=f(\rho)\, ,\qquad R=K(\rho)\varphi \, ,
\end{equation}
for certain functions $K$ and $f$, and where $\varphi$ is the new radial function. When we perform these transformations, the equation \req{radialeqrot1} becomes
\begin{align}\label{radialeqrot2}
A f^2 K \frac{d^2\varphi}{dy^2}+ \frac{d\varphi}{dy}\left[A(f'K+2fK')+BfK\right]+\varphi(CK+BK'+AK'')=0\, ,
\end{align}
where the primes denote derivatives with respect to $\rho$, $f'=df/d\rho$ and so on. Then, in order to remove the term $\frac{d\varphi}{dy}$, we see that  $K$ and $f$ must satisfy
\begin{equation}\label{eq:fK2}
fK^2=e^{-\int d\rho B/A}\, .
\end{equation}
Thus, we have the freedom to fix one of these functions, the other one will then be determined by the relation above. We find that it is convenient to choose the function $f(\rho)$ so that the coordinate $y$ has an appropriate definition. We make the following choice
\begin{equation}
f(\rho)=1+k+\frac{2M}{\rho-\rho_{+}}\, ,\quad \rho_{+}=M\left(1+\sqrt{1-\chi^2}\right)\, ,
\end{equation}
where $k$ is a constant of the order of the corrections ($\mathcal{O}(\ell^4)$ or $\mathcal{O}(\ell^6)$) that we will fix later. Thus, we get the following relation between $y$ and $\rho$,
\begin{equation}
y=\rho(1+k)+2M\log\left(\frac{\rho}{\rho_{+}}-1\right)\, ,
\end{equation}
so that $\rho\rightarrow\rho_{+}$ corresponds to $y\rightarrow-\infty$. Finally, dividing the equation \req{radialeqrot2} by $A f^2 K$, we can write it in the standard form
\begin{equation}\label{radialeqrot2}
 \frac{d^2\varphi}{dy^2}+\left(\omega^2-V(y;\omega)\right)\varphi=0\, ,
\end{equation}
where the potential is given by
\begin{equation}\label{eq:Vyomega}
V(y;\omega)=\omega^2-\frac{(CK+BK'+AK'')}{A f^2 K}\, .
\end{equation}
By expanding everything in a power series of $\chi$ and $c$ and linearly in $\ell^4$ (or $\ell^6$), it is possible to calculate the integral \req{eq:fK2} analytically and to obtain the potential explicitly. 
The constant $k$ is then chosen in a way such that $V(y;\omega)\sim\mathcal{O}(1/y^2)$ when $y\rightarrow\infty$.\footnote{An arbitrary choice of $k$ leads to $V(\infty;\omega)\neq 0$.} We note however that the potential generically tends to a non-zero constant in the opposite limit, $y\rightarrow-\infty$. This could be removed by considering a more complicated choice of coordinate $y$, but at the end of the day we think our choice is more convenient because it allows us to invert easily the relation $\rho(y)$.  In fact, this relationship can be written in terms of the Lambert function $W$ as
\begin{equation}
\rho=\rho_{+}+\frac{2M}{1+ k}W\left(\frac{\rho_{+}(1+ k)}{2M}e^{\frac{y-\rho_{+}(1+ k)}{2M}}\right)\, .
\end{equation} 

As functions of $\rho$, the potentials have lengthy but otherwise simple expressions that can be verified in the ancillary files of this paper. For illustrative purposes let us just show the general structure of these potentials. First, for each value of $l$ and $m$ they can be decomposed as the sum of a zeroth-order part plus linear corrections,
\begin{equation}
V=V^{(0)}+\frac{\ell^4}{M^4}\left(\alpha_1^2V^{(\alpha_1)}+\alpha_2^2V^{(\alpha_2)}+\lambda_{\rm ev}V^{(\rm ev)}\right)\, .
\end{equation}
Then, each of these parts $V^{(i)}$ is decomposed in a series in $\chi$ and $c=M\omega\chi$:
\begin{equation}
V^{(i)}=\sum_{q,p} \chi^{q} c^{p} V^{(i)}_{qp}\, .
\end{equation}
Finally, each of the terms $V^{(i)}_{qp}$ is a finite polynomial in $1/\rho$ where the coefficients are quadratic polynomials in $\omega$, hence they have the form
\begin{equation}
V^{(i)}_{qp}=\frac{1}{M^2}\sum_{j=0}^{j_{\rm max}} \frac{M^j}{\rho^j}\left(v^{(i)}_{qpj}+u^{(i)}_{qpj}M\omega +w^{(i)}_{qpj}(M\omega)^2\right)\, ,
\end{equation}
where $v^{(i)}_{qpj}$, $u^{(i)}_{qpj}$, $w^{(i)}_{qpj}$ are constant, dimensionless coefficients.

\section{Quasinormal frequencies of rotating black holes}\label{sec:QNFRBH}
Since the potentials $V$ do not vanish in the limit $y\rightarrow-\infty$, the quasinormal modes correspond to the solutions of \req{radialeqrot2} satisfying the following boundary conditions
\begin{equation}
\varphi\propto\begin{cases}
e^{i\omega y}\quad &\text{when} \quad y\rightarrow\infty\\
e^{-i\sqrt{\omega^2-V(-\infty;\omega)}y}\quad &\text{when} \quad y\rightarrow-\infty\, .
\end{cases} 
\end{equation}
We realize that if for some $\omega$ we had $\omega^2-V(-\infty;\omega)=0$ we could have a problem. However, it turns out that $V(-\infty;\omega)=p\omega^2$ for a certain constant $p$. One can check that $p<0$ as long as the corrections are small, hence we never run into a zero in the square root. 

\subsection{$\mathcal{O}(\ell^4)$ corrections}
The first thing that we note when computing the effective potentials for the theories in \req{Action} is that they are independent of the parameters $\theta_m$ and $\lambda_{\rm odd}$, so parity-breaking terms in the effective action do not contribute to the scalar QNMs. 
Since we are only interested in the linear corrections to the quasinormal frequencies, we can write
\begin{equation}\label{qnfrotation}
M\omega_{l,m,n}= M\omega^{(0)}_{l,m,n}+\frac{\ell^4}{M^4}\left(\alpha_{1}^{2}\Delta\omega^{(1)}_{l,m,n}+\alpha_{2}^{2}\Delta\omega^{(2)}_{l,m,n}+\lambda_{\rm ev}\Delta\omega^{(\rm ev)}_{l,m,n}\right)\, ,
\end{equation}
where $n$ labels the overtone. The main difference with respect to the static case is that now the degeneracy in $m$ is broken, and that the coefficients $\Delta\omega^{(i)}_{l,m,n}$ are not constant anymore, but functions of the spin $\chi$. In order to obtain the values of these coefficients along with the Kerr quasinormal frequencies $\omega^{(0)}_{l,m,n}$, we compute the quasinormal frequencies of the effective potentials derived in the previous section for various values of the couplings $\alpha_1^2$, $\alpha_2^2$ and $\lambda_{\rm ev}$, and then we perform a linear fit of the formula \req{qnfrotation} to the data. 
Given the complicated form of these potentials, we are not able to employ standard methods valid for the Kerr case, such as Leaver's continued fraction method \cite{Leaver:1985ax,doi:10.1063/1.527130,Leaver:1986gd}, which seems to be the most powerful one --- see \cite{Berti:2009kk} and references therein. We find that the most accurate way to obtain the quasinormal frequencies --- at least when $l$ is not large --- consists in performing a numeric integration of the wave equation \req{radialeqrot2} --- see appendix \ref{app:methods}. This method allows us to compute the fundamental quasinormal frequencies ($n=0$) with high precision --- we have checked that our results for the Kerr QNFs $\omega^{(0)}_{l,m,n}$ agree at least up to 5 digits with the values provided in \cite{WebBerti} --- but obtaining the overtones proves to be much more difficult. Thus, we will focus only on the fundamental modes, which, on the other hand, are the most relevant ones during the ringdown. 
 
Besides the intrinsic errors of the method used to compute the quasinormal frequencies --- which we can always keep small if we are careful enough ---, we have to account for the error committed due to the $\chi$-expansion of the potentials. If the spin is small, we expect that only a few terms suffice to produce an accurate result. The higher we want to go in the spin, the more terms we need to add. For instance, when the corrections are set to zero, we have checked that an expansion up to order $\chi^{14}$ yields an approximation to the Kerr QNFs with an error below 0.1\% for $\chi=0.8$ (and an increasingly better approximation for smaller $\chi$). However, it turns out that in order to correctly capture the corrections to the Kerr QNFs we need more accuracy, and indeed, we observe that for an expansion up to order $\chi^{14}$ the results for the coefficients $\Delta\omega^{(i)}_{l,m,n}$ are only good enough for $\chi\le 0.6 - 0.7$, depending on the case.
We will restrict ourselves to those values of the spin, but it would be desirable --- with higher computational power --- to go to larger values of $\chi$. The analysis will nevertheless break near extremality, $\chi\sim 1$, in whose case one would need to use a different approach in order to obtain the corrections to the Kerr metric. 

\noindent
Let us now present our results.

\subsubsection{Low $l$}
It is expected that the mode with $l=m=n=0$ is the dominant one after the scalar field is perturbed in the vicinity of the horizon, and hence it deserves special attention. This mode is also the one less affected by rotation. In Table~\ref{table:l0m0} we show the values of these quasinormal frequencies for various values of $\chi$.  Together with those values we find, as usual, the symmetric frequencies by changing the sign of the real parts. As a consistency check, we see that the coefficients $\Delta\omega^{(i)}_{0,0,0}$ for $\chi=0$ agree (up to a $1\%$ of discrepancy) with those shown in Table~\ref{table:sphericn0} corresponding to the static case.  This is indeed a good test of our computations, since the potentials used in each case are actually different due to the different choice of ``tortoise coordinate''. 

\bgroup
\def\arraystretch{1.2}
\begin{table*}[t]
	\centering
	\begin{tabular}{|c|c|c|c|c|}
		\hline
		$\chi$ & $M\omega^{(0)}_{0,0,0}$& $\Delta\omega^{(1)}_{0,0,0}$& $\Delta\omega^{(1)}_{0,0,0}$&$\Delta\omega^{(\rm ev)}_{0,0,0}$\\
		\hline
		0  &$0.110454- 0.104897\, i$ &$0.0516+ 0.00135\, i$ &$0+ 0\, i $&$0.0261+ 0.0153\, i $\\
		\hline
		0.1&$0.110532 -0.104801\, i$ &$0.0517+ 0.00161\, i$ &$0.00031+ 0.00000\, i $&$0.0260+ 0.0153\, i $\\
		\hline
		0.2&$0.110767- 0.104512\, i$ &$0.0517+ 0.00239\, i$ &$0.00126+ 0.00000\, i $&$0.0255+ 0.0154\, i $\\
		\hline
		0.3&$0.111157- 0.104008\, i$ &$0.0517+ 0.00374\, i$ &$0.00296+ 0.00006\, i $&$0.0247+ 0.0155\, i $\\
		\hline
		0.4&$0.111699- 0.103253\, i$ &$0.0516+ 0.00570\, i$ &$0.00564+ 0.00023\, i $&$0.0235+ 0.0155\, i $\\
		\hline
		0.5&$0.112381- 0.102183\, i$ &$0.0513+ 0.00837\, i$ &$0.00967+ 0.00070\, i $&$0.0218+ 0.0154\, i $\\
		\hline
		0.6&$0.113172- 0.100698\, i$ &$0.0508+ 0.01183\, i$ &$0.01574+ 0.00180\, i $&$0.0183+ 0.0152\, i $\\
		\hline
	\end{tabular}
	\caption{Fundamental quasinormal frequencies $(n=0)$ for $l=m=0$ and several values of the spin $\chi$. Changing the sign of the real parts produces another quasinormal frequency.}
	\label{table:l0m0}
\end{table*}
\egroup

\begin{figure}[t!]
	\begin{center}
		\includegraphics[width=0.48\textwidth]{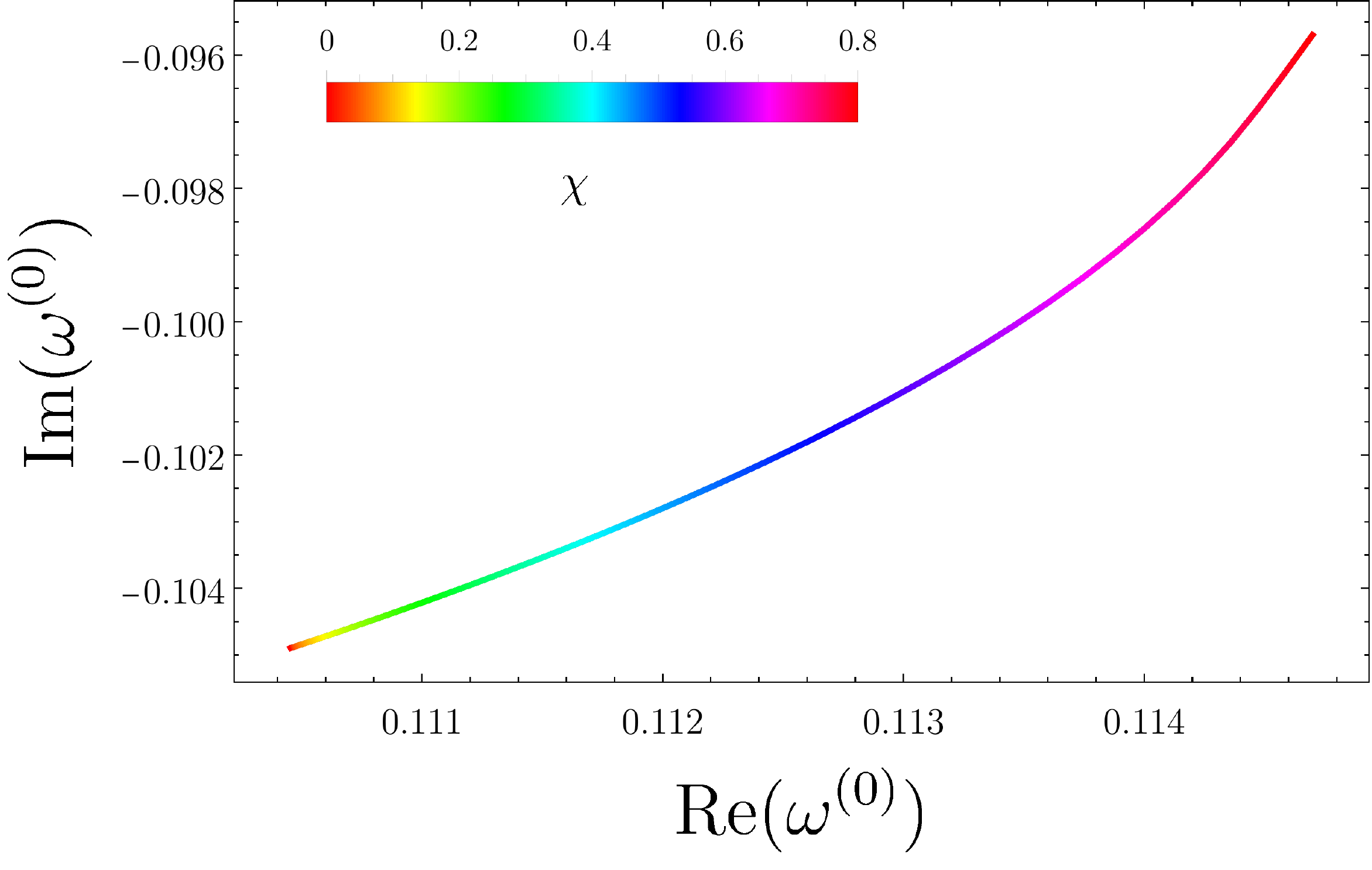} 
	    \includegraphics[width=0.48\textwidth]{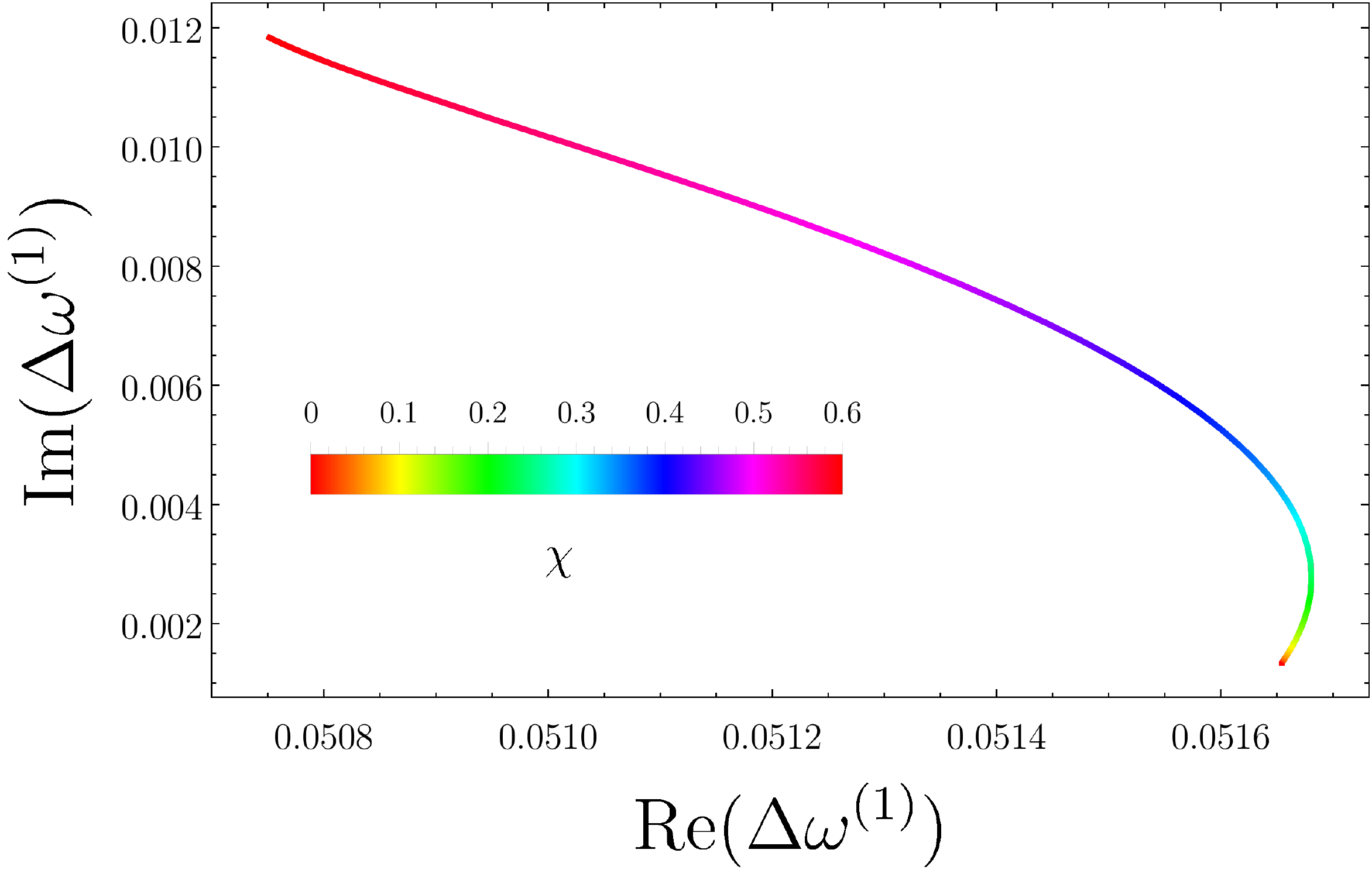}
	    \includegraphics[width=0.48\textwidth]{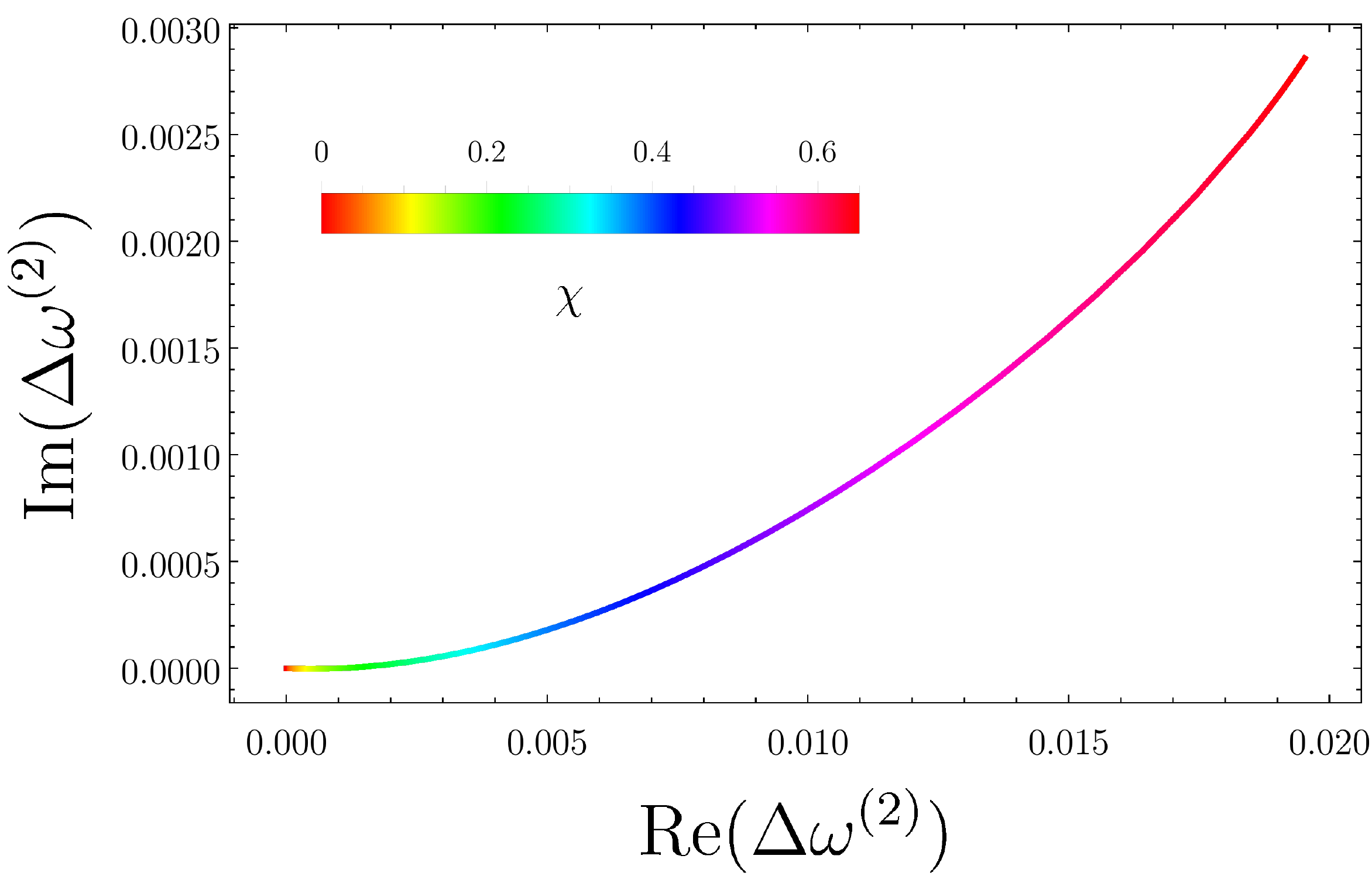}
	    \includegraphics[width=0.48\textwidth]{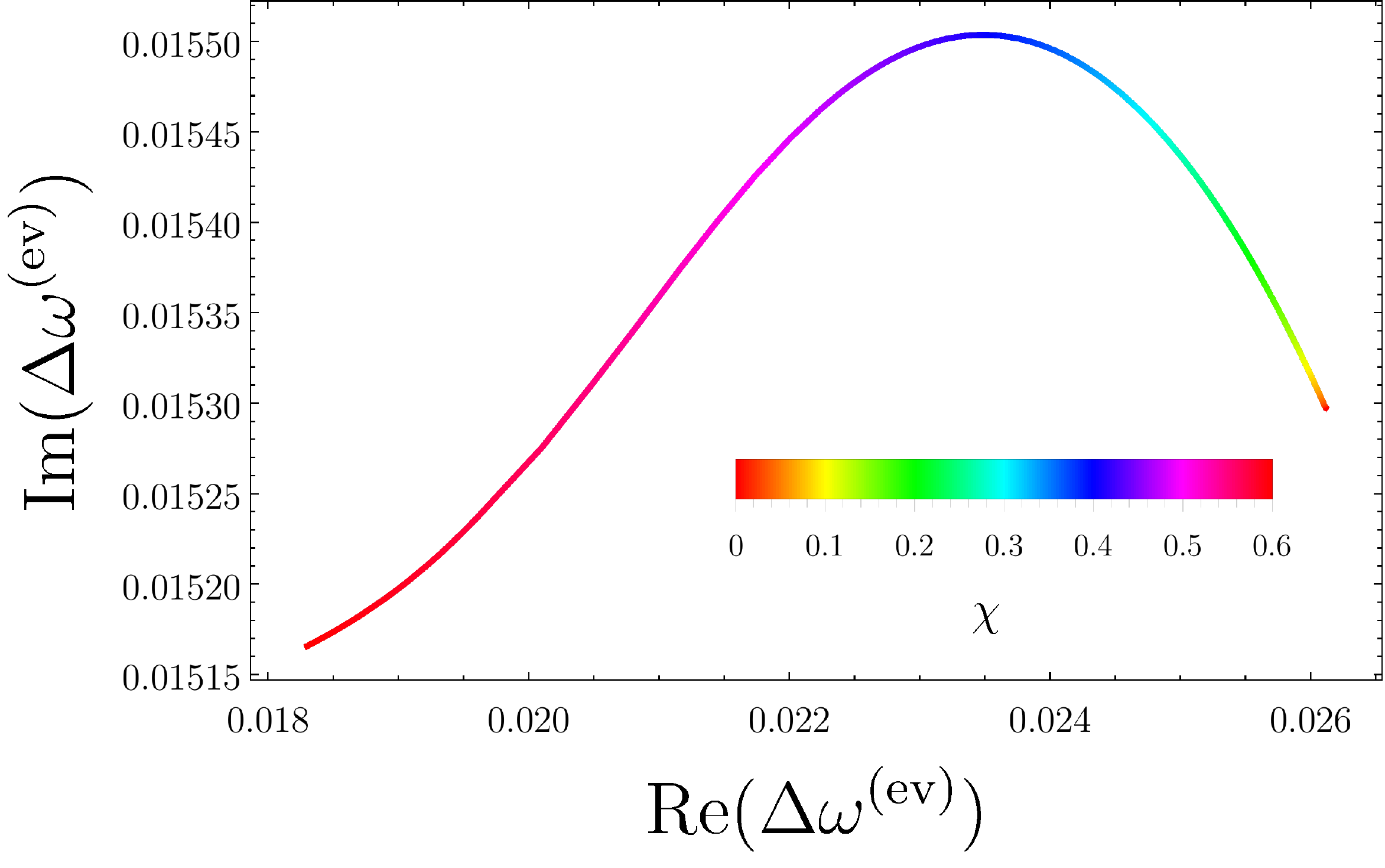}
		\caption{Trajectories in the complex plane of the Kerr quasinormal frequency $\omega^{(0)}_{0,0,0}$ and of the correction coefficients $\Delta\omega^{(i)}_{0,0,0}$. The dependence in the spin is labeled by the color code (note that the maximum spin represented in each case is different).}
		\label{fig:qnfl0m0}
	\end{center}
\end{figure}

In order to understand the behaviour of these frequencies it is most useful to plot them. In Fig.~\ref{fig:qnfl0m0} we show the trajectory in the complex plane of the uncorrected quasinormal frequency $\omega^{(0)}_{0,0,0}$ and of the correction coefficients $\Delta\omega^{(i)}_{0,0,0}$. As we can observe, the zeroth-order value does not vary much as we increase the spin, while the effect can be more important on the corrections. In particular, the correction associated to $\alpha_2$ becomes non-vanishing when $\chi\neq 0$, and both its real and imaginary parts are monotonically increasing with $\chi$.  In the case of the $\alpha_1$ correction, we see that its real part barely changes, while its imaginary part is increased by a factor of 9 when we go from $\chi=0$ to $\chi=0.6$.  As for $\Delta\omega^{(\rm ev)}_{0,0,0}$, its imaginary part remains practically constant, while the the real part does have a significant variation. Thus, depending on the spin the effect of the corrections can be more or less relevant.
Let us also note that, in the case of the corrections coming from the quadratic curvature terms the sign is fixed, since the corrections are proportional to $\alpha_{1,2}^2$. Therefore, we see that they always tend to make the quasinormal modes longer-lived and of higher frequency. If we could extrapolate these results to black holes of small enough masses, this would imply that at some point the imaginary part of $\omega_{0,0,0}$ would become positive, hence we would find an instability. Of course, this would be well beyond the perturbative regime we are considering, so we cannot conclude that such instability will occur. 
On the other hand, the effect of the cubic curvature terms can be the same as for scalars ($\lambda_{\rm ev}>0$) or the opposite one ($\lambda_{\rm ev}<0$).

It is also interesting to fit the numeric results to polynomial functions in $\chi$, so that we can obtain a compact expression for the quasinormal frequencies valid for arbitrary values of the spin. We note the effect of the spin is quadratic, this is, the different quantities behave as $\omega(\chi)=\omega_0+\chi^2\omega_2+\ldots$, when $\chi$ is small --- linear terms in $\chi$ will generically appear when $m\neq 0$ --- so it seems appropriate to fit the results to a polynomial in $\chi^2$.  Taking this into account, we obtain the following fits including three terms

\begin{align}
\notag
M\omega^{(0)}_{0,0,0}=&(0.110452\, -0.104892 i)+(0.007977\, +0.009137 i) \chi ^2-(0.001109\, -0.006895 i) \chi ^4\, ,\\
\notag
\Delta\omega^{(1)}_{0,0,0}=&(0.05166\, +0.001347 i)+(0.001030\, +0.02572 i) \chi ^2-(0.0101\, -0.00949 i) \chi ^4\, ,\\
\notag
\Delta\omega^{(2)}_{0,0,0}=&0.03043\chi ^2+(0.02432\, +0.005049 i) \chi ^4+(0.0350\, +0.0243 i) \chi ^6\, ,\\
\Delta\omega^{(\rm ev)}_{0,0,0}=&(0.02605\, +0.01529 i)-(0.01121\, -0.002620 i) \chi ^2-(0.0257\, +0.00835 i) \chi ^4\, .
\end{align}
In the case of $\Delta\omega^{(2)}_{0,0,0}$ we have forced the constant term to vanish and we have also made explicit that the imaginary part behaves as  $\sim\chi^4$ for small $\chi$.
These functions approximate very well the numeric values of the quasinormal frequencies in the interval $0\le\chi\le 0.6$ (the maximum value of the residuals is of the order of $1\%$), and they probably yield a good estimate for slightly higher values.

The case of $l=0$ is relevant becase, as we remarked, it would correspond to the dominant mode of scalar perturbations. Nevertheless, it is interesting to study also the quasinormal modes for $l>0$, since they develop a rich structure which is not appreciated in the case of $l=0$. In particular,  the frequencies with different $m$ split, and in addition the effect of rotation becomes increasingly relevant for larger $l$ and $m$. Besides, the results for $l>0$ can serve as an approximation to the quasinormal frequencies of higher spin fields. 

\begin{figure}[t!]
	\begin{center}
		\includegraphics[width=0.49\textwidth]{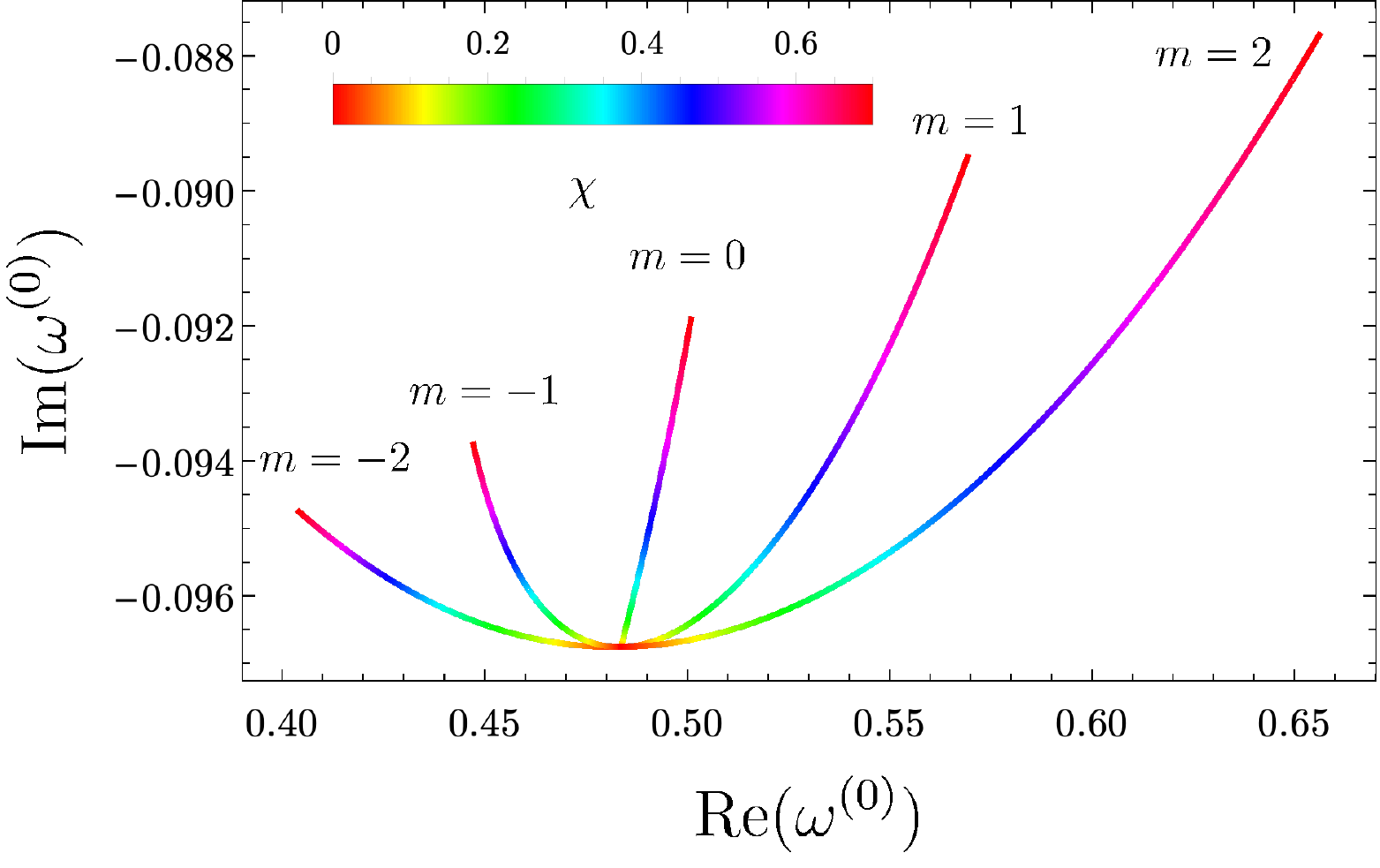} 
	    \includegraphics[width=0.49\textwidth]{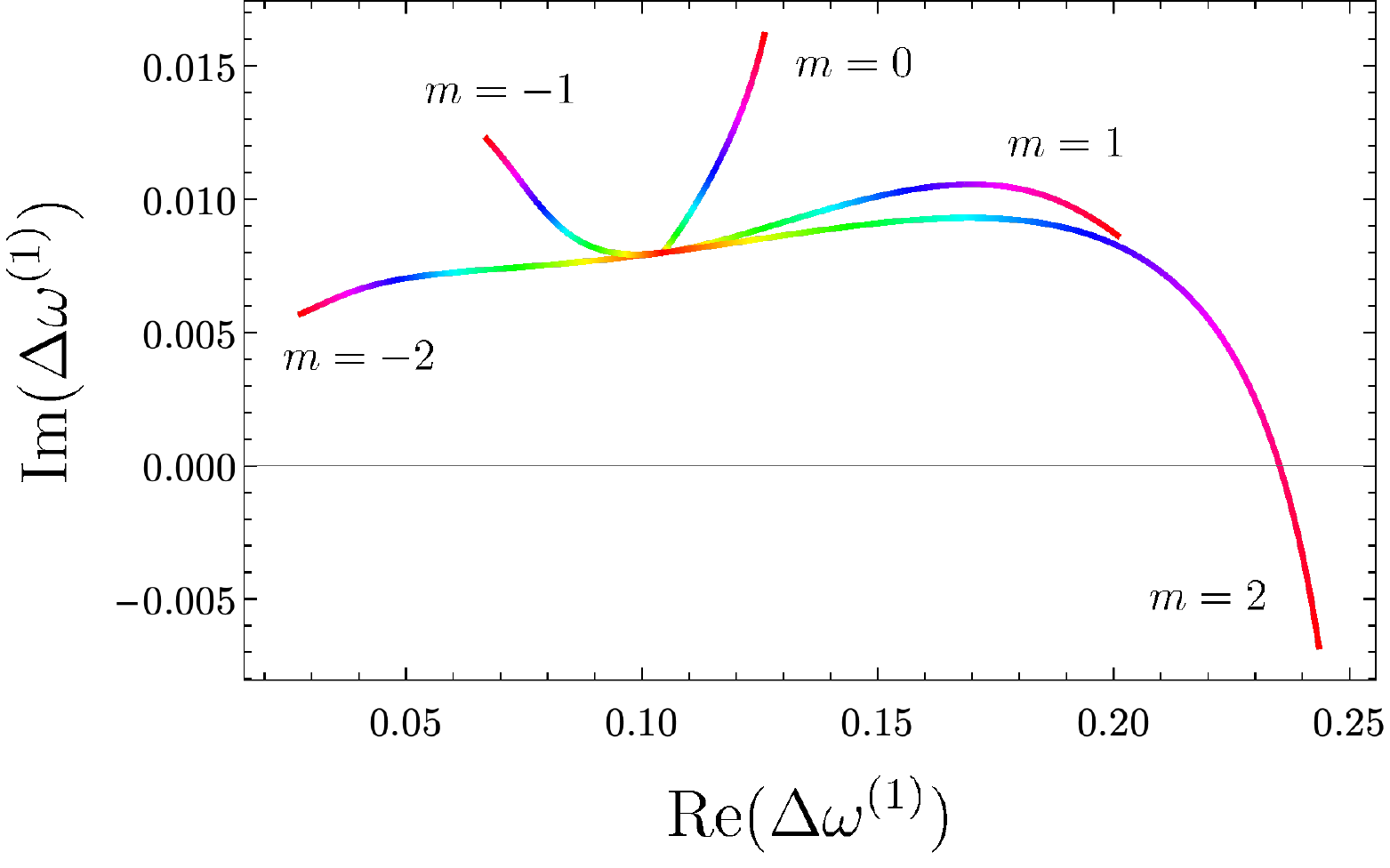}
	    \includegraphics[width=0.49\textwidth]{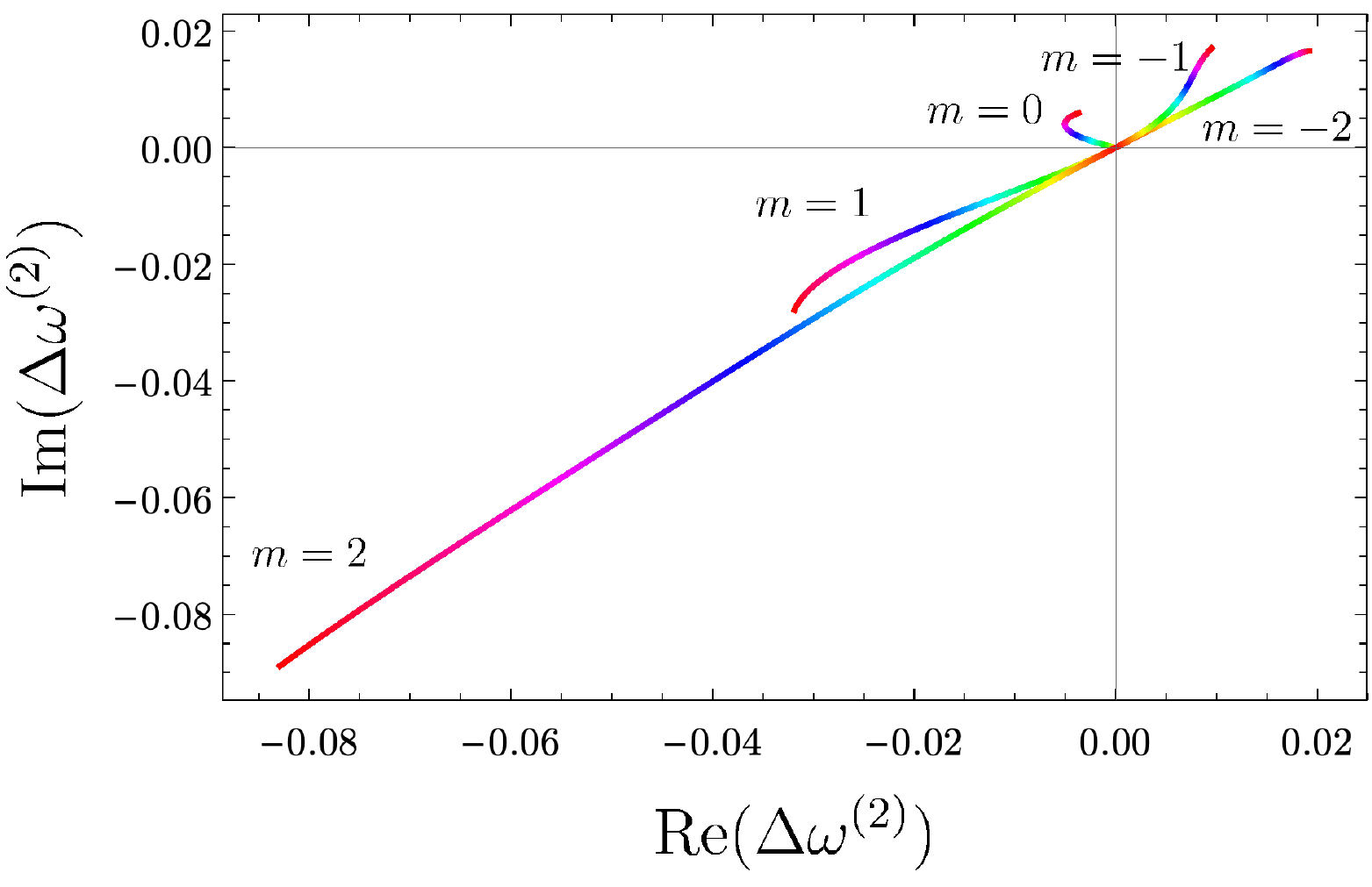}
	    \includegraphics[width=0.49\textwidth]{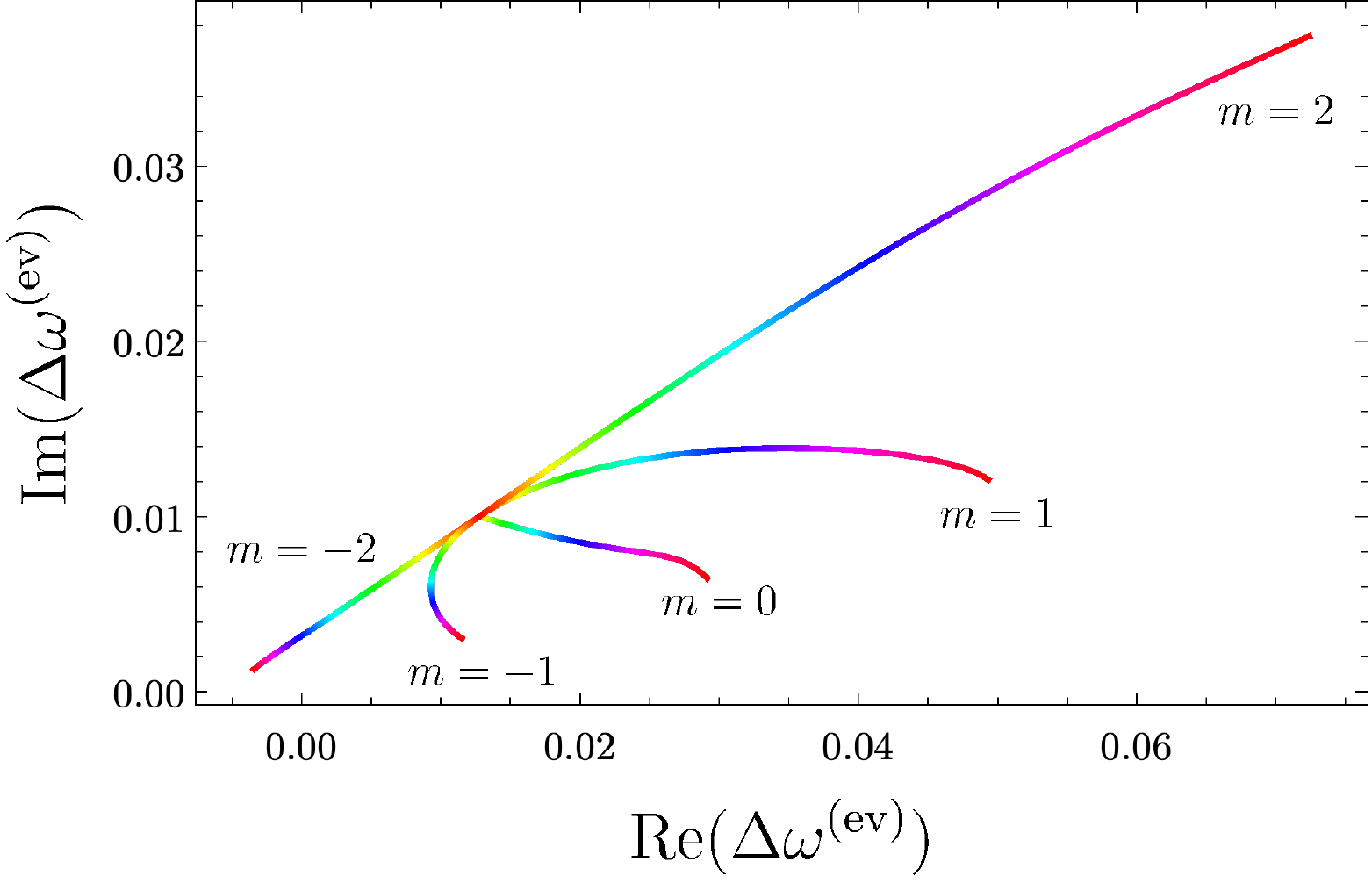}
		\caption{Trajectories in the complex plane of the Kerr quasinormal frequencies $\omega^{(0)}_{l,m,0}$ for $l=2$ and $m=-2,-1,0,1,2$, and of the correction coefficients $\Delta\omega^{(i)}_{l,m,0}$. The dependence in the spin is labeled by the color code.}
		\label{fig:qnfl2}
	\end{center}
\end{figure}

The corrected quasinormal frequencies $\omega_{l,m,n}$ still satisfy the symmetry
\begin{equation}
\omega_{l,m,n}(\chi)=\omega_{l,-m,n}(-\chi)\, .
\end{equation}
This symmetry is expected to be broken in parity-violating theories, but we have seen that such effects never appear in the case of scalar perturbations. On the other hand, as in the Kerr case, there there is a set of frequencies $\omega_{l,m,n}$ with positive real part and another set $\hat{\omega}_{l,m,n}$  with negative real part, but both are related by the simple relation
\begin{equation}
\hat{\omega}_{l,m,n}=-\omega^{*}_{l,-m,n}\, .
\end{equation}
Therefore, we can restrict ourselves to $\chi\ge0$ and to QNFs with positive real part. 
\begin{figure}[t!]
	\begin{center}
		\includegraphics[width=0.49\textwidth]{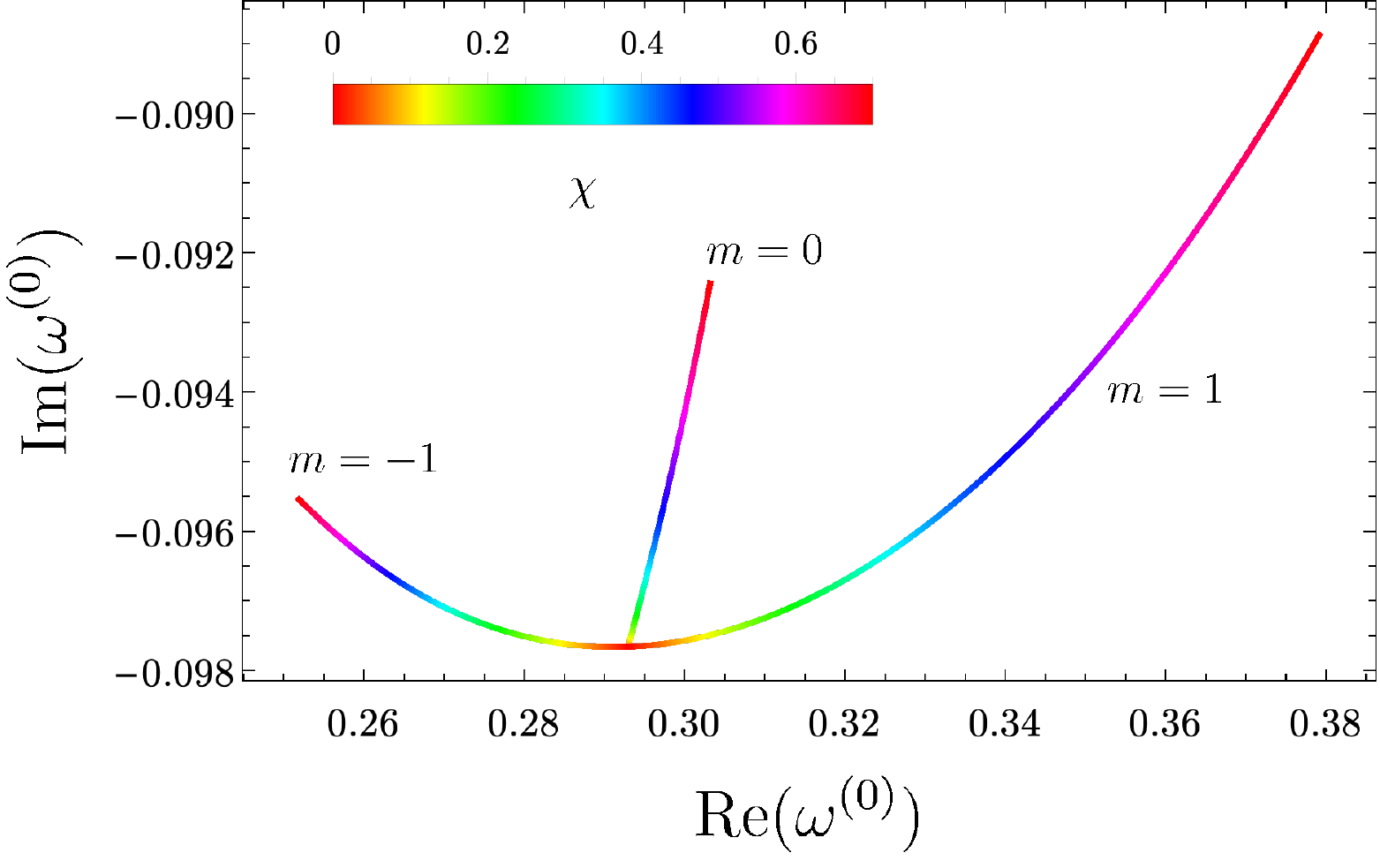} 
	    \includegraphics[width=0.49\textwidth]{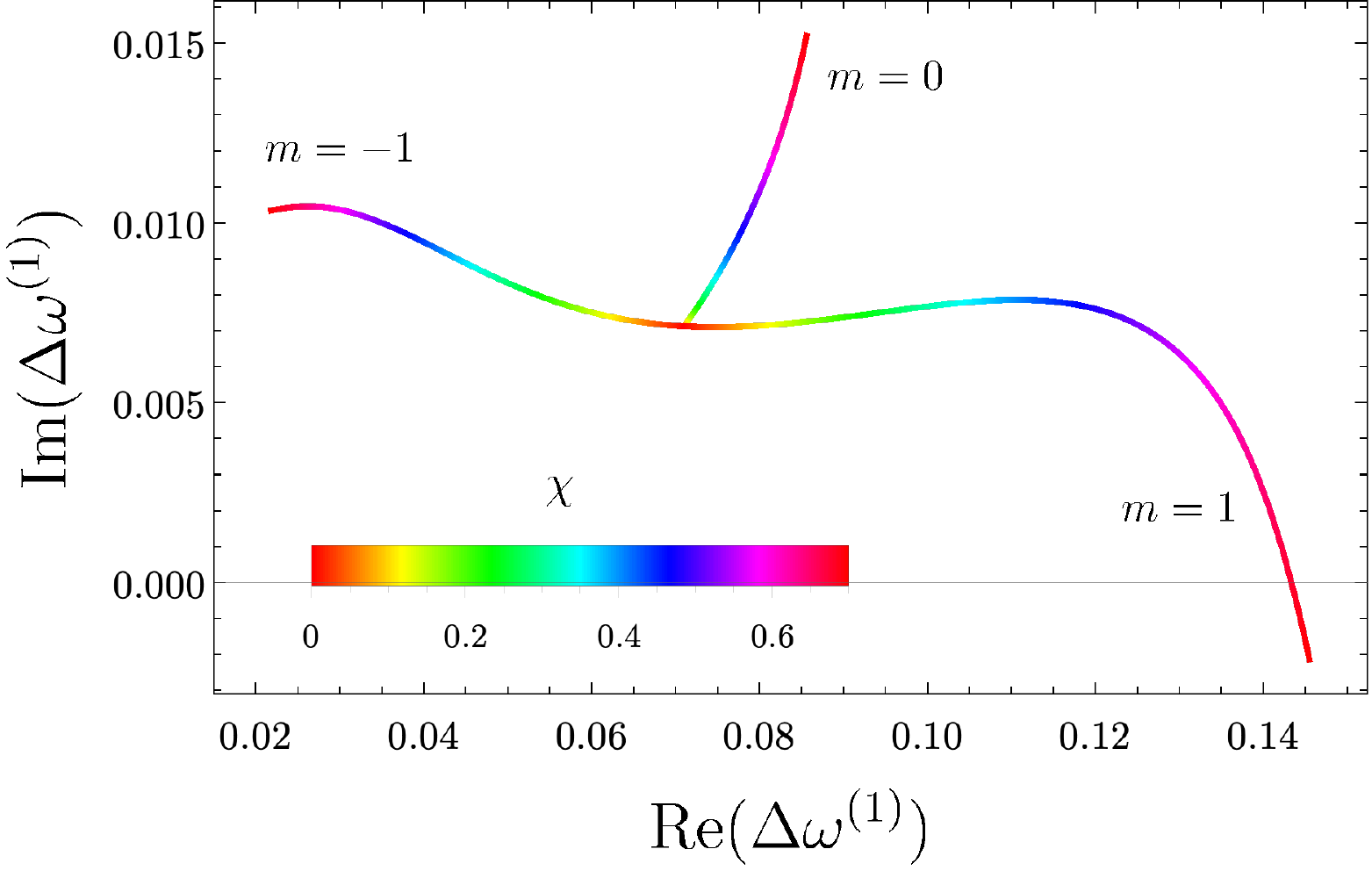}
	    \includegraphics[width=0.49\textwidth]{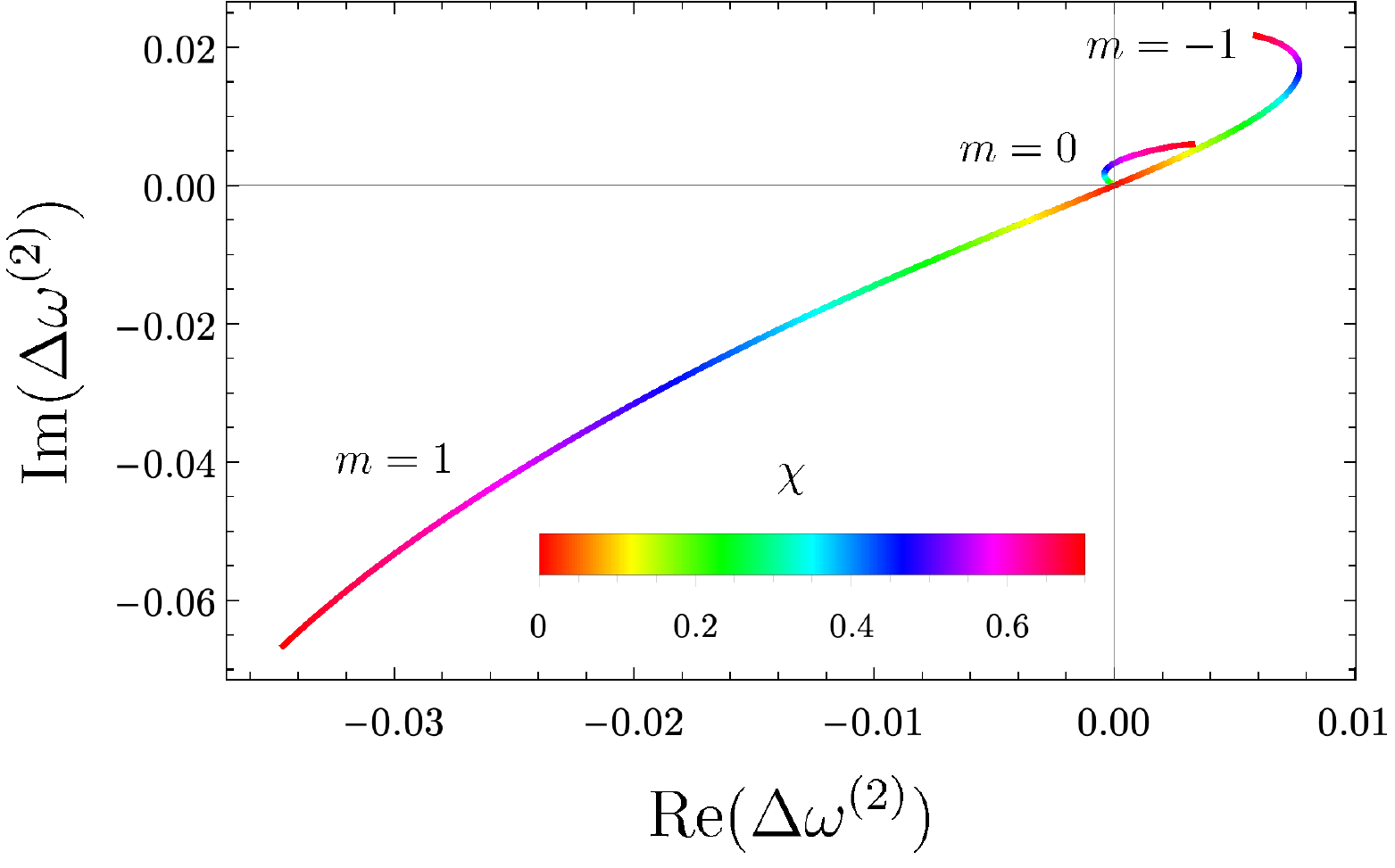}
	    \includegraphics[width=0.49\textwidth]{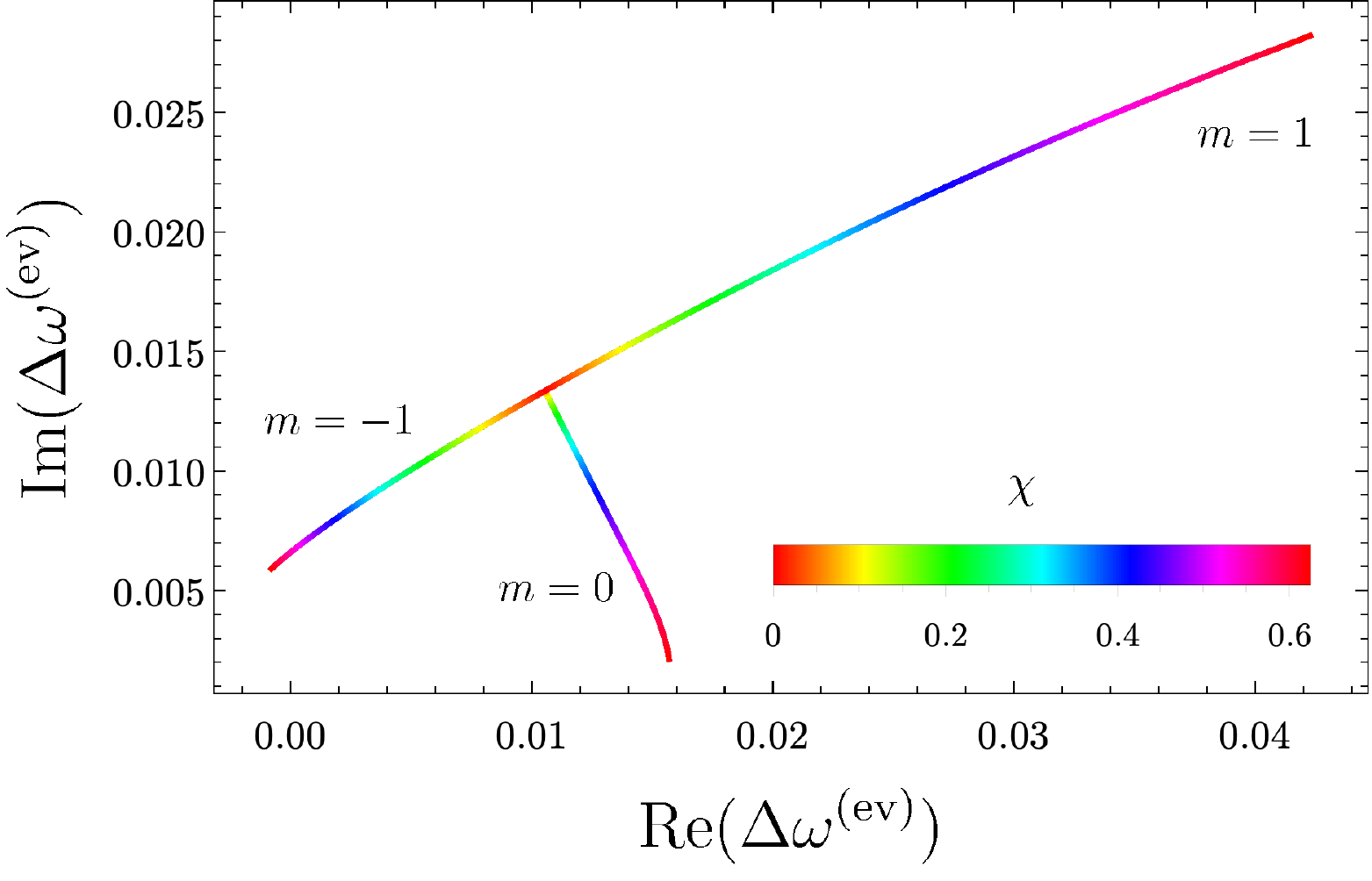}
		\caption{Trajectories in the complex plane of the Kerr quasinormal frequencies $\omega^{(0)}_{l,m,0}$ for $l=1$ and $m=-1,0,1$, and of the correction coefficients $\Delta\omega^{(i)}_{l,m,0}$. The dependence in the spin is labeled by the color code (note that the maximum spin represented in each case is different).}
		\label{fig:qnfl1}
	\end{center}
\end{figure}
In Fig.~\ref{fig:qnfl1} we plot the trajectories in the complex plane of the zeroth-order quasinormal frequencies and of the correction coefficients for $l=1$, while in Fig.~\ref{fig:qnfl2} we show the corresponding plots for $l=2$. Here the $\mathcal{O}(\chi^{14})$ expansion of the potentials gives us a good result up to $\chi\sim 0.7$, except for $\Delta\omega^{(\rm ev)}_{1,m,0}$ which seems to be accurate enough only for $\chi\le 0.625$.  On the other hand, polynomial fits of the quasinormal frequencies  are provided in  appendix \ref{app:fit}.

As we can see from the plots, we correctly reproduce the behaviour of the uncorrected Kerr frequencies with the characteristic splitting of the different $m$ modes.  On the other hand, the shift introduced in the quasinormal frequencies due to the the corrections  has a great variability as a function of $m$ and $\chi$. Thus, the effect of the corrections can be several times larger or smaller depending on the value of the spin, and in general, the mode with $m=l \operatorname{sign}{(\chi)}$ seems to be the most affected one. 
We can appreciate that the curves followed in the complex plane by each type of correction have an overall pattern that does not change much as we vary $l$. By increasing the value of $l$, we have observed that the trajectories have a similar behaviour to those in Figs.~\ref{fig:qnfl1} and \ref{fig:qnfl2} but with an increasing number of lines in between (corresponding, naturally, to the increasing number of values of $m$).

\begin{figure}[t!]
	\begin{center}
		\includegraphics[width=0.49\textwidth]{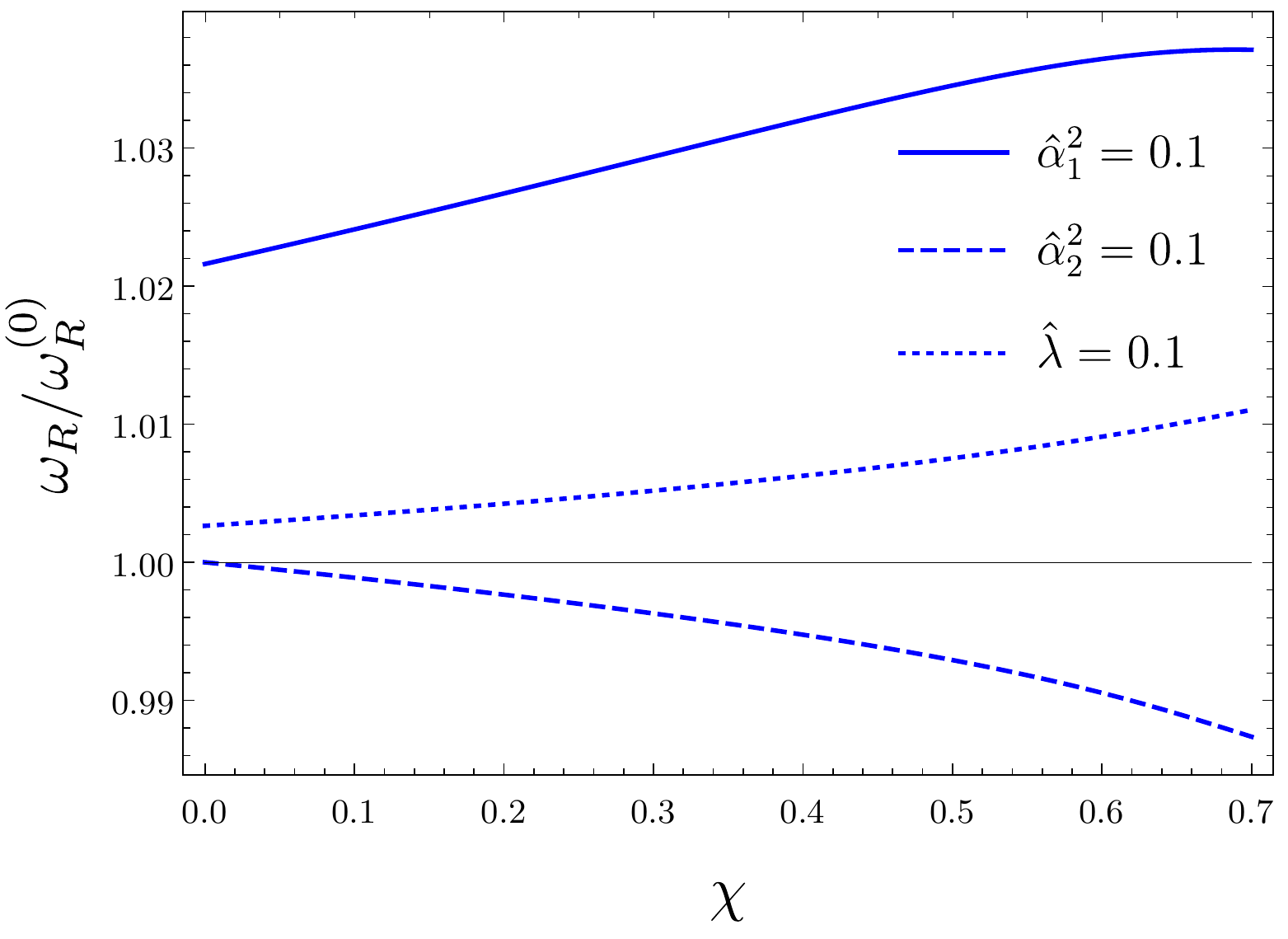} 
	    \includegraphics[width=0.49\textwidth]{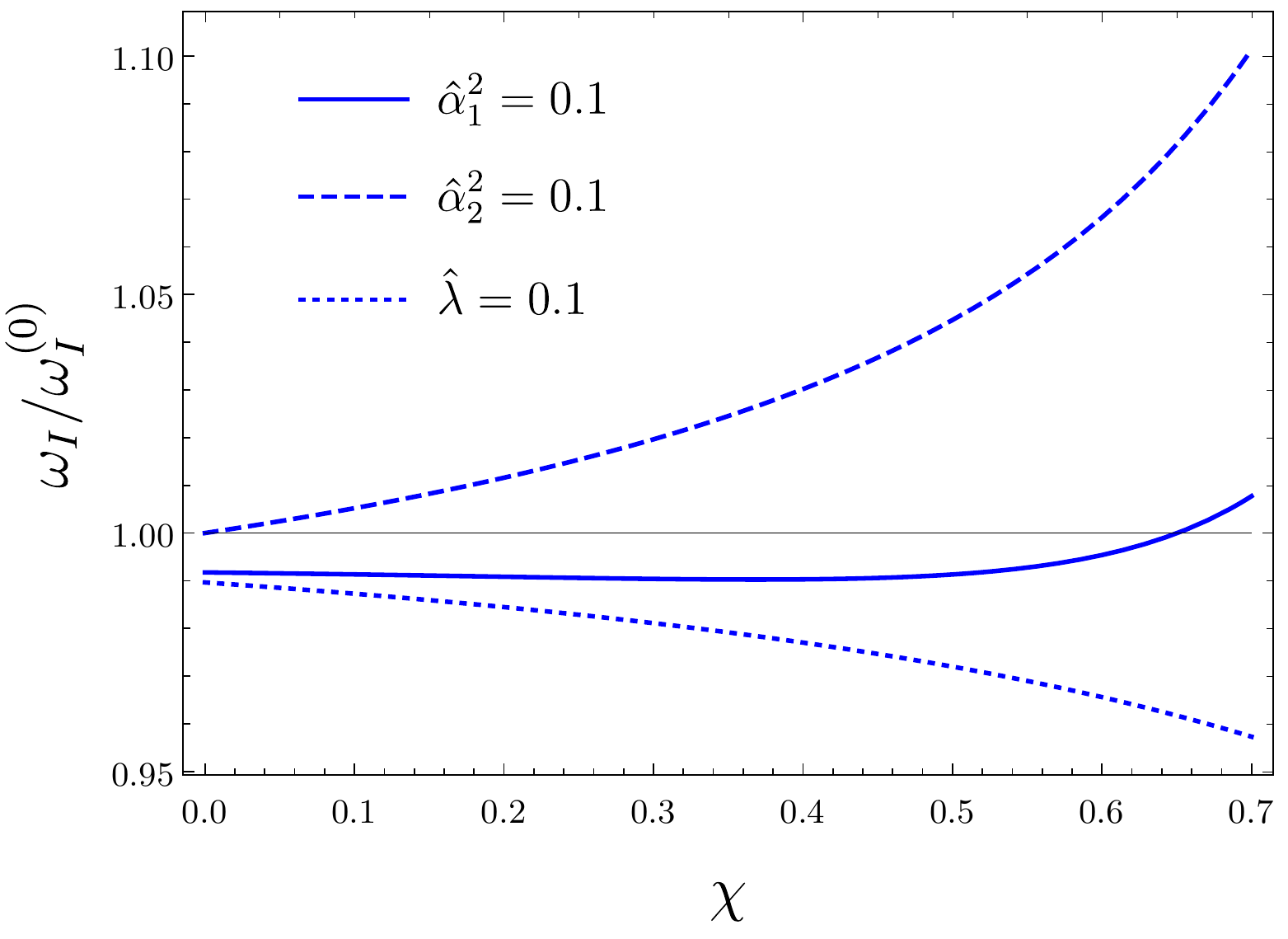}
		\caption{Ratios between the corrected quasinormal frequencies $\omega_{l,m,0}$ and the corresponding Kerr values $\omega^{(0)}_{l,m,0}$ for $l=m=2$. Left: ratios of the real parts. Right: ratios of the imaginary parts. Each line represents the result when only the indicated coupling has a non-vanishing value. In the general case, the corresponding curve is a linear combinations of the ones we represent.}
		\label{fig:ratio}
	\end{center}
\end{figure}

In the case of $l=2$, the Kerr's gravitational quasinormal frequencies are not too different from the scalar ones, and indeed they have a similar trajectory in the complex plane. Therefore, we might expect that the corrections to the gravitational quasinormal frequencies will follow trajectories similar to those shown in Fig.~\ref{fig:qnfl2}, at least in qualitative terms. In regard to this, it can be interesting to analyze the ratios $\frac{\omega_R}{\omega^{(0)}_R}$, $\frac{\omega_I}{\omega^{(0)}_I}$, between the real and imaginary parts of the quasinormal frequencies and their corresponding values for Kerr. Assuming that these ratios do not change much with the spin of the perturbations, this would be a way to obtain a rough estimate for the gravitational quasinormal frequencies.  Focusing on the dominant mode $l=2$, $m=2$, we find the following expressions for these ratios after fitting the numerical data:

\begin{align}
\notag
\frac{\omega_R}{\omega^{(0)}_R}&=1+0.0264 \hat{\lambda }+0.2160 \hat{\alpha }_1^2+\left(0.0745\hat{\lambda }+0.2434 \hat{\alpha
   }_1^2-0.1101 \hat{\alpha }_2^2\right) \chi \\ \notag
   &+\left(-0.0109 \hat{\lambda }+0.0893 \hat{\alpha
   }_1^2+0.0357 \hat{\alpha }_2^2\right) \chi ^2+\left(0.3387 \hat{\lambda }-0.2379 \hat{\alpha
   }_1^2-0.6590\hat{\alpha }_2^2\right) \chi ^3\\ \notag
   &+\left(-1.0002 \hat{\lambda }+0.5816 \hat{\alpha
   }_1^2+2.0828 \hat{\alpha }_2^2\right) \chi ^4+\left(1.4728 \hat{\lambda }-0.6561 \hat{\alpha
   }_1^2-3.0440 \hat{\alpha }_2^2\right) \chi ^5\\ 
   &+\left(-0.7340 \hat{\lambda }-0.0579 \hat{\alpha
   }_1^2+1.4510 \hat{\alpha }_2^2\right) \chi ^6\, , 
\label{ratioR}
   \end{align}
   
   \begin{align}
   \notag
\frac{\omega_I}{\omega^{(0)}_I}&=1-0.1034 \hat{\lambda }-0.0823 \hat{\alpha }_1^2+\left(-0.2178 \hat{\lambda }-0.0586
   \hat{\alpha }_1^2+0.4619 \hat{\alpha }_2^2\right) \chi \\ \notag
   &+\left(-0.1653 \hat{\lambda }+0.4022
   \hat{\alpha }_1^2+0.7111 \hat{\alpha }_2^2\right) \chi ^2+\left(-0.2023 \hat{\lambda }-3.4773
   \hat{\alpha }_1^2-1.8926 \hat{\alpha }_2^2\right) \chi ^3\\ \notag
   &+\left(0.1605 \hat{\lambda }+12.337
   \hat{\alpha }_1^2+8.9607 \hat{\alpha }_2^2\right) \chi ^4+\left(-0.3331 \hat{\lambda }-19.8908
   \hat{\alpha }_1^2-14.7249 \hat{\alpha }_2^2\right) \chi ^5\\ 
   &+\left(-0.0233 \hat{\lambda }+13.4036
   \hat{\alpha }_1^2+11.1683 \hat{\alpha }_2^2\right) \chi ^6\, ,
   \label{ratioI}
\end{align}
where we have introduced the hatted quantities
\begin{equation}
\hat\lambda=\lambda_{\rm ev} \frac{\ell^4}{M^4}\, ,\quad \hat\alpha_1^2=\alpha_1^2\frac{\ell^4}{M^4}\, ,\quad \hat\alpha_2^2=\alpha_2^2\frac{\ell^4}{M^4}\, .
\end{equation}
These expressions reproduce very well the numeric results obtained for $0\le\chi\le 0.7$, and they probably yield a reasonably good approximation for even larger values of $\chi$. A profile of these ratios for particular values of the couplings is shown in Fig~\ref{fig:ratio}. Now the question is whether the quasinormal frequencies of gravitational perturbations, compared to their respective Kerr counterparts, will yield similar ratios. If this were the case, then we could simply estimate the gravitational frequencies by using $\omega^{\rm grav}_R\sim\omega_R^{\rm grav, (0)}\times (\omega_R^{\rm scalar}/\omega^{\rm scalar, (0)}_R)$, and similarly for the imaginary part. We cannot provide a definitive answer to this question, but nevertheless we can try to test this formula for particular cases. Unfortunately, to the best of our knowledge, gravitational quasinormal modes of rotating black holes have not been studied yet in any of the models that our theory \req{Action} contains. On the other hand, the quasinormal frequencies of static black holes in Einstein-dilaton-Gauss-Bonnet gravity --- which is contained in \req{Action} --- are known.  The perturbative and non-perturbative regimes of these quasinormal modes were studied in Ref.~\cite{Blazquez-Salcedo:2016enn}  for both axial and polar perturbations. In the case of axial perturbations,\footnote{Polar ones behave differently because they couple to the scalar field.} the following result was found for the $l=2$, $m=2$ fundamental mode,\footnote{In our conventions we have $\zeta=4\hat\alpha_1$.}\footnote{We believe there is a typo in Table I of \cite{Blazquez-Salcedo:2016enn}: the entry corresponding to $R_2$ and $l=2$  should be $1.002 \times 10^{-2}$. 
Note that the values of $\omega_R^{\rm axial}$ and $\omega_I^{\rm axial}$ that we write are the same ones as in \cite{Carson:2020cqb}. }
\begin{equation}\label{axialEdGB}
\frac{\omega_R^{\rm axial}}{\omega^{\rm axial, (0)}_R}=1+0.1603\hat\alpha_1^2\, ,\quad \frac{\omega_I^{\rm axial}}{\omega^{\rm axial, (0)}_I}=1-0.0828\hat\alpha_1^2\, .
\end{equation}
Comparing with our formulas \req{ratioR} and \req{ratioI}, we see that the result is indeed quite close --- for the imaginary part it is almost coincident. Therefore, it would seem that these ratios can be used to get, at least, an order-of-magnitude estimate for the corrections to the gravitational quasinormal frequencies.

\subsubsection{High $l$: analytic approximations}\label{subsec:analytic}

As discussed for a non-rotating black hole, approximate analytic methods such as using an effective P\"oschl-Teller approximation or the WKB method become more effective for large $l$. On the one hand, an analytic estimate may serve as a check of our numerical computations. On the other, these methods allow to compute higher overtones easily, although here we will only focus on the fundamental modes. 
The WKB and PT methods provide explicit expressions for the quasinormal frequencies in terms of the parameters of the potential, but in the case of a rotating black hole obtaining these frequencies becomes more difficult as the potential has a further nontrivial $\omega$ dependence. For instance, in \eqref{eqn:PT} this means $V_0$ and $\alpha$ are still $\omega$-dependent, and the remaining algebraic equation still needs to be solved. Nevertheless, it is still possible to solve that equation explicitly by performing a perturbative expansion in $\chi$. This is in fact a consistent strategy taking into account that our effective potentials are expressed as a series expansion in the spin. 

As an example, replacing the $l=5$, $m=5$ potential by a P\"oschl-Teller one, we get the following approximation for corresponding $n=0$ quasinormal frequency (up to order $\chi^3$):
\begin{align}
\notag
&M \omega_{5,5,0}\Big|_{\rm PT}   =  (1.061-0.097 i)+(0.381\, -0.001 i) \chi +(0.195\, + 0.007 i) \chi
^2+(0.120\, +0.007 i) \chi ^3 \\\notag
&+\frac{\alpha_1^2 \ell^4}{M^4}
\left((0.221+0.007 i)+(0.367\, +0.005 i) \chi +(0.213\, +0.002
i) \chi^2+(0.076\, -0.007 i) \chi ^3\right)  \\\notag
&+ \frac{\alpha_2^2 \ell^4}{M^4}
\chi \left(-(0.142+0.0472 i)  -(0.104\, +0.059 i) \chi
-(0.053\, +0.054 i) \chi^2\right) \\
&+ \frac{\lambda_{\rm ev} \ell^4}{M^4}
\left((0.029+0.007 i)+(0.0881\, +0.021 i) \chi +(0.055\, +0.023
i) \chi ^2+(0.023\, +0.018 i) \chi ^3\right)\, .
\label{eqn:PT55} 
\end{align}
It is possible to obtain the analytic values of the coefficients in the expression above, but for practical purposes we are showing only the numeric values with a few digits. 

We note that although this corresponds roughly to \eqref{PTstatic55} for $\chi =0$, it is not exactly equal because of the different choice of ``tortoise coordinate''. This P\"oschl-Teller approximation can be generalized to use instead the Rosen-Morse potential, reviewed in Appendix \ref{app:methods}, to accommodate better the structure of the effective potentials. The Rosen-Morse potential contains one additional parameter that allows to set different asymptotic values of the potential. We recall that our effective potentials \req{eq:Vyomega} generically have a non-vanishing value at $y\rightarrow-\infty$, and hence in principle the RM method should provide a better approximation of the quasinormal frequencies. Solving \eqref{eqn:RMQNMs} perturbatively for $l=5$, $m=5$ one finds

\begin{align}\notag
&M \omega_{5,5,0}\Big|_{\rm RM} =  (1.061-0.097 i)+0.370 \chi +(0.190\, +0.007 i) \chi
^2+(0.117\, +0.007 i) \chi ^3 \\\notag
&+ \frac{\alpha_1^2 \ell^4}{M^4} \left((0.215+0.008 i)+(0.354\, +0.008 i)
\chi +(0.205\, +0.003 i) \chi ^2+(0.077\, -0.010 i) \chi ^3\right) \\\notag
&+\frac{\alpha_2^2 \ell^4}{M^4} \chi  \left(-(0.126+0.049 i)
-(0.089\, +0.062 i) \chi-(0.043\, +0.053 i) \chi
^2\right)\\
&+ \frac{\lambda_{\rm ev} \ell^4}{M^4} \left((0.028+0.008 i)+(0.082\, +0.022 i)
\chi +(0.049\, +0.024 i) \chi ^2+(0.021\, +0.017 i) \chi
^3\right) \, .
\label{eqn:RM55}
\end{align}

\noindent
As a final approximation, we also calculate the leading order WKB approximation (effectively the "Bohr-Sommerfeld quantization rule") \cite{Berti:2009kk}, which yields the following equation for the QNFs
\be
\frac{\omega^2-V_0}{\sqrt{-2 V_0''}}=i(n+\frac{1}{2}) \, ,
\ee
where $V_0''$ is the second derivative of the potential at the peak, and $n=0,1,2,...$ is the overtone index. 
In the same example $l=5$, $m=5$, this leads to
\begin{align}\notag
&M \omega_{5,5,0}\Big|_{\rm WKB}= (1.070-0.096 i)+0.374 \chi +(0.194\, +0.008 i) \chi
^2+(0.119\, +0.007 i) \chi ^3 \\\notag
&+\frac{\alpha_1^2 \ell^4}{M^4} \left((0.217+0.008 i)+(0.361\, +0.007 i)
\chi +(0.215\, +0.003 i) \chi ^2+(0.081\, -0.010 i) \chi ^3\right) \\\notag
&+ \frac{\alpha_2^2 \ell^4}{M^4} \chi  \left(-(0.132-0.048 i)
-(0.010\, +0.062 i) \chi-(0.051\, +0.053 i) \chi
^2\right)\\
&+\frac{\lambda_{\rm ev} \ell^4}{M^4} \left((0.028+0.008 i)+(0.084\, +0.021 i)
\chi +(0.054\, +0.024 i) \chi ^2+(0.023\, +0.017 i) \chi
^3\right)  \, ,
\label{eqn:WKB55}
\end{align}
for the fundamental mode, but higher overtones can be easily computed as well.\footnote{Note that due to the non-trivial dependence of the potential on $\omega$, the solution depends on $n$ in a complicated way.}

Although all these approximations are roughly in correspondence with each other, they nevertheless have differences in excess of 10\% in some of the coefficients, especially for higher orders of $\chi$. In Fig.~\ref{fig:analyticapprox} (in the appendix), we illustrate the relative error of $M\omega^{(0)}_{5,5,0}$, $\Delta\omega^{(1)}_{5,5,0}$, $\Delta\omega^{(2)}_{5,5,0}$, $\Delta\omega^{(\rm ev)}_{5,5,0}$ based on the analytic estimates in \eqref{eqn:PT55}, \eqref{eqn:RM55}, \eqref{eqn:WKB55} compared to the numerical results. It can be seen that the Rosen-Morse approach performs best overall, with the WKB giving comparably good results for the corrections, while the P\"oschl-Teller approximation is significantly worse.

These approximate methods become more precise as we increase $l$. For comparison with the approach of the next section, let us also show the perturbative solution in the case of $l=10$, $m=10$. Based on the conclusion that the Rosen-Morse potential is most effective, we restrict ourselves to this approximation, which yields
\begin{align}\notag
&M \omega_{10,10,0} \Big|_{\rm RM}=(2.022-0.096 i)+0.741 \chi +(0.385\, +0.007 i) \chi^2+(0.240\, +0.007 i) \chi^3 \\\notag
&+\frac{\alpha_1^2 \ell^4}{M^4} \left((0.408+0.008 i)+(0.700\, +0.009 i)\chi +(0.410\, +0.002 i) \chi^2+(0.138\, -0.012 i) \chi^3\right) \\\notag
&+\frac{\alpha_2^2 \ell^4}{M^4} \chi \left(-(0.255+0.050 i)
-(0.171\, +0.069 i) \chi -(0.078\, +0.062 i) \chi^2\right)\\
&+ \frac{\lambda_{\rm ev} \ell^4}{M^4}\left((0.055+0.007 i)+(0.163\, +0.021 i)
\chi +(0.088\, +0.025 i) \chi^2+(0.019\, +0.019 i) \chi^3\right) \, .
\label{RMl10m10}
\end{align}

\subsubsection{Eikonal limit and lightring geodesics}

In the eikonal limit $l\rightarrow\infty$, the quasinormal frequencies are expected to be related to the unstable null geodesics around the black hole. Thus, the orbital frequencies of these geodesics control the real frequencies, while the damping time is determined by the Lyapunov exponents. The quasinormal ringing/null geodesics correspondence for general $m$ modes in Kerr black holes is quite complicated --- see Ref.~\cite{Yang:2012he}. However, the $m=\pm l$ cases are much simpler, as they only involve geodesics in the equatorial plane. In particular, for the $m=l$ mode, the frequencies are given by \cite{PhysRevD.30.295,PhysRevD.82.104003,Cardoso:2008bp}
\begin{equation}\label{eikfreq}
\omega_{l,l,n}=\left(l+\frac{1}{2}\right)\Omega_{+}-i\left(n+\frac{1}{2}\right)|\lambda|\, ,
\end{equation}
where $\Omega_{+}$ is the orbital frequency of the equatorial null geodesic rotating in the positive direction, while $\lambda$ is the corresponding Lyapunov exponent measuring the instability timescale of the geodesic. The details on how to obtain these quantities can be found \textit{e.g.} in \cite{Cardoso:2008bp}. The corresponding value for the $m=-l$ mode can be found simply by exchanging the sign of $\chi$ in the expression above, $\omega_{l,-l,n}(\chi)=\omega_{l,l,n}(-\chi)$. The orbital frequency and the Lyapunov exponent for the black holes in the theory \req{Action} can be found analytically performing an expansion in $\chi$ \cite{Cano:2019ore}. The result can be written as
\begin{align}
\Omega_{+}&=\Omega_{+}^{(0)}+\frac{\ell^4}{M^5}\left(\alpha_1^2\Omega_{+}^{(1)}+\alpha_2^2\Omega_{+}^{(2)}+\lambda_{\rm ev}\Omega_{+}^{(\rm ev)}\right)\, ,\\
\lambda&=\lambda^{(0)}+\frac{\ell^4}{M^5}\left(\alpha_1^2\lambda^{(1)}+\alpha_2^2\lambda^{(2)}+\lambda_{\rm ev}\lambda^{(\rm ev)}\right)\, ,
\end{align}
where the different quantities have the following expansions in $\chi$ (up to order $\chi^8$) 
{\allowdisplaybreaks 

\begin{align}
\notag
M\Omega_{+}^{(0)}=&\frac{1}{3 \sqrt{3}}+\frac{2 \chi }{27}+\frac{11 \chi ^2}{162 \sqrt{3}}+\frac{2 \chi ^3}{81}+\frac{523 \chi ^4}{17496
	\sqrt{3}}+\frac{254 \chi ^5}{19683}+\frac{16543 \chi ^6}{944784 \sqrt{3}}+\frac{4354 \chi ^7}{531441}\\
&+\frac{2408051
	\chi ^8}{204073344 \sqrt{3}}\, ,\\ \notag
\Omega_{+}^{(1)}=&\frac{4397}{65610 \sqrt{3}}+\frac{20596 \chi }{295245}+\frac{1028803 \chi ^2}{14467005 \sqrt{3}}+\frac{45262543 \chi
	^3}{3906091350}-\frac{3685587061 \chi ^4}{328111673400 \sqrt{3}}\\ \notag &-\frac{110632797883 \chi
	^5}{5413842611100}-\frac{910228742414947 \chi ^6}{17151053391964800 \sqrt{3}}-\frac{15449837941866829 \chi
	^7}{401334649371976320}\\
&-\frac{11134828406941279099 \chi ^8}{143477137150481534400 \sqrt{3}}\, ,\\ \notag
\Omega_{+}^{(2)}=&-\frac{131 \chi }{5103}-\frac{11047 \chi ^2}{381024 \sqrt{3}}-\frac{9491513 \chi ^3}{1388832480}-\frac{19022279 \chi
	^4}{925888320 \sqrt{3}}-\frac{353193404087 \chi ^5}{23099061807360}\\
&-\frac{2452581602509 \chi ^6}{63522419970240
	\sqrt{3}}-\frac{5958423964756267 \chi ^7}{222963694095542400}-\frac{37265277503432903 \chi ^8}{668891082286627200
	\sqrt{3}}\, ,\\ \notag
\Omega_{+}^{(\rm ev)}=&\frac{20}{2187 \sqrt{3}}+\frac{320 \chi }{19683}+\frac{26749 \chi ^2}{1928934 \sqrt{3}}-\frac{12967 \chi
	^3}{104162436}-\frac{4415651 \chi ^4}{1249949232 \sqrt{3}}-\frac{3101153 \chi ^5}{937461924}\\
&-\frac{33998483 \chi
	^6}{6629195034 \sqrt{3}}-\frac{18127693795 \chi ^7}{6682228594272}-\frac{601383641851 \chi ^8}{173737943451072
	\sqrt{3}}\, .
\end{align}

\begin{align}
M\lambda^{(0)}=&\frac{1}{3 \sqrt{3}}-\frac{2 \chi ^2}{81 \sqrt{3}}-\frac{10 \chi ^3}{729}-\frac{44 \chi ^4}{2187 \sqrt{3}}-\frac{191 \chi
	^5}{19683}-\frac{836 \chi ^6}{59049 \sqrt{3}}-\frac{61 \chi ^7}{8748}-\frac{16636 \chi ^8}{1594323 \sqrt{3}}\, ,\\ \notag
\lambda^{(1)}=&-\frac{1843}{65610 \sqrt{3}}-\frac{40 \chi }{2187}-\frac{58879 \chi ^2}{19289340 \sqrt{3}}+\frac{34963739 \chi
	^3}{1302030450}+\frac{101574788537 \chi ^4}{1312446693600 \sqrt{3}}\\ \notag
&+\frac{23802643939 \chi
	^5}{433107408888}+\frac{900673269063947 \chi ^6}{8575526695982400 \sqrt{3}}+\frac{1387887433593317 \chi
	^7}{22051354361097600}\\
&+\frac{4930958596510920137 \chi ^8}{44146811430917395200 \sqrt{3}}\, ,\\ \notag
\lambda^{(2)}=&\frac{230 \chi }{2187}+\frac{678569 \chi ^2}{2571912 \sqrt{3}}+\frac{196638109 \chi ^3}{1388832480}+\frac{5614994047 \chi
	^4}{24998984640 \sqrt{3}}+\frac{7708776117503 \chi ^5}{69297185422080}\\
&+\frac{3929424844884557 \chi
	^6}{22868071189286400 \sqrt{3}}+\frac{46142799714903611 \chi ^7}{535112865829301760}+\frac{362024510543473937 \chi
	^8}{2675564329146508800 \sqrt{3}}\, ,\\ \notag
\lambda^{(\rm ev)}=&-\frac{52}{2187 \sqrt{3}}-\frac{32 \chi }{729}-\frac{88097 \chi ^2}{964467 \sqrt{3}}-\frac{1548863 \chi
	^3}{34720812}-\frac{40153453 \chi ^4}{624974616 \sqrt{3}}-\frac{18431767 \chi ^5}{624974616}\\
&-\frac{15457985849 \chi
	^6}{371234921904 \sqrt{3}}-\frac{5846574233 \chi ^7}{303737663376}-\frac{52186447127 \chi ^8}{1909208169792 \sqrt{3}}\, .
\end{align}
}
It can be checked that these expansions are convergent for $|\chi|<1$, and the truncation up to order $8$ is a good approximation for $|\chi|\le 0.7$. Now we can compare these results with the expressions found for the quasinormal frequencies using approximate methods. Taking the numeric values for the $l=m=10$, $n=0$ mode, we see that the first terms in $\chi$ read

\begin{align}\notag
&M \omega_{10,10,0} \Big|_{\rm LR}=(2.021-0.096 i)+0.778 \chi +(0.412\, +0.007 i) \chi^2+(0.259\, +0.007 i) \chi^3 \\\notag
&+\frac{\alpha_1^2 \ell^4}{M^4} \left((0.406+0.008 i)+(0.732\, +0.009 i)\chi +(0.431\, +0.001 i) \chi^2+(0.122\, -0.013 i) \chi^3\right) \\\notag
&+\frac{\alpha_2^2 \ell^4}{M^4} \chi \left(-(0.270+0.053 i)
-(0.177\, +0.076 i) \chi -(0.072\, +0.071 i) \chi^2\right)\\
&+ \frac{\lambda_{\rm ev} \ell^4}{M^4}\left((0.055+0.007 i)+(0.171\, +0.022 i)
\chi +(0.084\, +0.026 i) \chi^2+(-0.001\, +0.022 i) \chi^3\right) \, .
\end{align}
We observe that the coefficients in this expansion are quite close to the ones in \req{RMl10m10}, obtained from the Rosen-Morse approximation of the effective potential. There are some differences as we go to higher orders in the $\chi$-expansion, but these can be explained on account of the various expansions and approximations involved in \req{RMl10m10} and in the fact that the convergence for $l\rightarrow\infty$ is slow. This example shows explicitly that the eikonal quasinormal frequencies are in fact linked to the photon-sphere geodesics and that they can be computed using \req{eikfreq}. Indeed, comparing with the numerics one can see that the eikonal formula already provides a reasonable approximation even for the $l=2$ modes. On the other hand, the fact that we obtain consistent results is a good test of the validity of the procedure explained in Sec.~\ref{sec:pertrotation}. 

However, the correspondence between quasinormal ringing and unstable lightring geodesics should be taken with care because it might not hold in general. For instance, we have seen that parity-odd interactions do not affect the effective potentials for scalar perturbations, and hence they do not modify the quasinormal frequencies. However, those terms do affect geodesics, and in fact there are no equatorial geodesics whenever they are non-vanishing --- in order to compute the lightring geodesics we have set the parity-breaking parameters to zero. It would be interesting to study what happens when such terms are non-vanishing. Hopefully they will modify the orbital frequency and Lyapunov exponents only at a non-linear level, so that the correspondence with quasinormal frequencies is still valid.

\subsection{$\mathcal{O}(\ell^6)$ corrections}
\begin{figure}[t!]
	\begin{center}
		\includegraphics[width=0.49\textwidth]{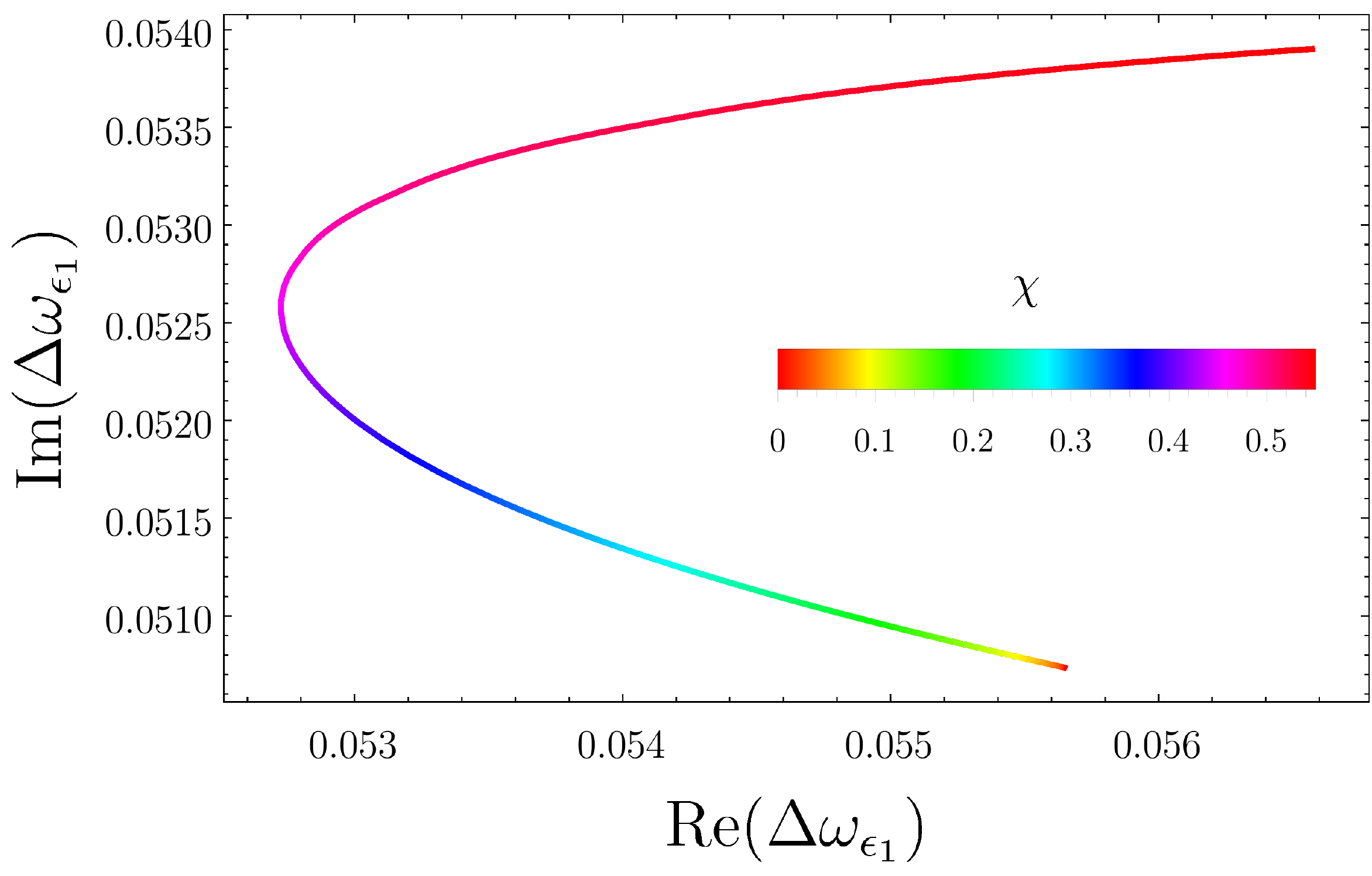}
		\includegraphics[width=0.49\textwidth]{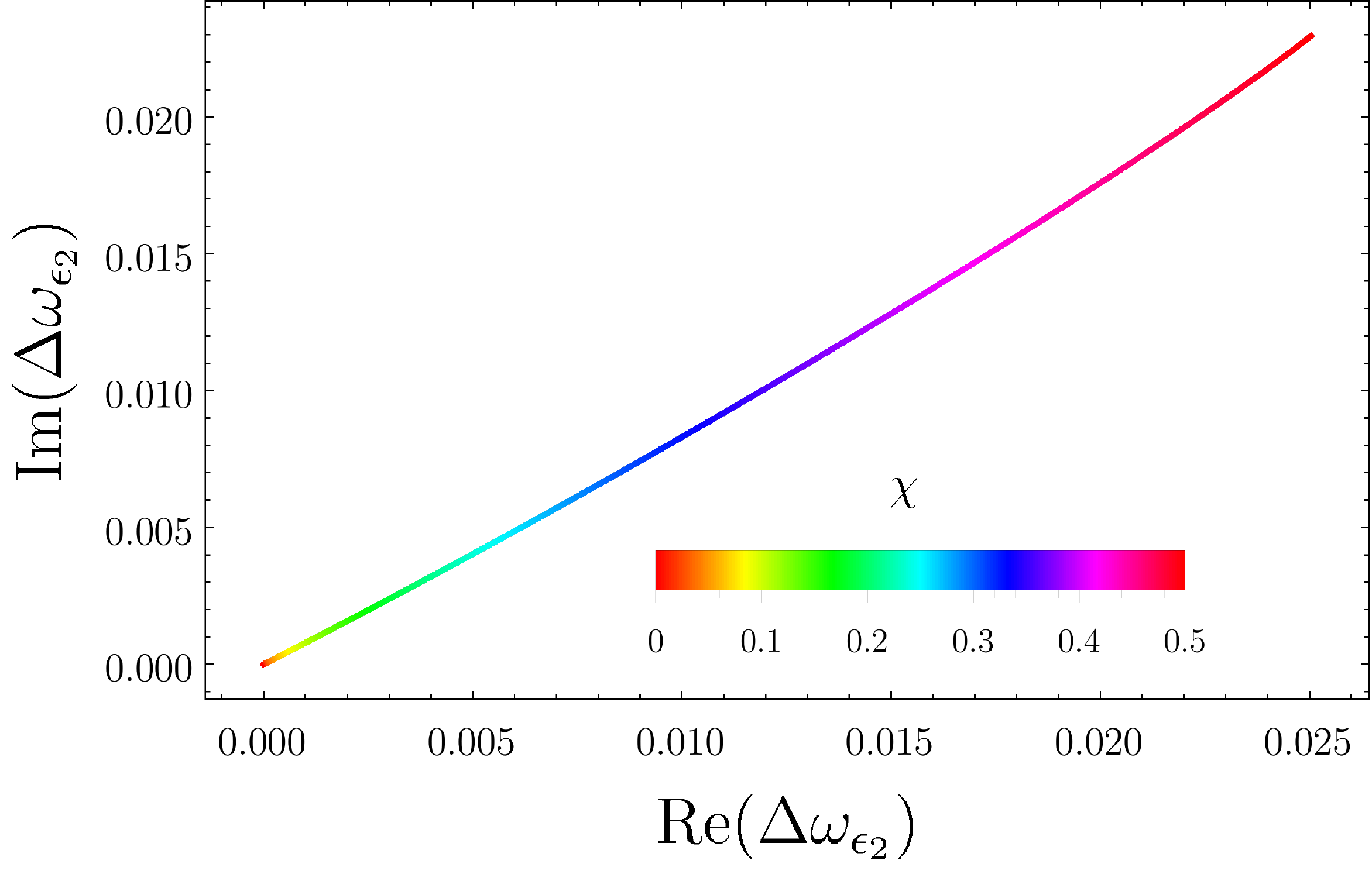}
		\includegraphics[width=0.49\textwidth]{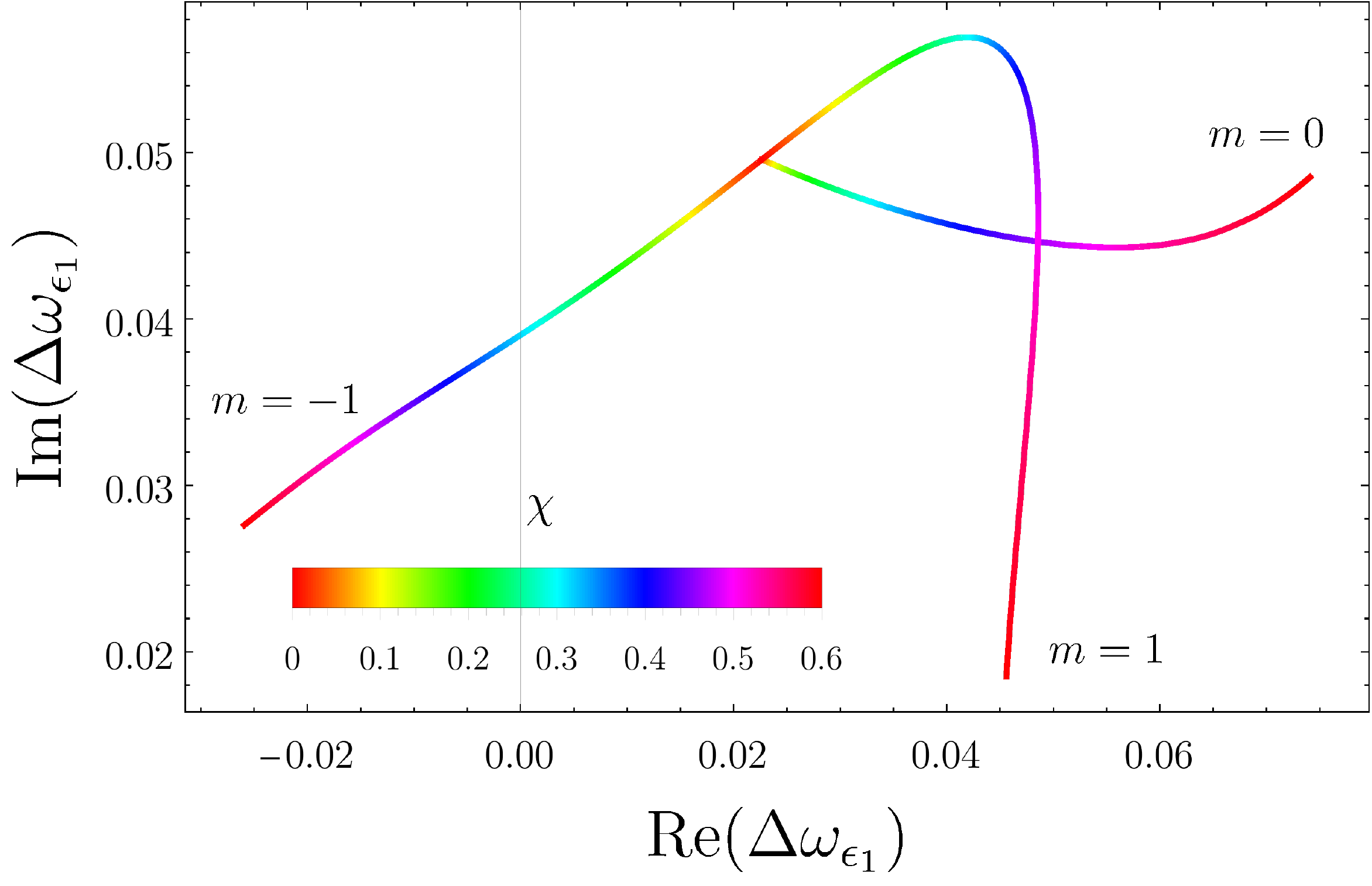}
		\includegraphics[width=0.49\textwidth]{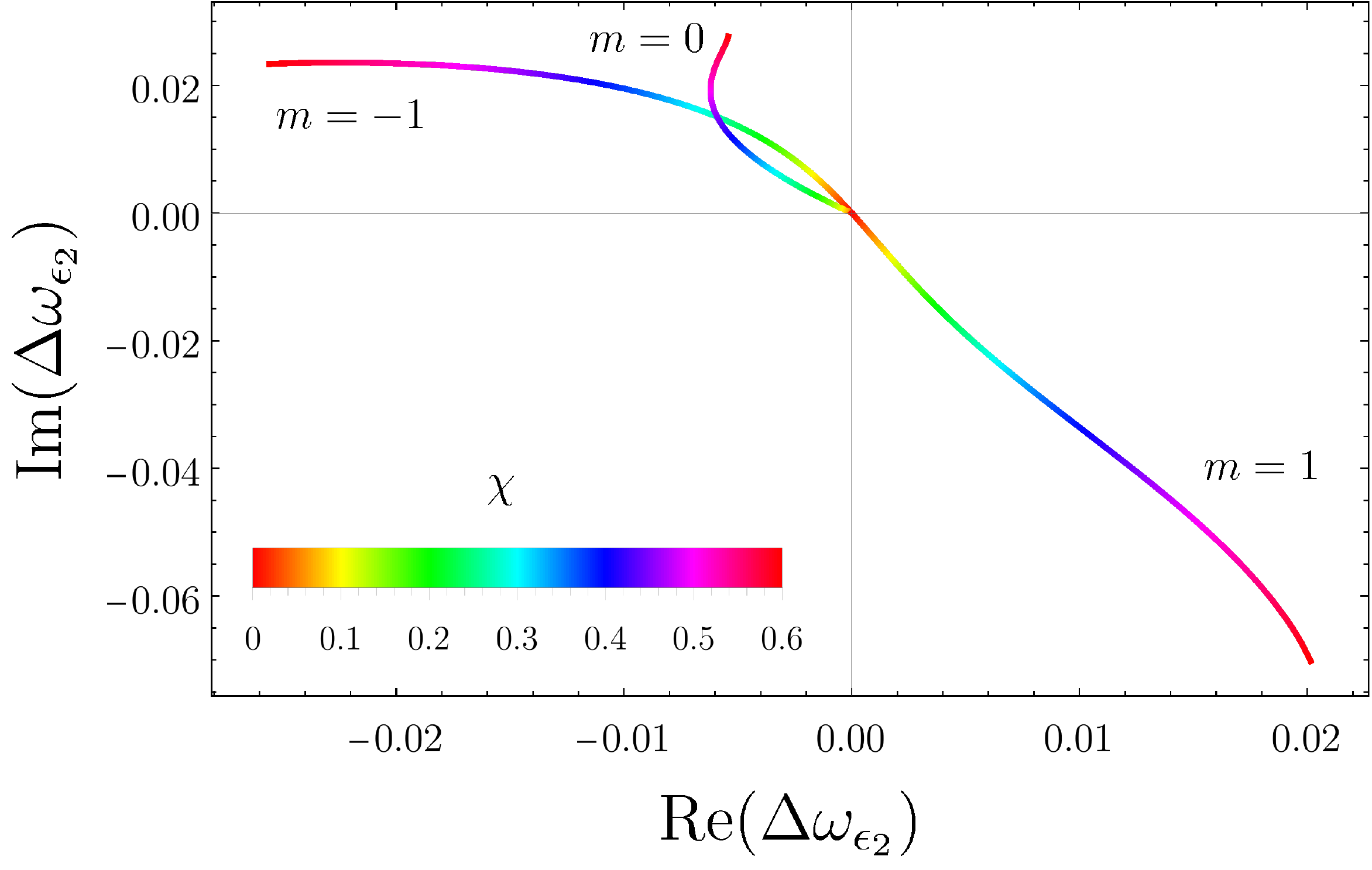}
		\includegraphics[width=0.49\textwidth]{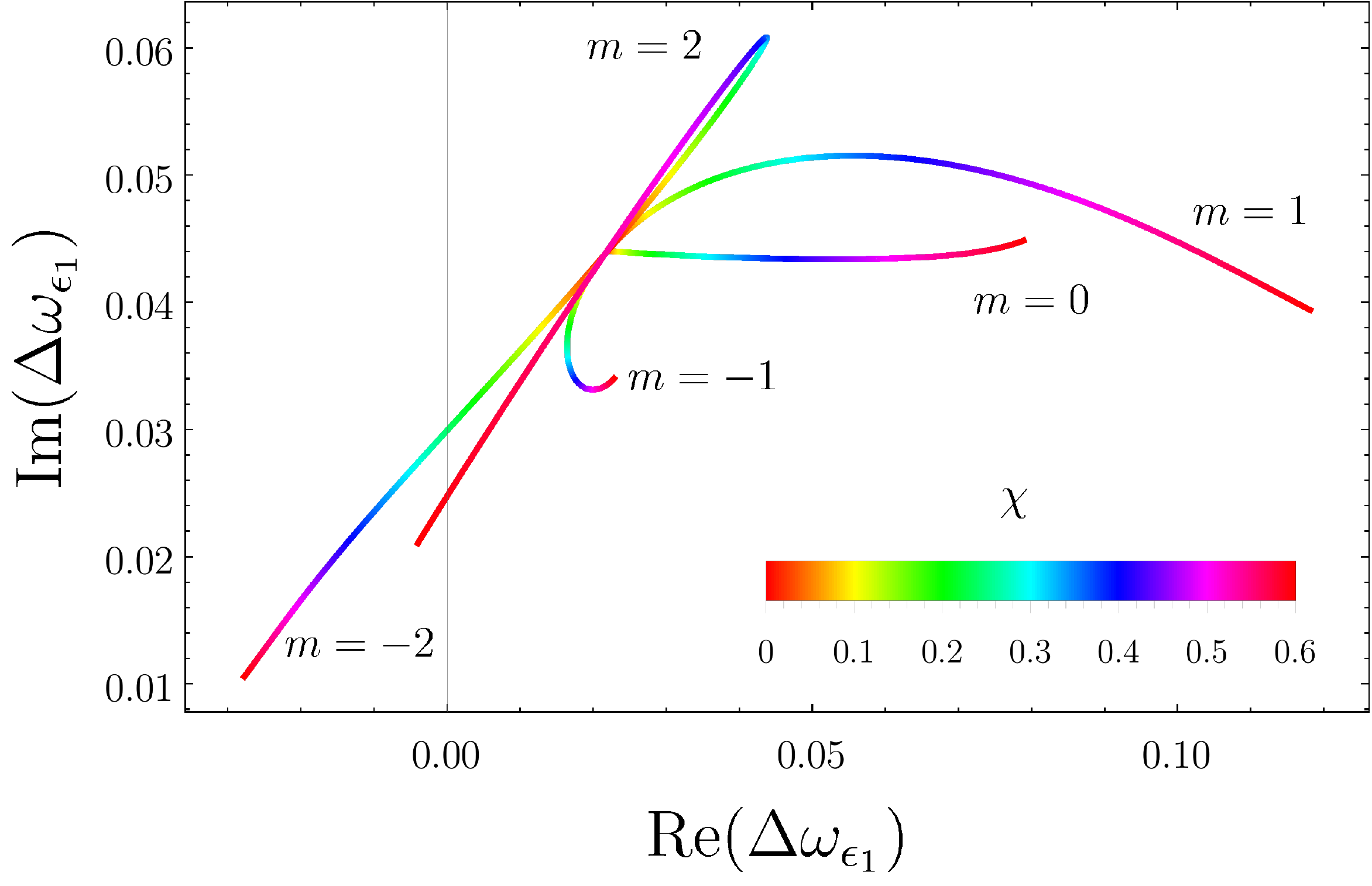}
		\includegraphics[width=0.5\textwidth]{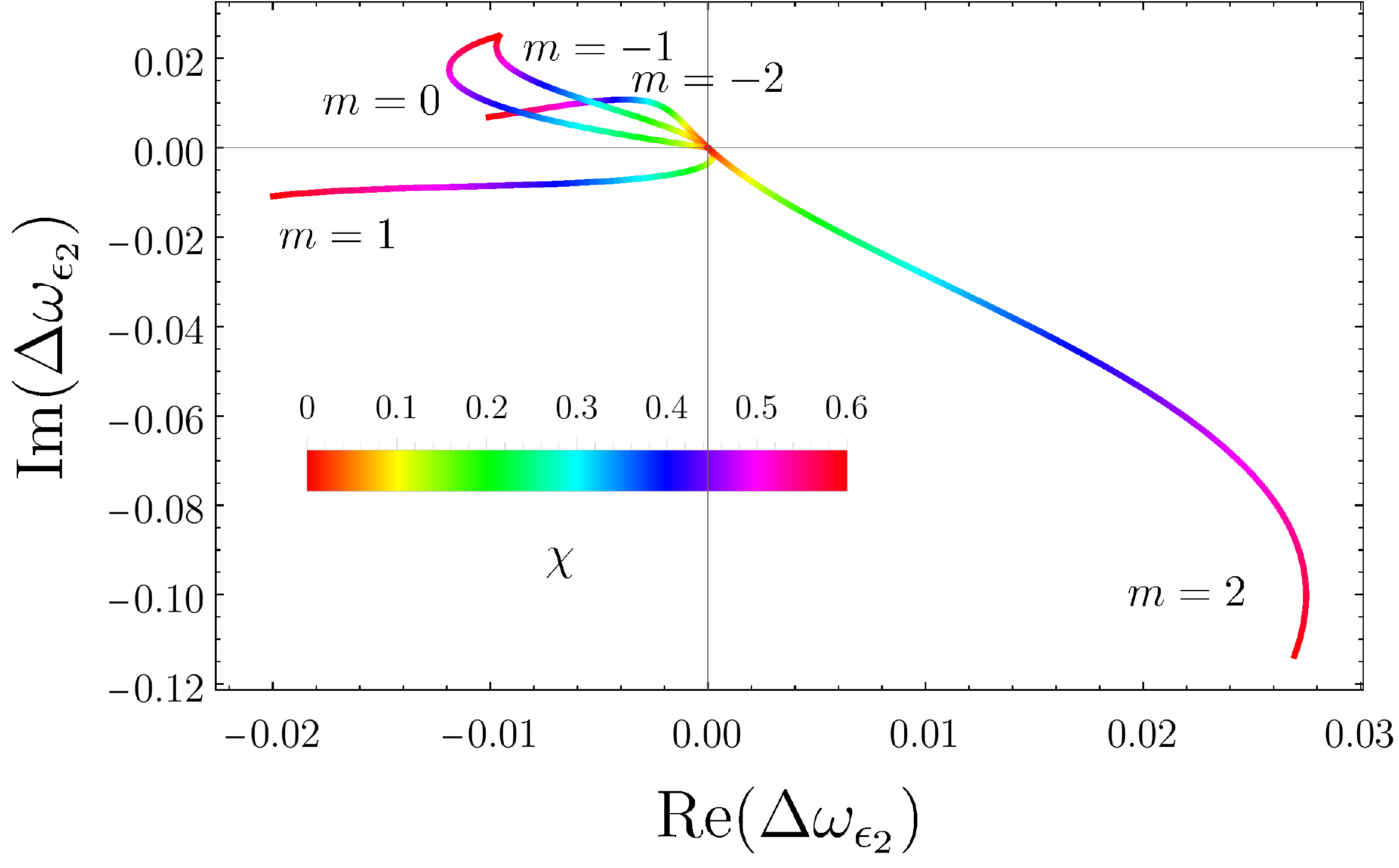}	
		\caption{Trajectories in the complex plane of the correction coefficients $\Delta\omega^{(\epsilon_1)}_{l,m,0}$ and $\Delta\omega^{(\epsilon_2)}_{l,m,0}$ as defined in \req{qnfrotationquartic}. From top to bottom we show the cases $l=0,1,2$ for all the values of $m$. The dependence in the spin is labeled by the color code.}
		\label{fig:qnfquartic}
	\end{center}
\end{figure}

Let us now present, in a less detailed way, the results for the quasinormal frequencies associated with the quartic corrections given in \req{eq:quarticL}. These appear at order $\mathcal{O}(\ell^6)$, and hence they are subleading whenever the corrections studied in the previous section are non-vanishing. In this case, the higher-curvature term controlled by the parameter $\epsilon_3$ is parity-breaking, and therefore it does not affect the scalar quasinormal frequencies. Thus, we can write 
\begin{equation}\label{qnfrotationquartic}
M\omega= M\omega^{(0)}+\frac{\ell^6}{M^6}\left(\epsilon_1\Delta\omega^{(\epsilon_1)}+\epsilon_2\Delta\omega^{(\epsilon_2)}\right)\, .
\end{equation}
For the few first values of $l$ we obtain the fundamental quasinormal frequencies by performing a numerical integration of the wave equation. We find that the convergence of the $\chi$-expansion of the effective potentials seems to be slower than in the case of the $\mathcal{O}(\ell^4)$ corrections, and thus, with an $\mathcal{O}(\chi^{14})$-expansion we have been able to obtain a reliable result only up to $\chi\sim0.6$, depending on the case. 
In Fig.~\ref{fig:qnfquartic} we show the trajectories in the complex plane of the coefficients $\Delta\omega^{(\epsilon_1)}$ and $\Delta\omega^{(\epsilon_2)}$ for  $l=0,1,2$ and all the values of $m$. We can observe that the corrections associated with $\epsilon_2$ are always vanishing for $\chi=0$. 

Polynomial fits of these coefficients are provided in appendix \ref{app:fit}, but for the sake of completeness let us study here the value of the quotients $\omega_R/\omega^{(0)}_R$, $\omega_I/\omega^{(0)}_I$ for the $l=m=2$ mode, analogously as we did for the $\mathcal{O}(\ell^4)$ corrections.
After fitting the numerical data, we find that these ratios are given by

\begin{align}
\notag
\frac{\omega_R}{\omega^{(0)}_R}=&1+0.0454 \hat{\epsilon }_1+\left(0.1723 \hat{\epsilon }_1+0.0289 \hat{\epsilon
}_2\right) \chi +\left(-0.1009 \hat{\epsilon }_1+0.0294 \hat{\epsilon }_2\right)
\chi ^2\\ \notag
&+\left(-0.1479 \hat{\epsilon }_1+0.5443 \hat{\epsilon }_2\right) \chi
^3+\left(-0.6853 \hat{\epsilon }_1-1.7375 \hat{\epsilon }_2\right) \chi ^4+\left(1.4074
\hat{\epsilon }_1+3.6973 \hat{\epsilon }_2\right) \chi ^5\\
&+\left(-2.3075 \hat{\epsilon
}_1-3.5189 \hat{\epsilon }_2\right) \chi ^6\, ,
\label{ratioRquartic}
\end{align}

\begin{align}
\notag
\frac{\omega_I}{\omega^{(0)}_I}=&1-0.4547 \hat{\epsilon }_1+\left(-0.6693 \hat{\epsilon }_1+0.5821 \hat{\epsilon
}_2\right) \chi +\left(0.8751 \hat{\epsilon }_1+1.653 \hat{\epsilon }_2\right) \chi
^2\\ \notag
&+\left(-8.1031 \hat{\epsilon }_1-6.0783 \hat{\epsilon }_2\right) \chi
^3+\left(36.0175 \hat{\epsilon }_1+28.78 \hat{\epsilon }_2\right) \chi ^4+\left(-66.4961
\hat{\epsilon }_1-53.8109 \hat{\epsilon }_2\right) \chi ^5\\ 
&+\left(54.9416 \hat{\epsilon
}_1+44.3289 \hat{\epsilon }_2\right) \chi ^6 \, ,
\label{ratioIquartic}
\end{align}
where we have introduced 
\begin{equation}
\hat\epsilon_{1}=\epsilon_{1} \frac{\ell^6}{M^6}\, ,\quad \hat\epsilon_{2}=\epsilon_{2} \frac{\ell^6}{M^6}\, .
\end{equation}
These expressions fit accurately the numerical values obtained in the interval $0<\chi<0.6$, and presumably provide a good approximation for somewhat larger values.  Now, in the static case we can compare these results with the ones obtained for gravitational perturbations in \cite{Cardoso:2018ptl}. The first thing we note is that the two different gravitational modes receive different corrections, but we may expect that at least one of them is corrected similarly as the scalar modes. This is indeed what we observed before in the case of the corrections to axial (parity-odd) quasinormal frequencies in Einstein-scalar-Gauss-Bonnet gravity --- see the discussion around Eq.~\req{axialEdGB}. For the quartic theories in Eq.~\req{eq:quarticL}, we could expect axial perturbations associated to $\epsilon_1$ and polar perturbations associated to $\epsilon_2$ to be relatively similar to the scalar ones.\footnote{The effect of the parameter $\epsilon_3$ in the perturbations is more involved since it mixes parity-odd and parity-even modes, and its effect on quasinormal frequencies has not been studied yet.} We observe that, in fact, the parameter $\epsilon_2$ does not modify polar modes, in analogy with the situation for scalar modes. However, the corrections to the axial quasinormal frequencies associated to $\epsilon_1$ do not resemble the scalar ones we show above. The discrepancy nevertheless seems to decrease as we increase $l$ --- see below --- so at least we will have some similarity for large enough $l$. 

Regarding eikonal modes, it is possible to perform again an analysis of lightring geodesics and to find the quasinormal frequencies using \req{eikfreq}. In this case the orbital frequencies and the Lyapunov exponents can be written as
\begin{align}
\Omega_{+}&=\Omega_{+}^{(0)}+\frac{\ell^6}{M^7}\left(\epsilon_1\Omega_{+}^{(\epsilon_1)}+\epsilon_2\Omega_{+}^{(\epsilon_2)}\right)\, ,\quad
\lambda=\lambda^{(0)}+\frac{\ell^6}{M^7}\left(\epsilon_1 \lambda^{(\epsilon_1)}+\epsilon_2 \lambda^{(\epsilon_2)}\right)\, ,
\end{align}
where, up to order $\chi^8$, the coefficients $\Omega_{+}^{(\epsilon_{1,2})}$ and $ \lambda^{(\epsilon_{1,2})}$ are given by

{\allowdisplaybreaks
\begin{align}
\notag
\Omega_{+}^{(\epsilon_{1})}=&\frac{832}{59049 \sqrt{3}}+\frac{159872 \chi }{5845851}-\frac{5606779 \chi ^2}{74401740
   \sqrt{3}}-\frac{5998600441 \chi ^3}{52665271659}-\frac{363654604519 \chi
   ^4}{1354249842660 \sqrt{3}}\\ \notag
   &-\frac{3392895395699 \chi
   ^5}{18485510352309}-\frac{5131021459454959 \chi ^6}{13309567453662480
   \sqrt{3}}-\frac{19543868951454137 \chi ^7}{78321685400398440}\\
   &-\frac{181533259063565863
   \chi ^8}{362020234739619456 \sqrt{3}}\, ,\label{Omegaepsilon1}\\\notag
\Omega_{+}^{(\epsilon_{2})}=&\frac{2048 \chi }{19683}+\frac{9489197 \chi ^2}{22733865 \sqrt{3}}+\frac{140916511 \chi
   ^3}{464420385}+\frac{1050592475 \chi ^4}{1350391581 \sqrt{3}}+\frac{183051051307 \chi
   ^5}{326023110270}\\
   &+\frac{299759777186281 \chi ^6}{246473471364120
   \sqrt{3}}+\frac{708676913356721 \chi ^7}{897867645683580}+\frac{57722661833164033 \chi
   ^8}{37710441118710360 \sqrt{3}}\,  ,
   \end{align}
    \begin{align}\notag
\lambda^{(\epsilon_{1})}=&-\frac{7232}{59049 \sqrt{3}}-\frac{3328 \chi }{19683}-\frac{77953591 \chi ^2}{409209570
   \sqrt{3}}+\frac{4635226837 \chi ^3}{112854153555}+\frac{27299185267 \chi
   ^4}{72921145374 \sqrt{3}}\\\notag
   &+\frac{4547096768374 \chi
   ^5}{13203935965935}+\frac{11075262022707251 \chi ^6}{13309567453662480
   \sqrt{3}}+\frac{13678829006217853 \chi ^7}{24242426433456660}\\
   &+\frac{769883777737335671\chi ^8}{678787940136786480 \sqrt{3}}\, ,\label{Lyapepsilon1}\\\notag
   \lambda^{(\epsilon_{2})}=&\frac{2048 \chi }{19683}+\frac{9489197 \chi ^2}{22733865 \sqrt{3}}+\frac{140916511 \chi
   ^3}{464420385}+\frac{1050592475 \chi ^4}{1350391581 \sqrt{3}}+\frac{183051051307 \chi
   ^5}{326023110270}\\
   &+\frac{299759777186281 \chi ^6}{246473471364120
   \sqrt{3}}+\frac{708676913356721 \chi ^7}{897867645683580}+\frac{57722661833164033 \chi
   ^8}{37710441118710360 \sqrt{3}} \, .
\end{align}
}
Again one can check that these expressions agree with different estimates of the quasinormal frequencies for sufficiently large $l$. 
Let us then return to the comparison with gravitational perturbations in the static case. Using the results for the master equations in Ref.~\cite{Cardoso:2018ptl}, we have checked that, in the case of axial perturbations, the corrections associated to $\epsilon_1$ in the eikonal limit match our formulas \req{Omegaepsilon1} and \req{Lyapepsilon1} with $\chi=0$. Therefore, we expect that, for large $l$ (and perhaps not necessarily very large), the scalar quasinormal frequencies will resemble the axial gravitational ones, also for the spinning case. As for the corrections to the polar (axial) modes associated to $\epsilon_1$  ($\epsilon_2$), their dependence on $l$ is quite unusual, as noted in \cite{Cardoso:2018ptl}, and the eikonal limit does not seem to make sense in those cases --- we will come back to this in the discussion section.

\section{Observing deviations from GR}\label{sec:obsdev}

In the last section we have presented an extensive analysis of the quasinormal frequencies of rotating black holes in the general set of higher-derivative extensions of GR given by \req{Action} and \req{eq:quarticL}. We should now use these results to quantify the observational implications of this. Of course, we are working only with scalar perturbations, which are different from gravitational ones. However, as we have seen, the relative corrections to the gravitational and scalar quasinormal frequencies may not be very different in some cases, even for $l=2$, and in other cases we can expect that they become similar for higher $l$. Since the tests that we will perform depend on the relative corrections to the QNFs, we can at least get some degree of approximation to the real problem by using the scalar QNFs. Otherwise, the analysis we carry out in this section can be considered as a simulacrum of the type of study one could perform if the corrections to the gravitational frequencies were available. 

We want to focus on two key aspects of the ringdown signal. First, by measuring one (complex) quasinormal frequency, one is able to infer the mass and the spin of the black hole, upon the assumption that this is given by the Kerr metric. Thus, one relevant question is what is the error in the mass and spin estimates if the black hole is actually non-Kerr, \textit{i.e.} a solution of one of the higher-derivative theories we consider.  Second, precisely because we can always fit a given QNF to a Kerr value, one QNF is not enough to test GR unless we have an independent estimate of the mass and the spin. We need at least two different QNFs in order to test GR using only ringdown measurements, and in that case it is important to determine to what extent the QNM spectrum of the corrected black holes is distinguishable from the Kerr spectrum. Next we analyze these two questions in more detail. 

\subsection{Inferring mass and spin from ringdown}
One way of determining the final mass and spin of the black hole resulting from the merger of a BH binary consists in measuring its dominant QNM. If there are corrections to GR, then the estimates of the black hole's charges using the quasinormal frequencies --- upon the assumption of the Kerr hypothesis --- will be off and they will not agree, for instance, with the estimates coming from the inspiral. Thus, these two independent estimates of $M$ and $\chi$ can be used to test GR. Of course, one would need take into account that the inspiral is also corrected, and so the mass and spin inferred in this way will also have deviations with respect their actual values. We will not worry about this and we will simply assume that somehow we have been able to determine the BH's properties using different measurements than the ringdown.\footnote{In the case of Einstein-dilaton-Gauss-Bonnet theory, the corrections to the final black hole mass and spin from inspiral observables have been recently computed \cite{Carson:2020ter}. It would be interesting to extend these results to the more general class of theories \req{Action}, which would allow us to perform a more accurate comparison between inspiral-based and ringdown-based estimations of the remnant black hole's mass and spin.} 

Let us then consider that we have a black hole of mass $M$ and spin $\chi$ with higher-derivative corrections, so that its quasinormal frequencies can generically be written as
\begin{equation}
\omega(M,\chi)=\omega_{(0)}(M,\chi)+\Delta\omega(M,\chi)\, ,
\end{equation}
where $\omega_{(0)}$ is the uncorrected, Kerr value, and $\Delta\omega$ is the correction. But then, suppose that we ignore that these deviations are there and we try to fit one of these quasinormal frequencies to a Kerr one. The ``equivalent'' Kerr black hole will have a certain mass $\tilde M$ and spin $\tilde\chi$ which will be different from the real ones. Thus, we would be solving the problem
\begin{equation}\label{equivKerreq}
\omega_{(0)}(\tilde M,\tilde \chi)=\omega(M,\chi)\, .
\end{equation}
If the corrections are (perturbatively) small then $(\tilde M,\tilde \chi)$ will be close to $(M, \chi)$ and we can write
\begin{equation}
\tilde M=M+\delta M\, ,\quad \tilde\chi=\chi+\delta\chi\, .
\end{equation}
Then, expanding linearly in $\delta M$ and $\delta\chi$, Eq.~\req{equivKerreq} can be written as
\begin{equation}
-\frac{\delta M}{M}\omega_{(0)}+\partial_{\chi}\omega_{(0)}\delta\chi \Big|_{(M,\chi)}=\Delta\omega(M,\chi)\, .
\end{equation}
Taking the real and imaginary parts we can then obtain $\delta M$ and $\delta \chi$. A compact way of expressing the result is as follows:
\begin{align}\label{generalshift}
\frac{\delta M}{M}=\frac{\text{Im}\left(\partial_{\chi}\omega^{*}_{(0)}\Delta\omega\right)}{\text{Im}\left(\omega^{*}_{(0)}\partial_{\chi}\omega_{(0)}\right)}\, ,\quad \delta\chi=\frac{\text{Im}\left(\omega^{*}_{(0)}\Delta\omega\right)}{\text{Im}\left(\omega^{*}_{(0)}\partial_{\chi}\omega_{(0)}\right)}\, ,
\end{align}
where $\omega^{*}_{(0)}$ is the complex conjugate. Before showing the values of these shifts for the set of corrections considered, let us illustrate in the general case how one can use experiments to constrain these corrections. Let us imagine that we have experimentally determined the mass of the black hole using the ringdown (the case of the spin is analogous), so that we get the estimate $M_1=M_{\rm ring}\pm \Delta M_{\rm ring}$. On the other hand, suppose that we have another independent estimate of the mass $M_2=M_{\rm other}\pm\Delta M_{\rm other}$. Then we can compute the relative difference between both estimates
\begin{equation}
\delta_M=\frac{M_1-M_2}{M_1}=\frac{M_{\rm ring}-M_{\rm other}}{M_{\rm ring}}\pm \frac{M_{\rm other}\Delta M_{\rm ring}+M_{\rm ring}\Delta M_{\rm other}}{M_{\rm ring}^2}\, .
\end{equation}
Now, assuming that the estimate $M_{\rm other}$ is faithful so that it yields the real value of the mass, then the expected value of $\delta_M$ is precisely $\langle\delta_M\rangle=\delta M/M$. If, eventually, the data is consistent with GR --- because 0 belongs to the confidence interval of  $\delta_M$ --- , then it means that the error is larger than $\delta M/M$. Thus, we get the following implication
\begin{equation}
\text{Data consistent with GR}\,\Rightarrow \,\left|\frac{\delta M}{M}\right|<\frac{M_{\rm other}\Delta M_{\rm ring}+M_{\rm ring}\Delta M_{\rm other}}{M_{\rm ring}^2}\, .
\end{equation}
Repeating the same analysis for the spin we get that the constraint is $|\delta\chi|<\Delta \chi_{\rm ring}+\Delta \chi_{\rm other}$\, where $\Delta \chi_{\rm ring}$ and $\Delta \chi_{\rm other}$ are the errors in the ringdown and alternative estimates of $\chi$. In the end, these are constraints on the possible corrections to GR. On the other hand, reversing the implications we get a lower bound on the precision we need in order to be able to detect these corrections. Let us now evaluate these formulas for the higher-derivative extensions of GR in Eqs.~\req{Action} and \req{eq:quarticL}.

\begin{figure}[t!]
	\begin{center}
		\includegraphics[width=0.49\textwidth]{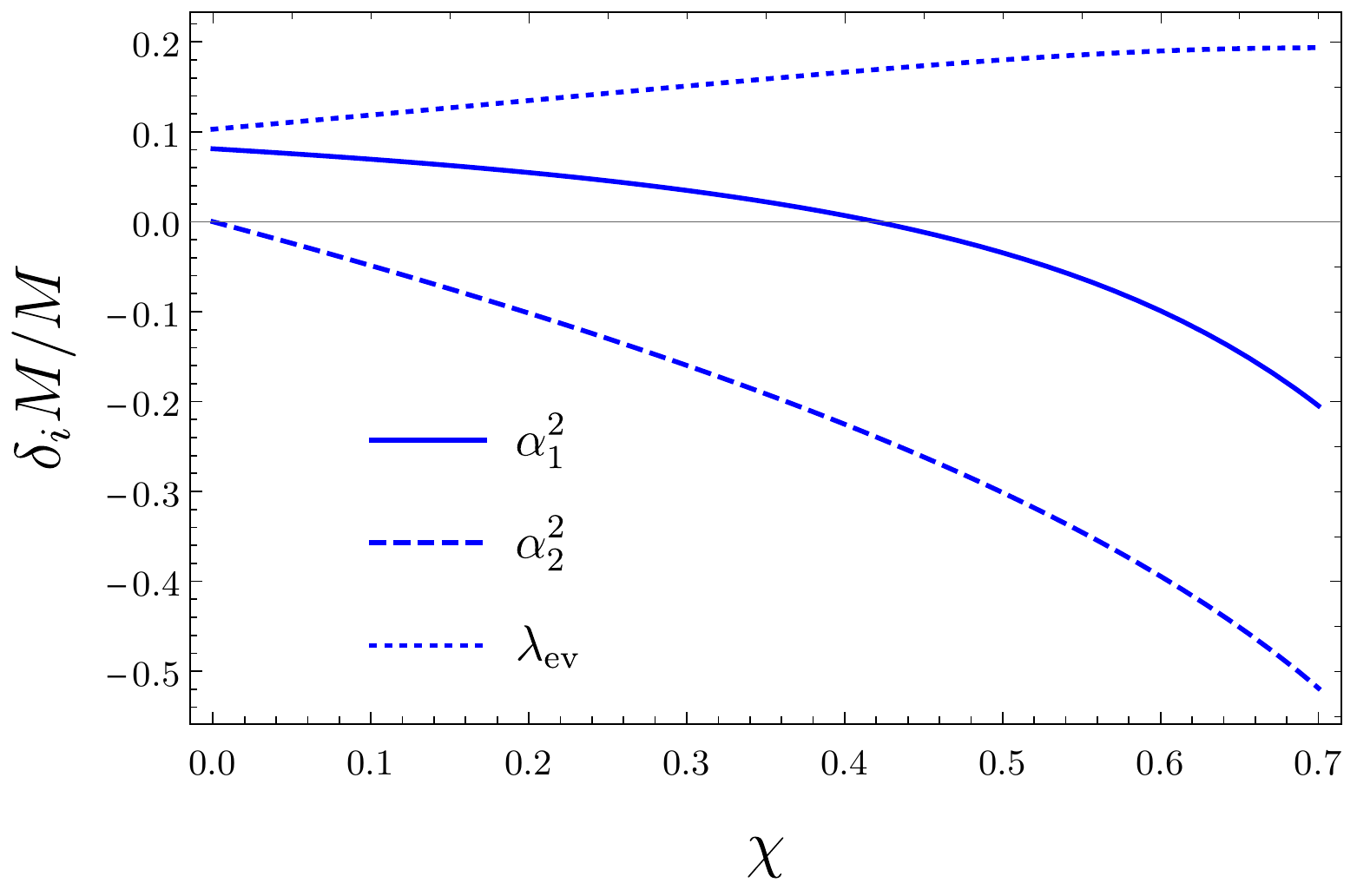}
		\includegraphics[width=0.49\textwidth]{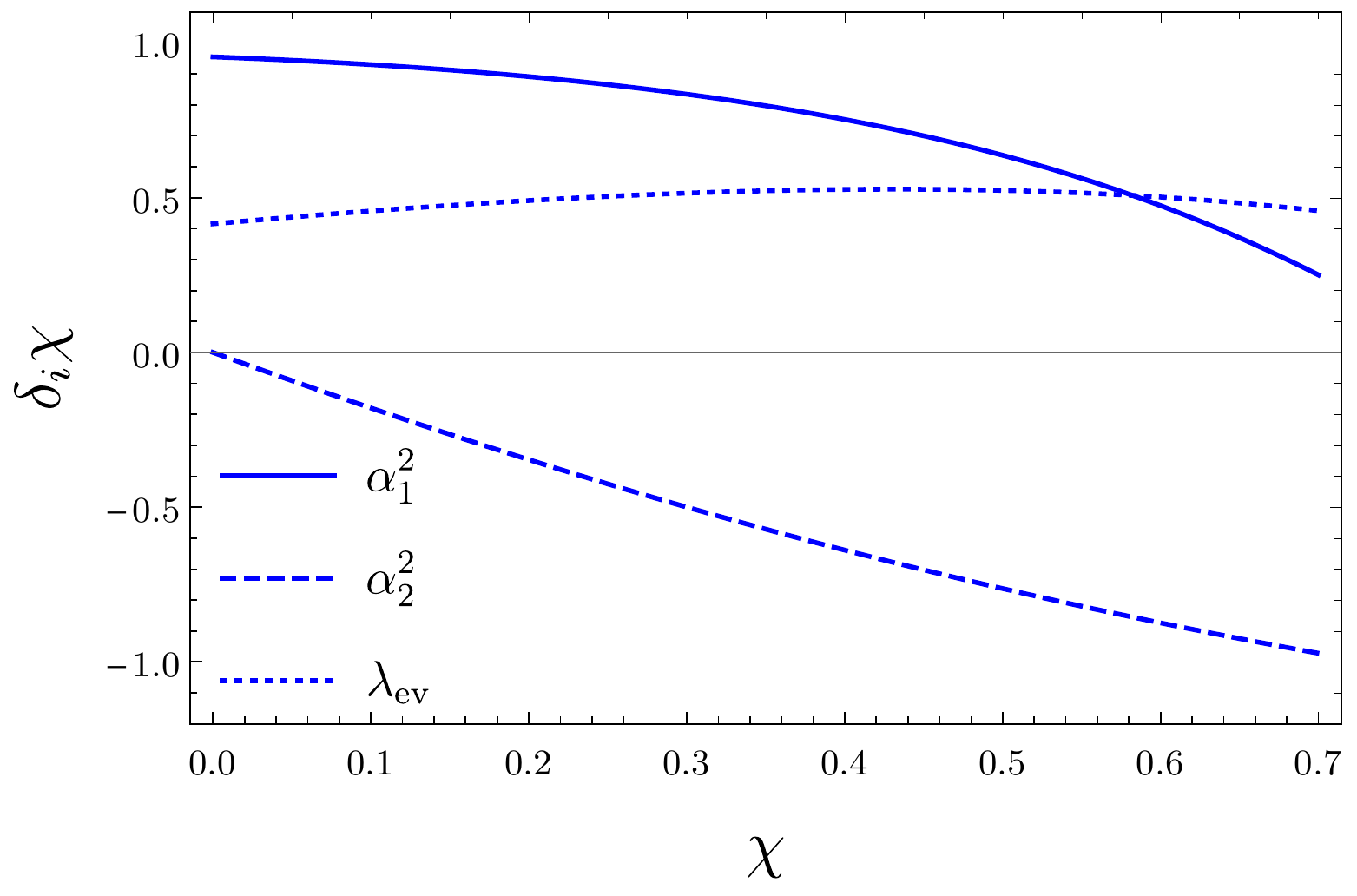}
		\includegraphics[width=0.49\textwidth]{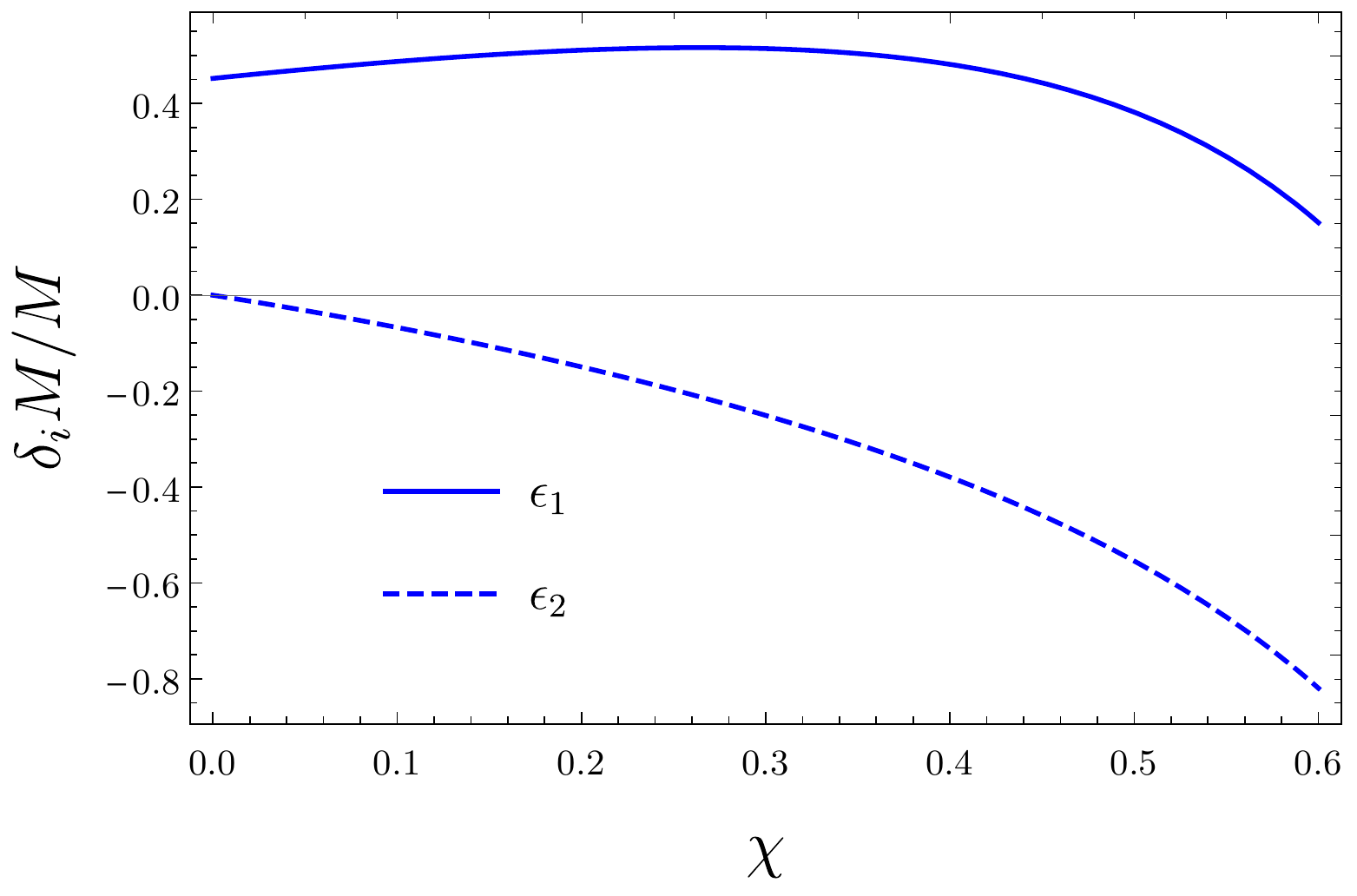}
		\includegraphics[width=0.49\textwidth]{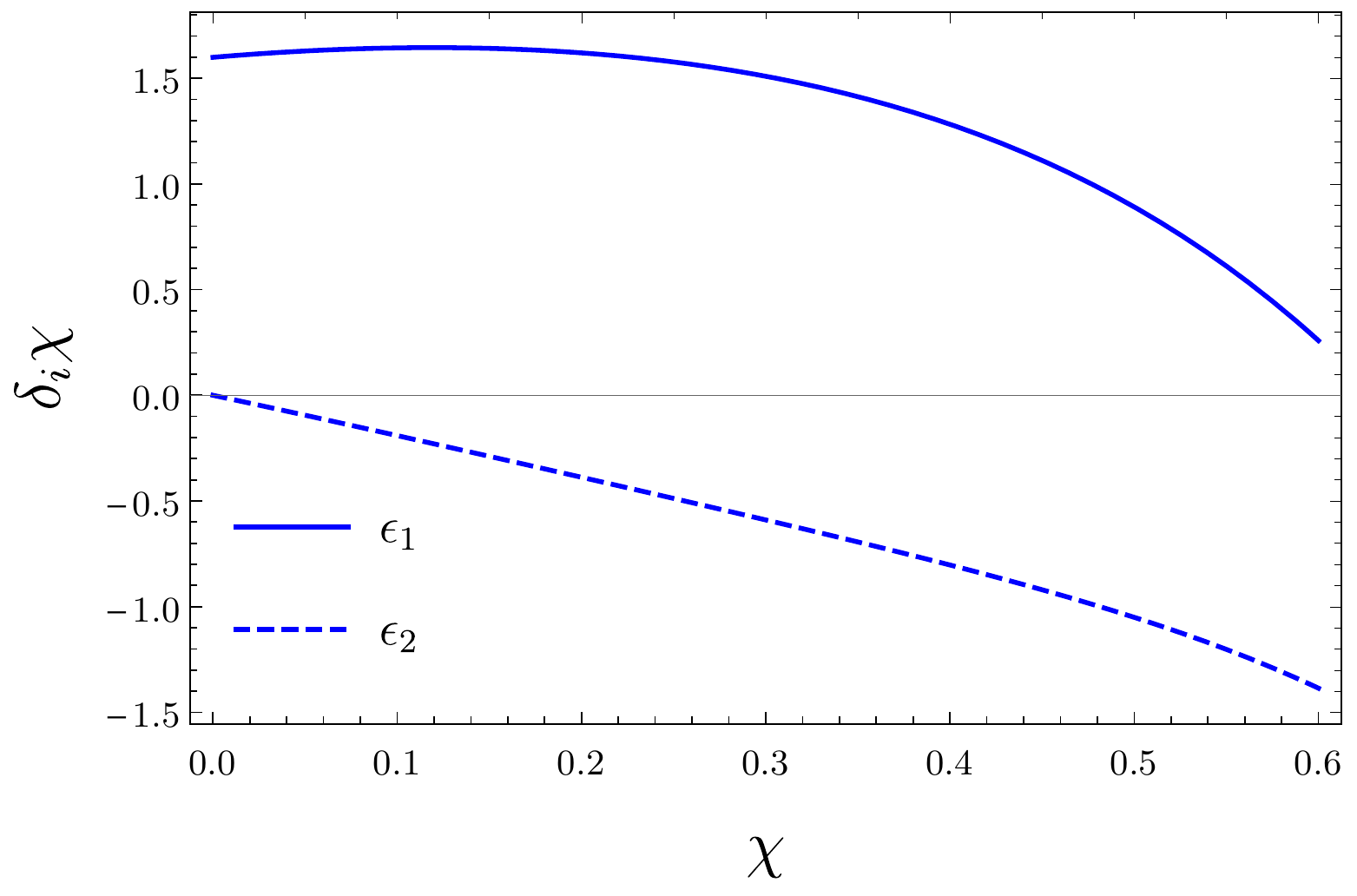}
		\caption{Error in the mass and angular momentum estimates when inferred from measurement of the $l=m=2$ quasinormal frequency, due to higher-derivative corrections. We show the coefficients entering in  Eqs.~\req{eq:dMcorr} and \req{eq:dchicorr}. Upper row:  $\mathcal{O}(\ell^4)$ corrections. Lower row: $\mathcal{O}(\ell^6)$ corrections.}
		\label{fig:dmdchi}
	\end{center}
\end{figure}

Each coupling modifies the equivalent Kerr parameters in a different way, so it is useful to write
\begin{align}\label{eq:dMcorr}
\frac{\delta M}{M}=&\frac{\ell^4}{M^4}\left(\alpha_1^2\frac{\delta_{\alpha_1^2} M}{M}+\alpha_2^2\frac{\delta_{\alpha_2^2} M}{M}+\lambda_{\rm ev}\frac{\delta_{\lambda_{\rm ev}} M}{M}\right)+\frac{\ell^6}{M^6}\left(\epsilon_1\frac{\delta_{\epsilon_1} M}{M}+\epsilon_2\frac{\delta_{\epsilon_2} M}{M}\right)\, ,\\
\label{eq:dchicorr}
\delta\chi=&\frac{\ell^4}{M^4}\left(\alpha_1^2\delta_{\alpha_1^2} \chi+\alpha_2^2\delta_{\alpha_2^2} \chi+\lambda_{\rm ev}\delta_{\lambda_{\rm ev}} \chi\right)+\frac{\ell^6}{M^6}\left(\epsilon_1\delta_{\epsilon_1} \chi+\epsilon_2\delta_{\epsilon_2} \chi\right)\, ,
\end{align}
where each of the $\delta_i M/M$ and $\delta_i\chi$ are dimensionless functions of the spin $\chi$. Also, it must be noted that the $\mathcal{O}(\ell^6)$ corrections are relevant only when the $\mathcal{O}(\ell^4)$ ones vanish, because if the $\mathcal{O}(\ell^4)$ terms are present there would be more $\mathcal{O}(\ell^6)$ terms in the effective action besides the ones considered in \req{eq:quarticL}.\footnote{For instance, a term such as $\ell^2\phi_1^2\mathcal{X}_4$ would modify the metric at order $\ell^6$ if $\alpha_1\neq 0$.}
The coefficients $\delta_i M/M$ and $\delta_i\chi$ can be easily obtained from the formulas \req{generalshift} using the fits to the QNFs provided in the appendix \ref{app:fit}. We show the profile of these quantities as functions of $\chi$ in Fig.~\ref{fig:dmdchi}. 

It is interesting to see what happens if one tries to determine the mass and spin from the measurement of a higher $l$ QNM, say one of the form $m=l$. It turns out that the shift in the mass and the spin does not change much with $l$. In fact, if one derives $\delta M/M$ and $\delta\chi$ for the eikonal QNFs, the corresponding curves are still quite similar to those shown in Fig.~\ref{fig:dmdchi}. Hence, we may also expect that these curves provide an approximation for the gravitational case. 

Let us write down an example in order to show the type of constraints we can obtain. For instance, let us suppose that we have measured the $l=m=2$ quasinormal frequency of a black hole of spin $\chi=0.7$ --- which is about the estimated value for the  LIGO/Virgo black holes \cite{LIGOScientific:2018mvr,Fishbach:2017dwv}. This black hole has a mass $M$, and for the purposes of this thought experiment we can assume $M_{\rm ring}\approx M_{\rm other}\approx M$. Then, we have the following implication:
\begin{equation}
\text{Data consistent with GR}\,\Rightarrow \begin{cases}\,\ell^4\left|0.20\alpha_1^2+0.52\alpha_2^2-0.19\lambda_{\rm ev}\right|&<M^3\left(\Delta M_{\rm ring}+\Delta M_{\rm other}\right)\, ,\\
\ell^4\left|0.25\alpha_1^2-0.97\alpha_2^2+0.46\lambda_{\rm ev}\right|&<M^4\left(\Delta \chi_{\rm ring}+\Delta \chi_{\rm other}\right)\, ,
\end{cases}
\label{eqn:constraints1}
\end{equation}
Thus, for smaller masses and higher precision, we can set more accurate bounds on these couplings. For GW150914, using the inspiral together with numerical relativity, one finds roughly $\frac{\Delta M_{\rm other}}{M} \approx 0.06$, $\Delta \chi_{\rm other} \approx 0.07$ and $\frac{\Delta M_{\rm ring}}{M} \approx 0.06$, $\Delta \chi_{\rm other} \approx 0.06$   \cite{TheLIGOScientific:2016src}. This cannot set a definitive bound on all the couplings since we have three of them and only two constraints. We would need to measure several black holes of sufficiently different spin in order to set a bound on all of them. Nevertheless, taking for example $\lambda_{\rm ev} = 0$, \eqref{eqn:constraints1} would imply approximately $\frac{\ell^4 \alpha_1^2}{M^4} <0.2$ and $\frac{\ell^4 \alpha_2^2}{M^4} <0.2$. This is hardly a constraint at all.

To get a better quantitative understanding for more promising third generation gravitational wave detectors, we consider detecting a number of $N$ binary black hole coalescences with the Einstein Telescope (ET) \cite{Punturo:2010zz}. The black holes in the binaries are drawn uniformly from a log-flat distribution between $[5, 95]M_{\odot}$ with a uniform spin distribution in $[-1, 1]$. We further limit ourselves to events for which the $l=2$, $m=2$ QNM is measured with SNR of 100 and assume this measurement error on the ringdown is dominant as compared to an independent mass and spin estimate from the inspiral. To determine the final black hole mass and spin we use semi-analytic formulae based on NR as determined in \cite{Healy:2017vuz}. The main motivation to make these choices is to easily compare against the framework presented in \cite{Maselli:2019mjd}. However, we shall see that in this mass range only the lightest few events contribute significantly to the constraints, as is expected by the $M^4$ suppression. As a result, we will limit ourselves further to only the subset of these low mass black hole binaries.  \\

\begin{figure}[t!]
	\begin{center}
		\includegraphics[width=0.48\textwidth]{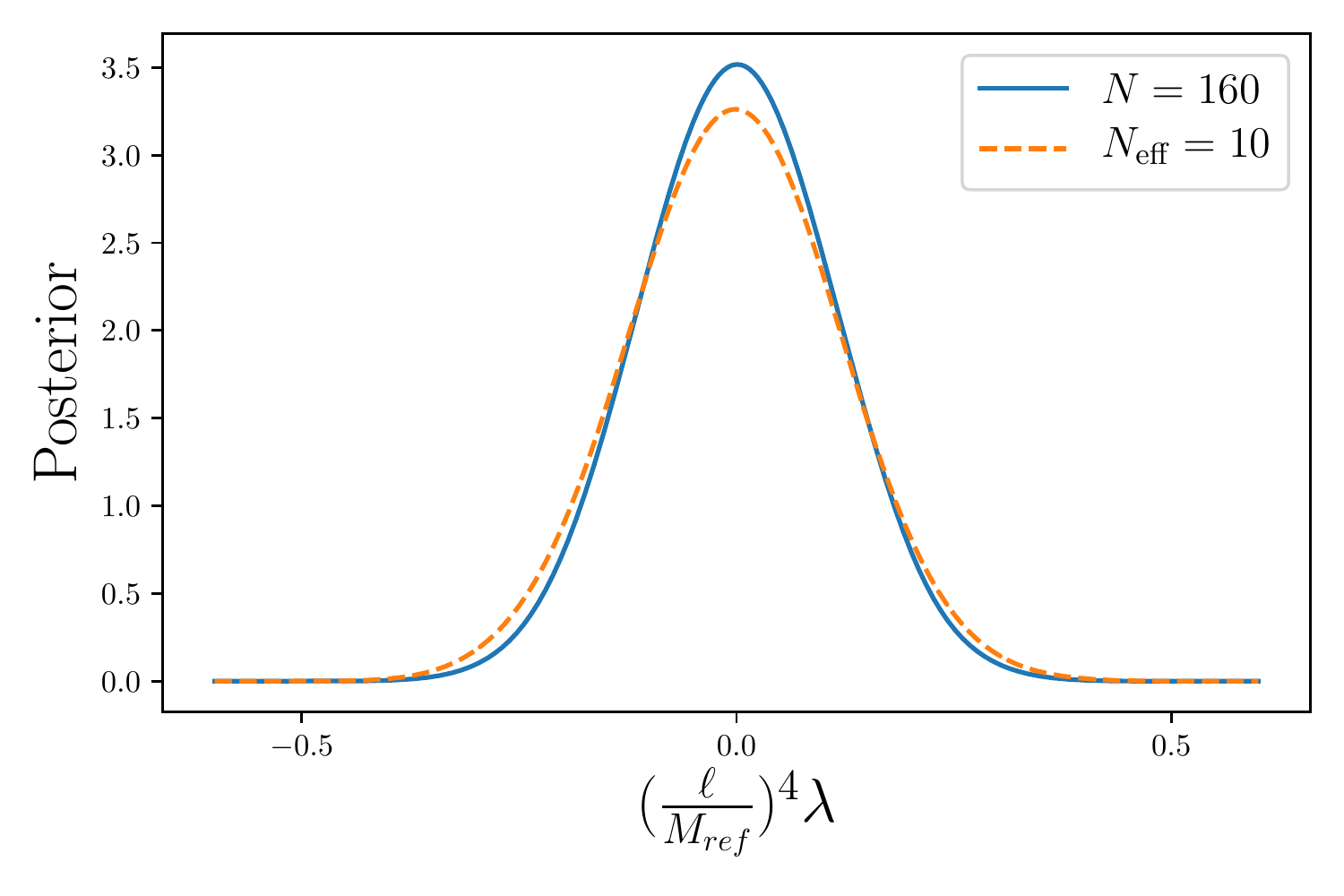}
		\includegraphics[width=0.48\textwidth]{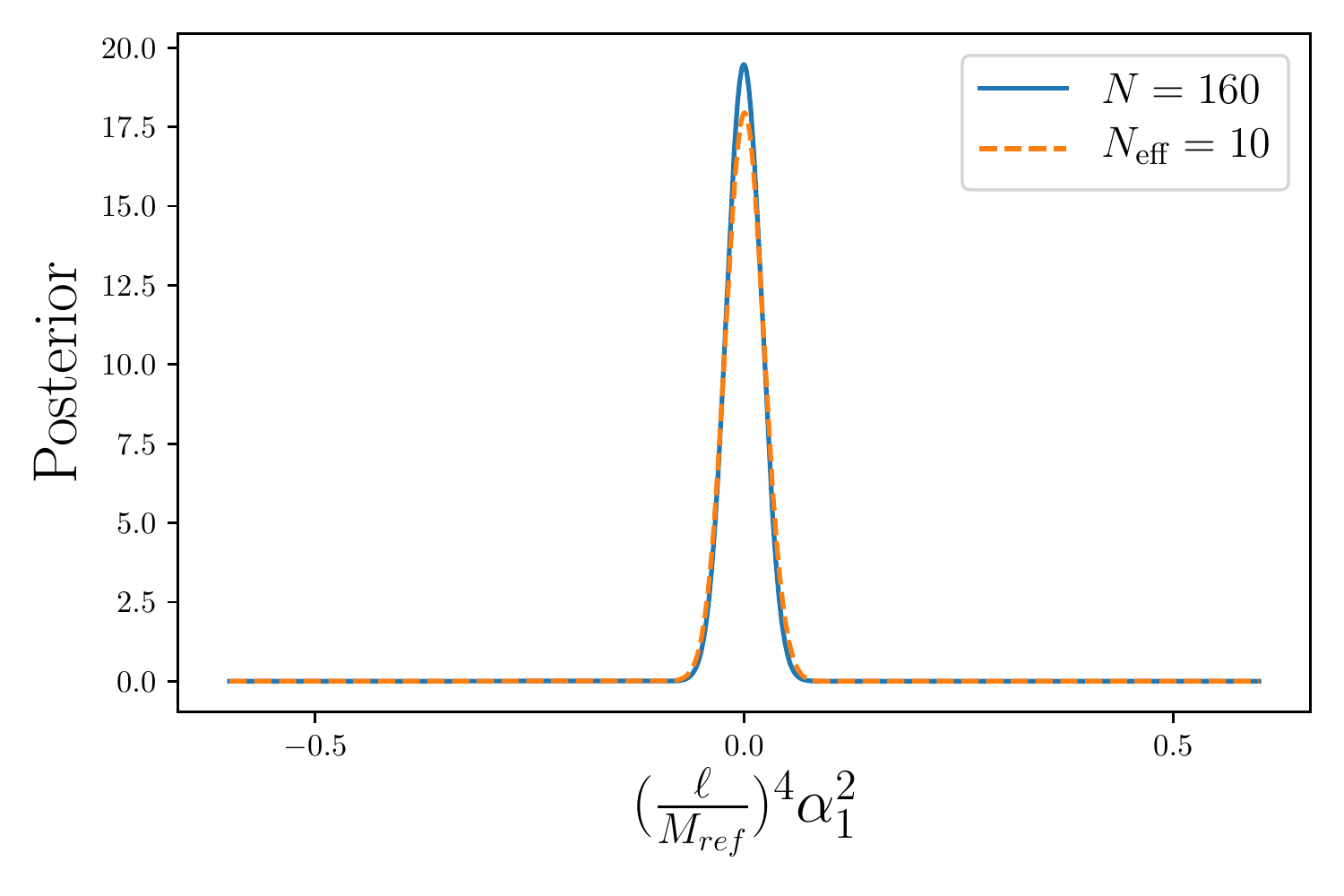}
		\includegraphics[width=0.48\textwidth]{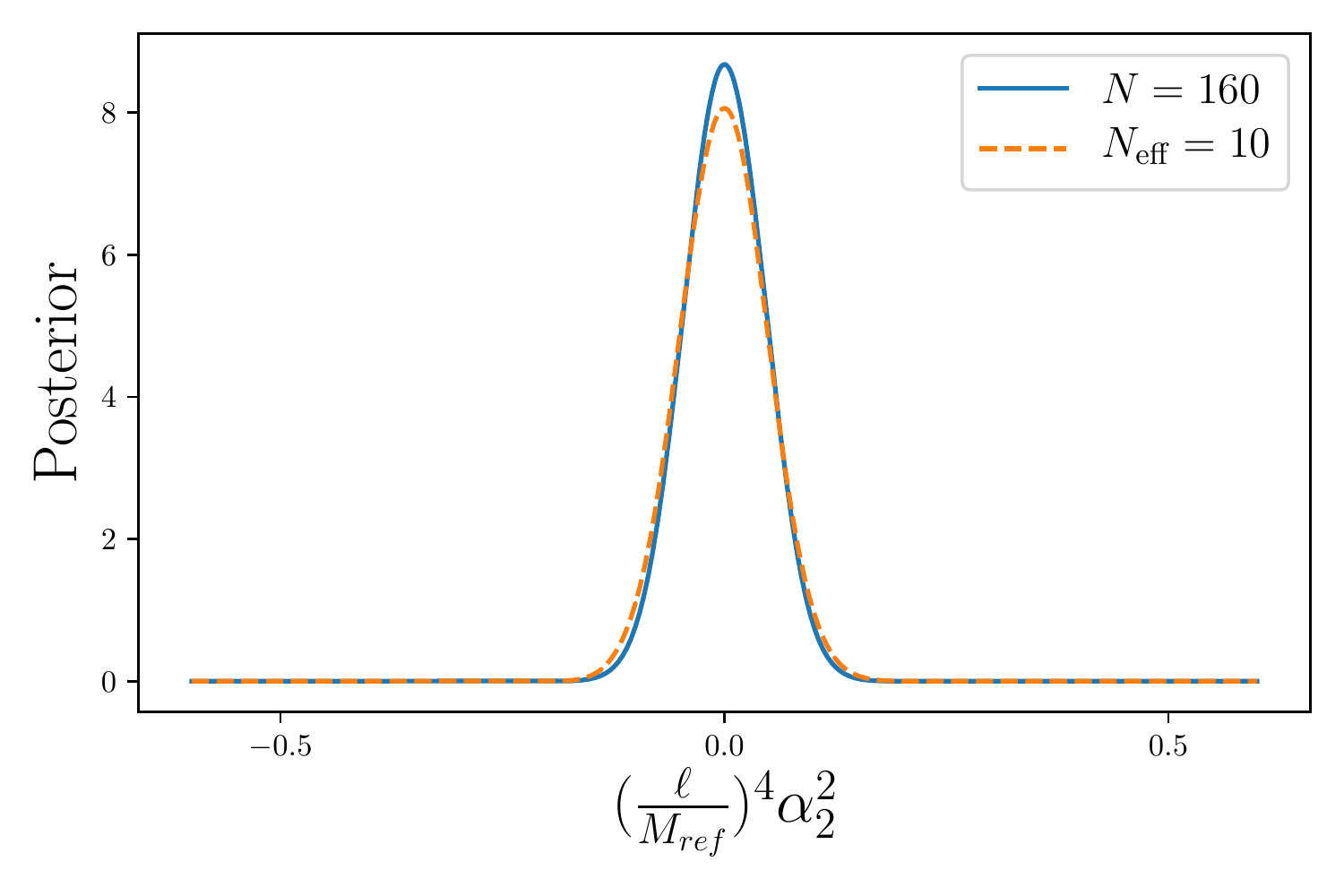}
		\caption{Posterior distributions for the dimensionless couplings $\hat \lambda$,  $\hat \alpha_1^2$, $\hat \alpha_2^2$ with respect to $M_{\rm ref} = 10 M_{\odot}$ after measuring $160$ events from a population in the range of masses $[5, 95]M_{\odot}$ with ET (full) as well as the posterior taking into account only the ten lowest mass events (dashed).}
		\label{fig:effectivepopulation}
	\end{center}
\end{figure}

The set of $N$ binary black hole detections will be analyzed using a Bayesian approach, sampling the posterior on our model parameters $\vec{\theta} = \lbrace \frac{\ell^4 \alpha_1^2}{M_{\rm ref}^4}, \frac{\ell^4 \alpha_2^2}{M_{\rm ref}^4}, \frac{\ell^4 \lambda}{M_{\rm ref}^4} \rbrace$ with uniform priors in the range $\lbrack-0.6, 0.6\rbrack$\footnote{For convenience we allow for the possibility that $\frac{\ell^4 \alpha_{1,2}^2}{M_{\rm ref}^4} < 0$.}. Here we have introduced the reference mass scale which will depend on the detector under consideration. The likelihood of event $i$ is given by a Gaussian centered around the expected value of the QNM $\omega_i$, computed using \req{ratioR} and \req{ratioI}, given mass $M_i$, spin $\chi_i$ and for model parameters $\vec{\theta}$,
\be
\cL_i(\omega^{\rm obs}_i|\vec{\theta}) = \cN(\omega^{\rm obs}_i; \omega_i, \Sigma_i)\, .
\ee 
Here $\omega_i$ includes the real as well as the imaginary part of the QNM. $\Sigma_i$ is the covariance matrix, including the correlation between these real and imaginary parts as inferred from a Fisher matrix analysis \cite{Berti:2005ys}. The full likelihood for all events is then simply given by 
\be \label{eqn:fulllikelihood}
\cL(\lbrace \omega^{\rm obs} \rbrace|\vec{\theta}) = \prod_{i=1}^{N} \cL_i(\omega^{\rm obs}_i|\vec{\theta})\, .
\ee

\begin{figure}[t!]
	\begin{center}
		\includegraphics[width=0.99\textwidth]{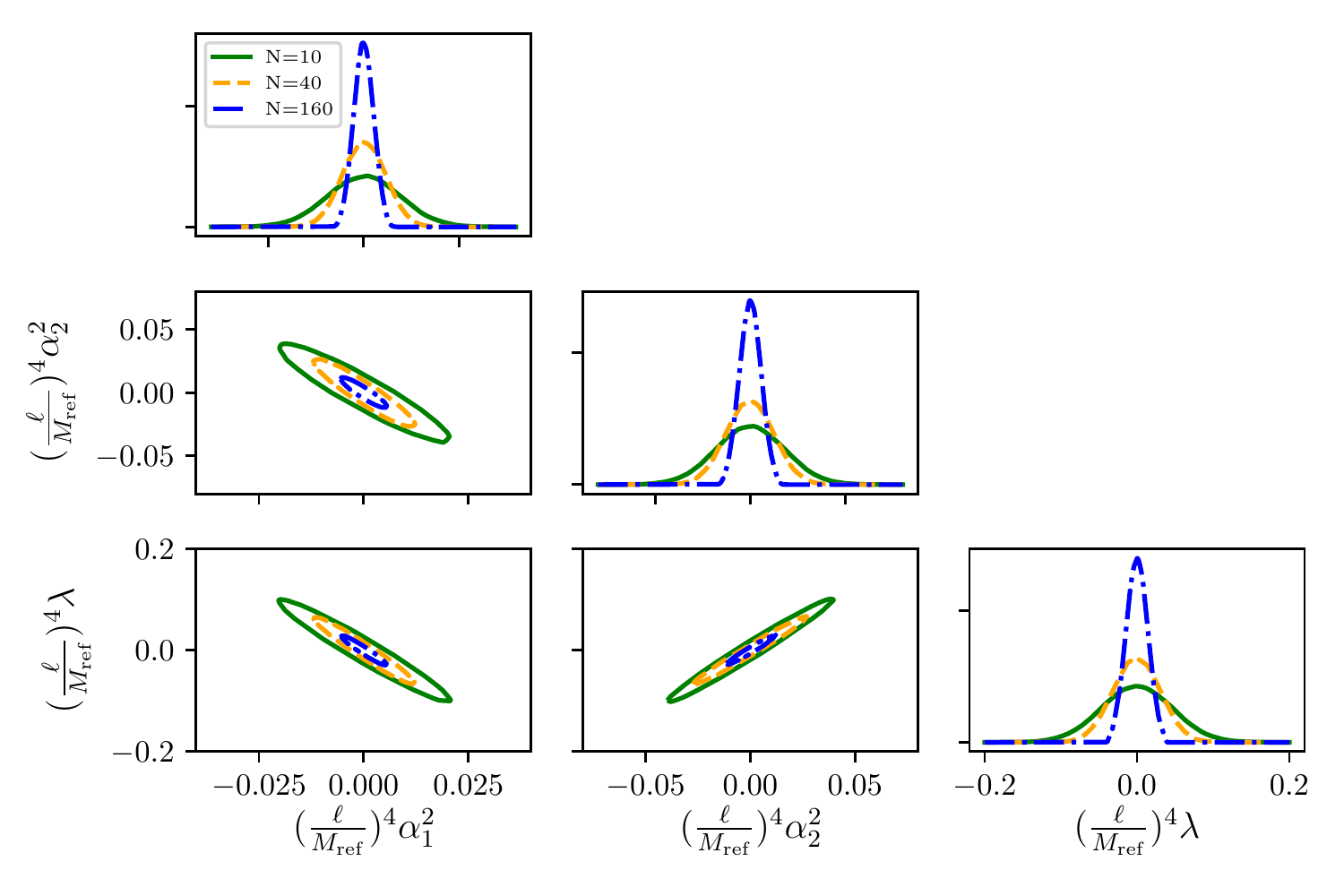}
		\caption{Corner plot for the posterior on the model parameters obtained for, from outer to inner, 10 (green), 40 (orange) and 160 (blue) ET observations of black hole binary coalescences from a restricted mass range $[5, 6.5]M_{\odot}$ if general relativity is correct. Here, $M_{\rm ref} = 10 M_{\odot}$. The diagonal elements show the marginalized posterior distributions while the off-diagonal elements show the 90\% credible intervals on the 2D joint distributions. Only deviations from general relativity outside these confidence intervals could be credibly detected.}
		\label{fig:cornerET}
	\end{center}
\end{figure}

In Fig.~\ref{fig:effectivepopulation}, the marginalized posteriors for $N=160$ following this approach are illustrated. In addition, it is shown how one recovers essentially the same posterior restricting to only the subset of lowest mass events (corresponding to $N_{\rm eff}=10$ events). We will continue considering only these events. As an example, we will more specifically restrict ourselves to the range $[5, 6.5]M_{\odot}$ but note that the results for the dimensionless ratios would not significantly vary when taking instead, say, $[15, 19.5]M_{\odot}$, as long as one appropriately adjusts $M_{\rm ref}$. Using the positive boundaries of the 90\% confidence intervals for the ($N=160$) posterior, we have made a comparison with the bounds obtained in \cite{Maselli:2019mjd}. Although there is no reason for them to match exactly, and indeed the comparison is made difficult due to the mass dependence, they are in qualitative agreement. In particular, our bound is similar for the zeroth order in spin contribution but tighter for higher orders.

\begin{figure}[t!]
	\begin{center}
		\includegraphics[width=0.99\textwidth]{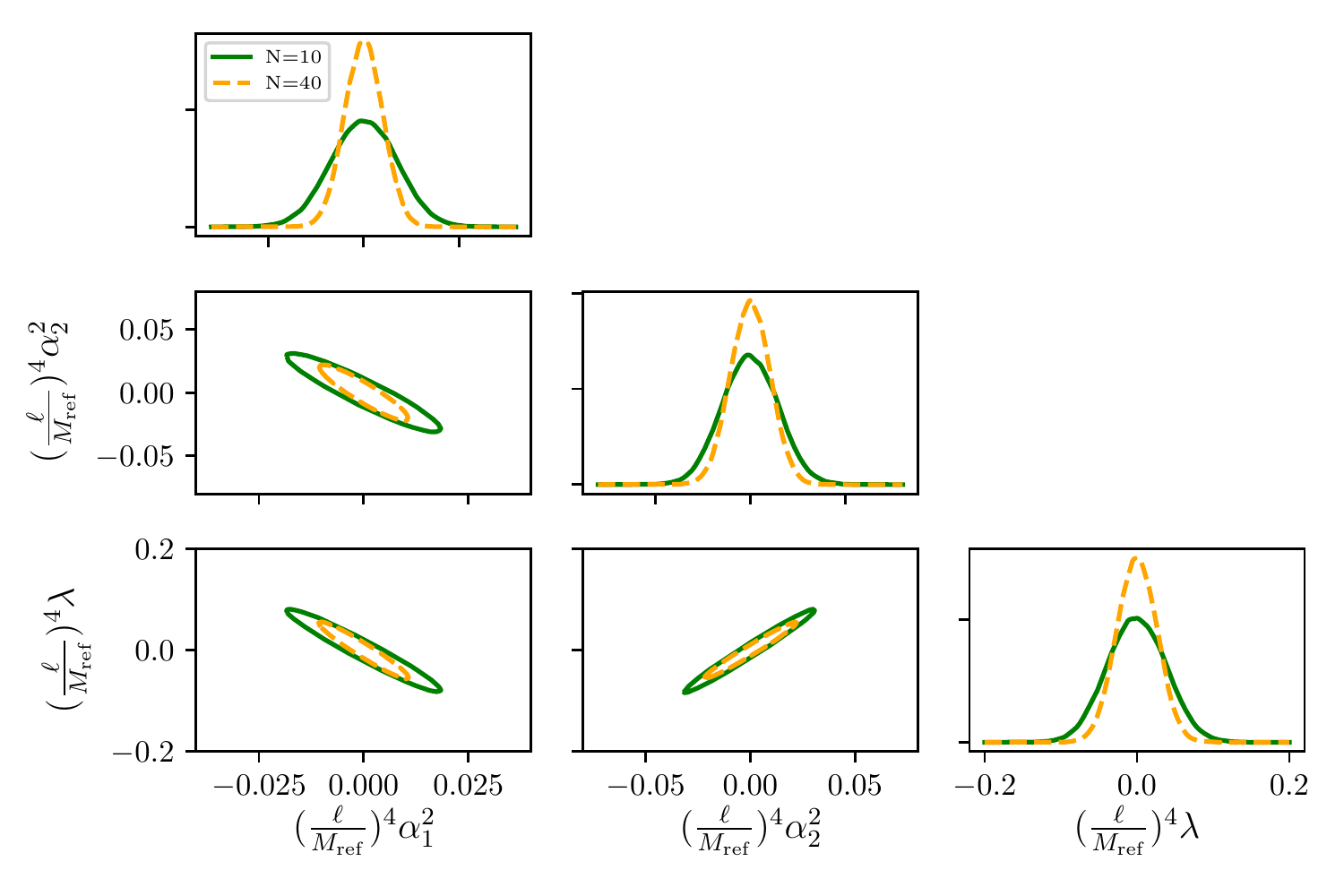}
		\caption{Corner plot for the posterior on the model parameters obtained for, from outer to inner, 10 (green), 40 (orange) LISA observations of black hole binary coalescences  if general relativity is correct. Here, $M_{\rm ref} = 10^6 M_{\odot}$. Note that since the reference mass is much larger than for ET, the constraints on the couplings are much weaker. The diagonal elements show the marginalized posterior distributions while the off-diagonal elements show the 90\% credible intervals on the 2D joint distributions. Only deviations from general relativity outside these confidence intervals could be credibly detected.}
		\label{fig:cornerLISA}
	\end{center}
\end{figure}

The posterior is found using a Markov Chain Monte Carlo method based on the Metropolis-Hastings algorithm from the open source software PyMC3 \cite{salvatier2016probabilistic}. The results for marginal and joint posteriors are shown in Fig.~\ref{fig:cornerET} for $N=10, 40, 160$. Those plots should be understood as the probability distribution for the corresponding parameters that we would obtain if GR is correct. Conversely, this means that we can expect to observe deviations with respect to GR only if the parameters lay outside the confidence intervals of the distributions in Fig.~\ref{fig:cornerET}.

Given the strong suppression for higher mass events, one might conclude that there is no added value in performing the same analysis for LISA, which would make such observations for massive black holes, about a million times heavier than the ET events. However, one should be open to the possibility that there is qualitative difference between these black hole populations, not only because the huge scale difference but also because many of the LISA mergers will occur at high redshift \cite{Berti:2016lat}. We consider only binary black hole coalescence between black holes of mass $M = 10^6 M_{\odot}$, the lower bound on the range of masses used in \cite{Maselli:2019mjd}, again with uniformly distributed spin. In addition, we assume a signal to noise of 1000 in the ringdown. The results for $N=10$, $N=40$  are given in Fig.~\ref{fig:cornerLISA}.

\subsection{Testing the QNF spectrum}
\begin{figure}[t!]
	\begin{center}
		\includegraphics[width=0.6\textwidth]{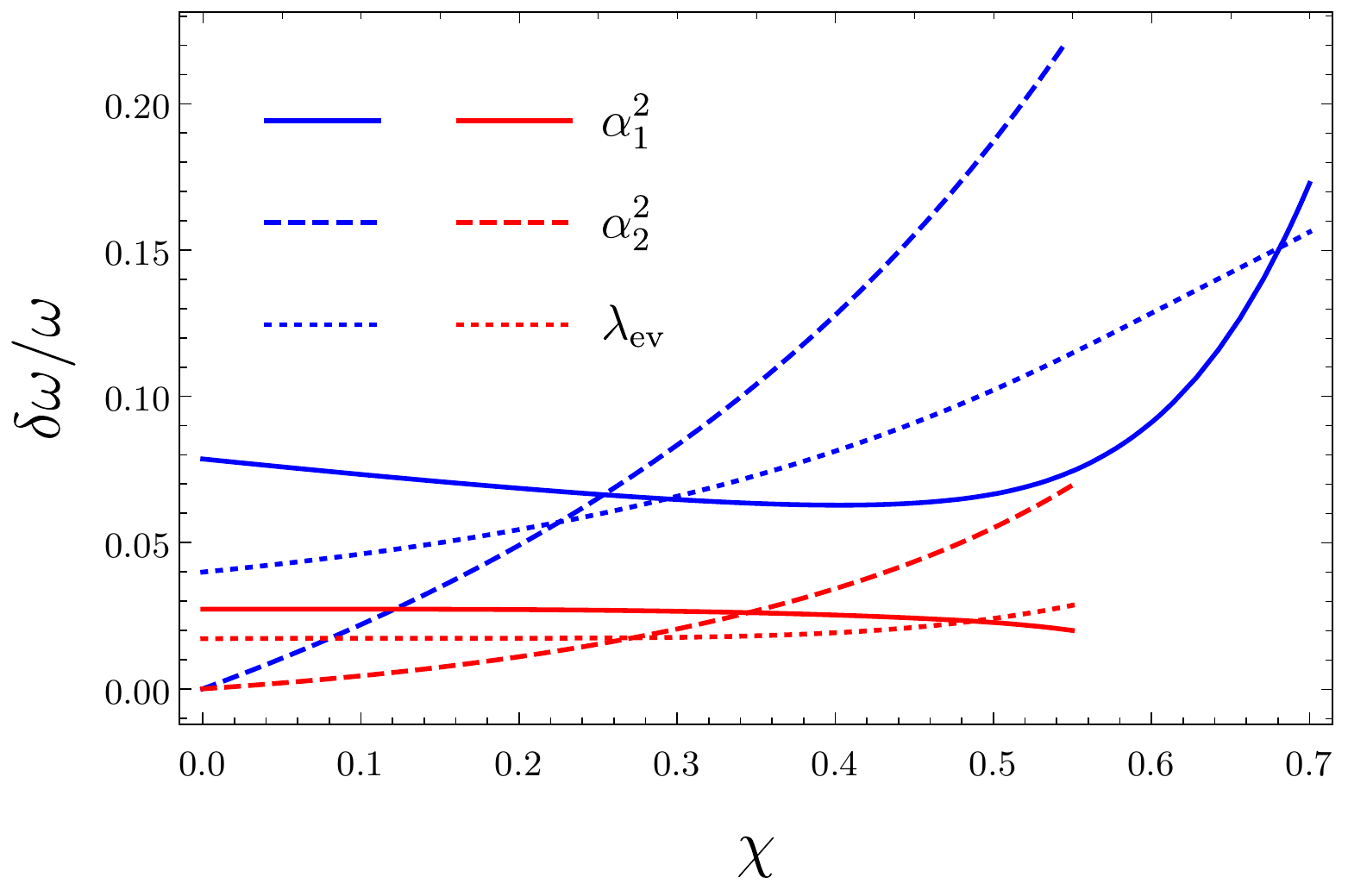}
		\caption{Relative difference (defined in \req{eq:reldif}) between the quasinormal frequencies of rotating black holes with higher-derivative corrections and the ones predicted by GR from measurement of the $l=m=2$ mode. In blue: $l=2$, $m=1$ mode. In red: $l=m=3$ mode. Each line represents the effect of the indicated correction and the value should be multiplied by the corresponding coupling times $\ell^4/M^4$.}
		\label{fig:relativediff}
	\end{center}
\end{figure}

We can test GR using ringdown measurements only if we are able to detect more than one quasinormal mode. In GR, the complete quasinormal mode spectrum is fully characterized by the mass and spin of the black hole. Hence, once the fundamental quasinormal frequency is determined, the frequencies of all modes can be predicted. However, if the geometry is not Kerr due to modifications of GR, then the predicted frequencies will be off and we could observe this deviation.  
Let us first show explicitly that such a deviation in fact occurs. To do so, let us assume that the $l=m=2$ mode has been measured, from which the mass $\tilde M_{2,2}$ and the spin $\tilde\chi_{2,2}$ predicted by GR have been estimated. This yields a GR prediction for the rest of the quasinormal frequencies, which would read
\begin{equation}
\omega^{\rm GR}_{l,m}=\omega_{(0)l,m}(\tilde M_{2,2},\tilde \chi_{2,2})\, .
\end{equation}
In our scenario, these predictions will not correspond to the real ones --- the error given by \req{generalshift}. 
The difference with respect to the actual quasinormal frequencies, taking into account that the black hole has a mass $M$ and spin $\chi$, is
\begin{equation}
\omega_{l,m}-\omega^{\rm GR}_{l,m}=\Delta\omega_{l,m}  +\frac{\delta M_{2,2}}{M_{2,2}}\omega_{(0)l,m}-\delta\chi_{2,2}\partial_{\chi}\omega_{(0)l,m}\Big|_{(M,\chi)}\, .
\end{equation}
It is useful to introduce the relative distance between the GR and the real QNFs defined as 
\begin{equation}\label{eq:reldif}
\frac{\delta\omega_{l,m}}{\omega_{l,m}}=\sqrt{\left(\frac{\text{Re }\omega_{l,m}}{\text{Re }\omega^{\rm GR}_{l,m}}-1\right)^2+\left(\frac{\text{Im }\omega_{l,m}}{\text{Im }\omega^{\rm GR}_{l,m}}-1\right)^2}\, .
\end{equation}
The value of this quantity determines the degree of distinguishability of the QNF spectrum with respect to the GR one. In Fig.~\ref{fig:relativediff} we show this difference for the $l=m=3$ and the $l=2$, $m=1$ modes, which are supposed the most relevant ones besides the $l=m=2$ one --- indeed, evidence for the $l=3$ multipole in the inspiral has been recently reported \cite{LIGOScientific:2020stg}. As we can see in Fig.~\ref{fig:relativediff}, the difference is greater for the $l=2$, $m=1$ QNF,  so that detecting this mode will provide a stronger constraint on the corrections than the $l=m=3$ mode. This suggests that measuring different $m$ modes rather than higher $l$ ones would be more efficient in order to test GR. On the other hand, those modes are probably less excited. 

\begin{figure}[t!]
	\begin{center}
				\includegraphics[width=0.48\textwidth]{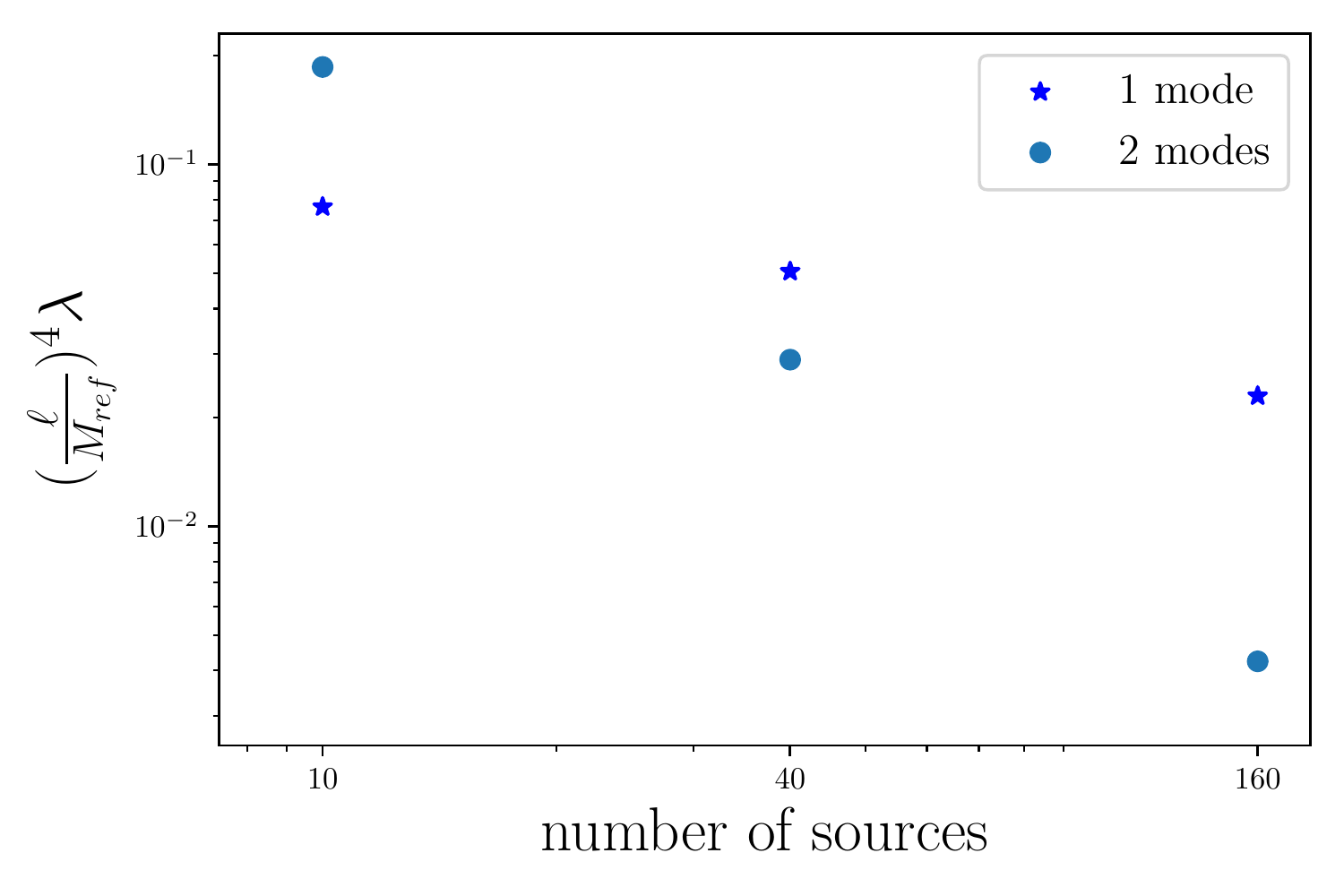}
				\includegraphics[width=0.48\textwidth]{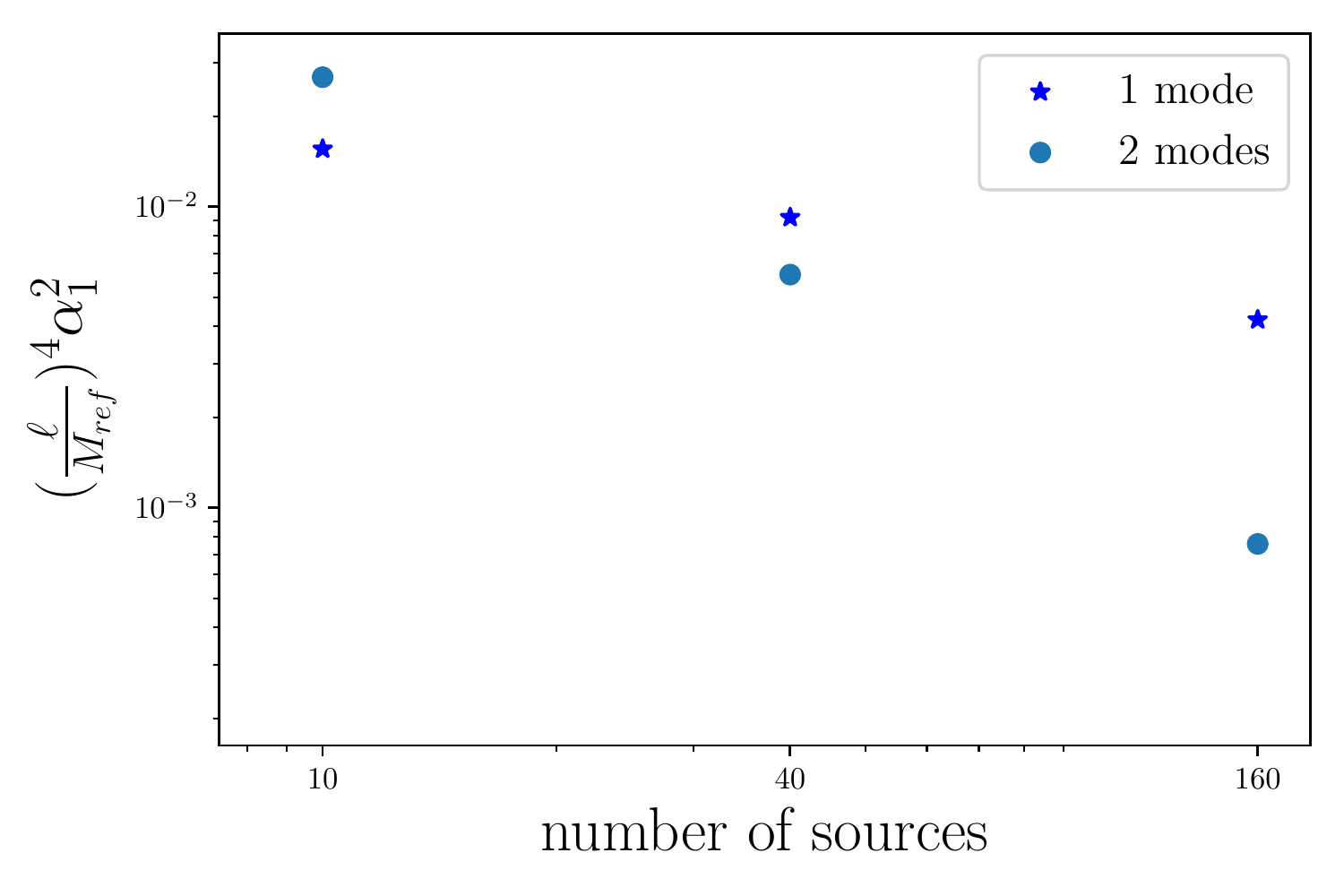}
				\includegraphics[width=0.48\textwidth]{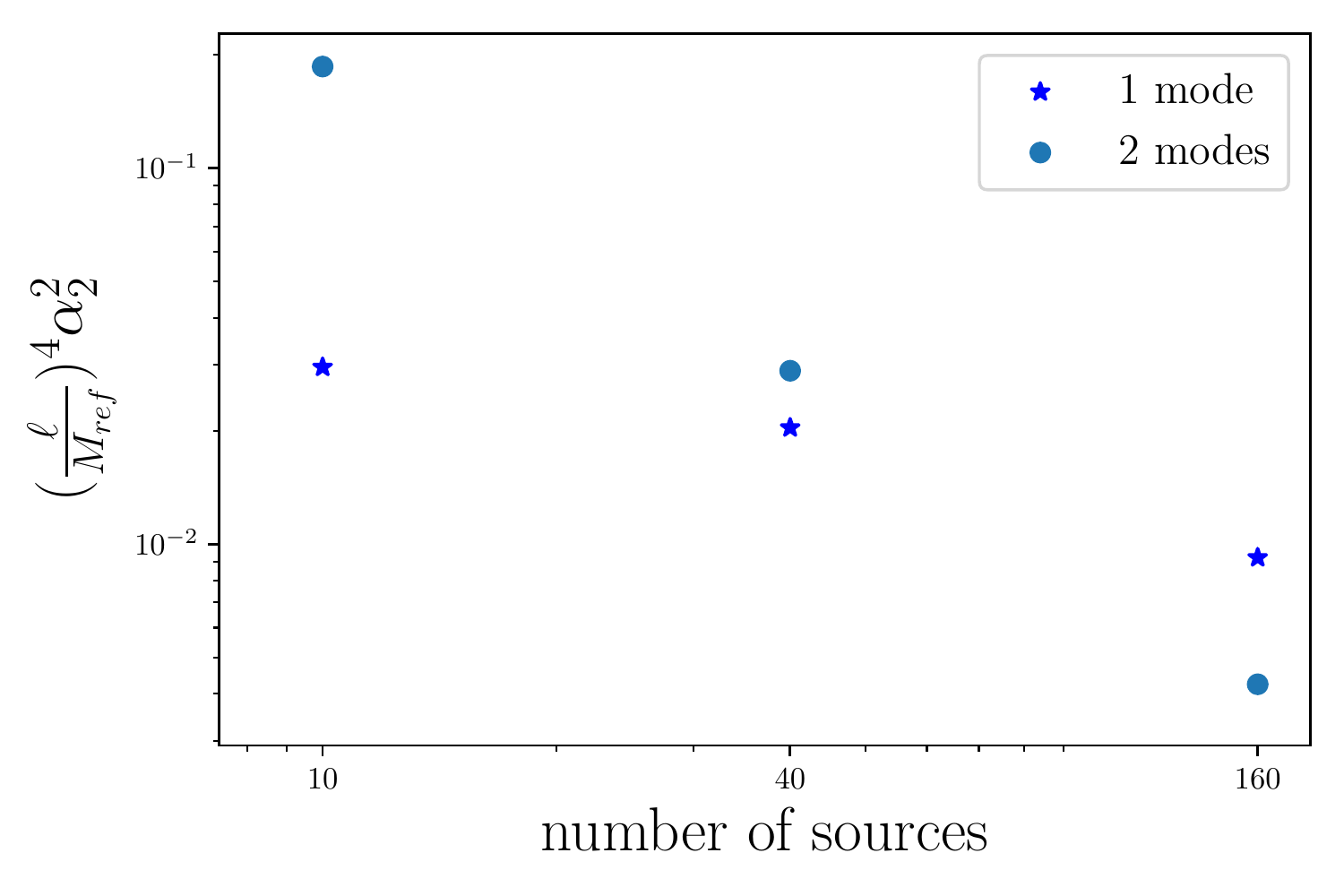}
		\caption{90\% confidence intervals for the dimensionless couplings $\hat \lambda$,  $\hat \alpha_1^2$, $\hat \alpha_2^2$ with respect to $M_{\rm ref} = 10 M_{\odot}$ from detecting two QNMs in the ringdown (circle) as well as only one QNM in addition to an (idealized) independent measurement of the mass and spin (star).}
		\label{fig:compare_2mode}
	\end{center}
\end{figure}

Finally we estimate the constraints on the parameters that follow from combining several QNM observations. We perform the same type of Bayesian inference for ET\footnote{See \cite{Cabero:2019zyt} for an expectation of what can be expected from the current generation of detectors.} as before but we now assume that in addition to the $l=2$, $m=2$ mode, an $l=3$, $m=3$ mode is measured. We further assume that this secondary mode has only 10\% of the energy of the $l=2$, $m=2$ mode \cite{Berti:2005ys}, again as in \cite{Maselli:2019mjd}. The price to pay is that now also the final mass and spin of the black hole needs to be inferred from the ringdown itself, as we no longer assume these to be determined from the inspiral. On the other hand, this renders the analysis self-contained and, in particular, independent of any assumptions on the inspiral-merger phase. Assuming we have independent measurements of the two modes, the full likelihood is

\be 
\cL(\lbrace \omega^{\rm obs} \rbrace|\vec{\theta}) = \prod_{i=1}^{N} \cL_i(\omega^{\rm obs}_{2,2 }{}_i|\vec{\theta}, M_i, \chi_i) \cL_i(\omega^{\rm obs}_{3,3 }{}_i|\vec{\theta}, M_i, \chi_i)\, ,
\ee 

analogous to \eqref{eqn:fulllikelihood}. Based on the previous observations about the strong $M$ suppression we again restrict ourselves to the lowest final masses in the population of detected black hole binary coalescences. In Fig.~\ref{fig:compare_2mode}, we compare the constraints coming from this ringdown-only approach with those from the (idealized) combination of an independent mass, spin determination and a ringdown measurement. With only a few available measurements, the ringdown-only approach performs worse on account of the combined uncertainty on both the theory and the masses and spins. Once more measurements become available this is compensated and it even performs better. However, it is not expected that this steep improvement will last but rather that it will asymptote to a $\sim N^{-1/2}$ behavior, similar to the combined method.

\section{Conclusions} \label{sec:con}

We have studied the scalar QNM frequencies in rotating black hole backgrounds in general well-motivated higher-derivative extensions of GR that reduce to GR in the weak field limit. The wave operator governing the QNMs in rotating black hole backgrounds deviating from Kerr is not separable, rendering the problem much harder than the calculation in Kerr geometries. However, we have shown that the projection of the wave operator onto the set of spheroidal harmonics yields a consistent second-order ODE for a single variable, from which one can extract the quasinormal frequencies applying standard methods. This has enabled us to obtain accurate results for the corrections to the scalar QNFs of rotating black holes with relatively large spin $\chi \sim 0.7$. Higher values of the spin can be straightforwardly probed with the software we provide --- it only requires more computational time. Our results are shown in detail in Section~\ref{sec:QNFRBH}, while some observational implications --- upon the assumption that the relative corrections to the gravitational and scalar QNM frequencies are similar --- are discussed in Section~\ref{sec:obsdev}. 

It is clearly of prime interest to extend our results to vectorial and, especially, gravitational quasinormal modes. 
This requires first and foremost the master equation for those perturbations, which on general grounds we expect to be non-separable.  Deriving the master equation for gravitational perturbations represents a remarkable challenge, but with the master equation at hand one could then apply the method given in Section~\ref{sec:pertrotation} to reduce the problem to an effective ODE for one variable. 

Let us conclude by discussing to what extent we can expect similarity between the corrections to scalar and gravitational QNMs. 

\subsubsection*{Scalar vs gravitational modes}

Only gravitational QNMs are relevant for gravitational wave astronomy, but these are much more complicated to study than scalar ones. 
It is thus an interesting question whether we can obtain an approximation to the gravitational quasinormal frequencies from the scalar modes. Comparing the $l=2$ modes for scalar and gravitational perturbations of the Kerr geometry, one observes that they are not too different, and they have a similar dependence on the spin. However, there is no reason to expect that the higher-derivative corrections to these QNFs will also be similar. Unfortunately we cannot check whether this is the case because the corrections to the gravitational QNFs of rotating BHs have not been computed yet. However, in Section~\ref{sec:QNFRBH} we provided two examples for static black holes. On the one hand, we observed that the (relative) correction to the axial $l=2$ gravitational mode is close to the correction to the scalar $l=2$ mode in Einstein-dilaton-Gauss-Bonnet (EdGB) gravity (and the same applies to the $l=3$ mode). On the other hand, we do not find agreement between the scalar and either axial or polar gravitational modes in the case of the quartic theories \req{eq:quarticL}. 
While all of this could be coincidental, let us offer a possible explanation. 
First, note that one of the main differences between scalar and gravitational perturbations is that, while in the former case we are keeping the operator fixed and we change the background, in the second case the wave operator also receives corrections\footnote{These are respectively referred to as ``background'' and ``dynamical'' modifications in \cite{Berti:2018vdi}.}. That is, Einstein's equations are modified, $G_{\mu\nu}+\mathcal{E}_{\mu\nu}=0$, and hence the linearized equations around the background of a black hole have corrections coming explicitly from $\mathcal{E}_{\mu\nu}$:
\begin{equation}
G^{L}_{\mu\nu}+\mathcal{E}^{L}_{\mu\nu}=0\, ,
\end{equation}
where $L$ denotes the linearized part. Since $\mathcal{E}^{L}_{\mu\nu}$ typically contains more than two derivatives, one should treat this object carefully. One possibility would be to directly truncate $\mathcal{E}^{L}_{\mu\nu}$, and hence considering that only the Einstein tensor is responsible for the dynamics of gravity. Perhaps a more rigorous approach is to reduce the higher-derivative terms in $\mathcal{E}^{L}_{\mu\nu}$ by using the zeroth-order equations, as in Ref.~\cite{Cardoso:2018ptl}. Even in this case one has to be careful with this operator since, for instance, it can become larger than $G^{L}_{\mu\nu}$ for large values of $l$ in which case the perturbative regime breaks down. Now, the equation $G^{L}_{\mu\nu}=0$ is the equivalent of $\nabla^2\psi=0$, so only when $\mathcal{E}^{L}_{\mu\nu}$ is truncated or is zero we can expect some resemblance between scalar and gravitational QNMs. 

To illustrate this let us consider the case of EdGB gravity. It turns out that axial (parity-odd) gravitational perturbations are decoupled from scalar perturbations in this theory, and thus it is easy to see that $\mathcal{E}^{L}_{\mu\nu}=0$ in that case. This explains why we get similar results for the scalar and axial gravitational QNFs for static black holes in this theory. In addition, since we expect $\mathcal{E}^{L}_{\mu\nu}=0$ for parity-odd perturbations in the rotating case too, we expect that the analogy with scalar QNFs also holds for rotating black holes. On the other hand, $\mathcal{E}^{L}_{\mu\nu}\neq 0$ for other higher-derivative terms --- in particular for the quartic ones --- so we cannot a priori expect agreement between the gravitational and scalar QNFs if $\mathcal{E}^{L}_{\mu\nu}$ is not truncated. 

Another possible example for which $\mathcal{E}^{L}_{\mu\nu}=0$ is the case of polar (parity-even) gravitational perturbations in dynamical Chern-Simons theory (dCS). In fact, due to parity, those perturbations decouple from scalar ones \cite{Cardoso:2009pk}, and due to the topological character of the Chern-Simons term, it does not contribute to the linearized equations in that case. Therefore, according to our argument above, one may expect that in dCS gravity the scalar QNFs of rotating BHs are similar to the polar gravitational ones.

Translating these results to our actions \req{Action} and \req{eq:quarticL}, the overall conclusion is that the corrections to the scalar QNFs associated to the parameters $\alpha_1$ and $\alpha_2$ may be similar to those of the parity-odd and parity-even gravitational modes, respectively. In the other cases, agreement is not expected unless the non-Einsteinian term $\mathcal{E}^{L}_{\mu\nu}$ is truncated from the linearized equations.

\acknowledgments
This work is supported by the C16/16/005 grant of the KU Leuven, the COST Action GWverse CA16104, and by the FWO Grant No. G092617N. K.F. is Aspirant FWO-Vlaanderen (ZKD4846-ASP/18). This work makes use of the Black Hole Perturbation Toolkit. 

\appendix

\section{Methods to compute quasinormal frequencies}\label{app:methods}
\subsection{Approximate methods}
For an overview of approximate methods to compute quasinormal modes we refer to \cite{Berti:2009kk}. This in particular contains a discussion on the WKB and P\"oschl-Teller approximations. The latter approach approximates the effective potential in \eqref{eq:perturbeq1}

\begin{equation}\label{radialeqrotapp}
\frac{d^2\varphi}{dy^2}+\left(\omega^2-V\right)\varphi=0\, ,
\end{equation}
by the exactly solvable P\"oschl-Teller potential
\be
V =\frac{V_0}{\cosh^2{\alpha y}}.
\ee
However, this potential goes to zero as $y\to \pm \infty$. We have found it more convenient to work with variables for which $V$ goes to a nonzero constant as $y \to - \infty$. To account for this difference, we extend the P\"oschl-Teller approximation using the exactly solvable Rosen-Morse potential \cite{rosen1932vibrations}
\be
V = \frac{\beta}{2 d^2} \left(\tanh{\frac{y}{d}}-1\right)-\frac{\gamma}{d^2\cosh^2{\frac{y}{d}}}.
\ee
Introducing
\bea
z&=&\frac{1}{2}\left(1+\tanh{\frac{y}{d}}\right), \\
a&=& \frac{i}{2}\left(-\sqrt{d^2 \omega^2+\beta}+d \omega\right), \\
b&=& -\frac{i}{2}\left(\sqrt{d^2 \omega^2+\beta}+d \omega\right), \\
\varphi &=& f(z) e^{\frac{ay}{d}}\cosh^{-b}{\frac{y}{d}},
\eea
the wave equation takes the hypergeometric form

\be
z(1-z)f''+(a+b+1-2(b+1)z)f'+(\gamma-b(b+1))f=0
\ee
with $z$ going from $0$ to $1$. When $f(z)$ is regular at $z=0,1$, the solution of the original wave equation behaves as $\varphi\sim e^{-i(y \sqrt{\omega^2+\frac{\beta}{d^2}})}$ at $y \to -\infty$, and $\varphi\sim e^{i \omega y}$ at  $y \to + \infty$, which are the correct QNM boundary conditions. This is the case for Jacobi polynomials, \textit{i.e.}, when the solution is given by

\be
f = {}_2F_1\left(b+\frac{1}{2}-\left(\gamma+\frac{1}{4}\right)^{1/2},b+\frac{1}{2}+\left(\gamma+\frac{1}{4}\right)^{1/2},a+b+1,z\right),
\ee
with

\be
b = \sqrt{\gamma + \frac{1}{4}}-n-\frac{1}{2},
\ee
and $n$ an integer. In terms of $\omega^2$ one finds\footnote{Using $\beta=-4ab$ or $a=-\frac{\beta}{2(\sqrt{4\gamma + 1}-2n-1)}$.}

\be
 \omega =  -\sqrt{-\frac{\gamma}{d^2} - \frac{1}{4d^2}}+\frac{\beta/d^2}{4(\sqrt{-\frac{\gamma}{d^2} - \frac{1}{4d^2}}+\frac{i}{d}(n+\frac{1}{2}))}-\frac{i}{d}(n+\frac{1}{2})
 \label{eqn:RMQNMs}
\ee
Now, given a potential $V_{l,m}$, we replace it by a RM potential that has the same asymptotic behaviour and whose maximum value and second derivative at the maximum agree with those of the original potential. This leads to the following identifications of the constants, 
\begin{figure}[t]
	\begin{center}
		\includegraphics[width=0.49\textwidth]{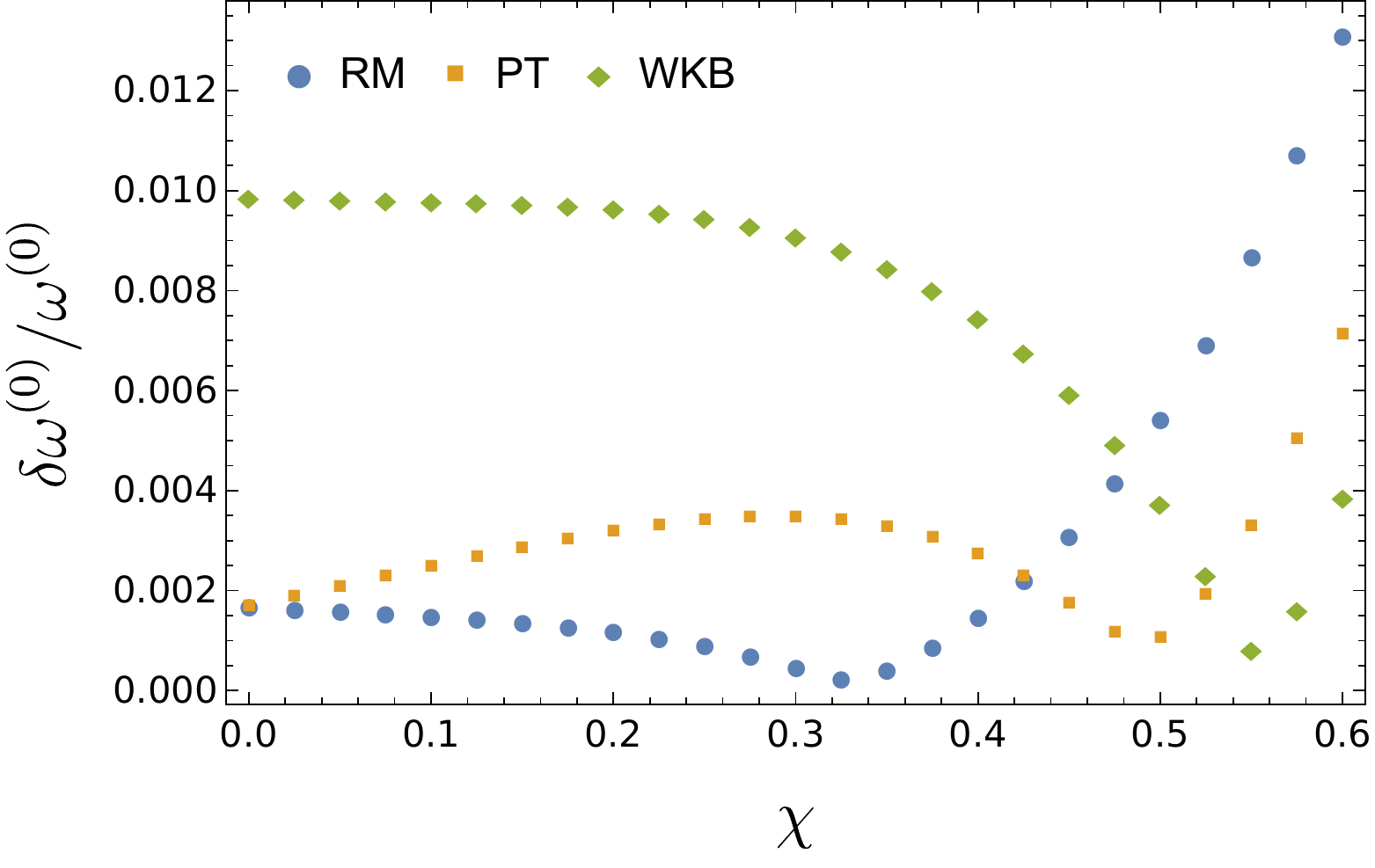} 
		\includegraphics[width=0.49\textwidth]{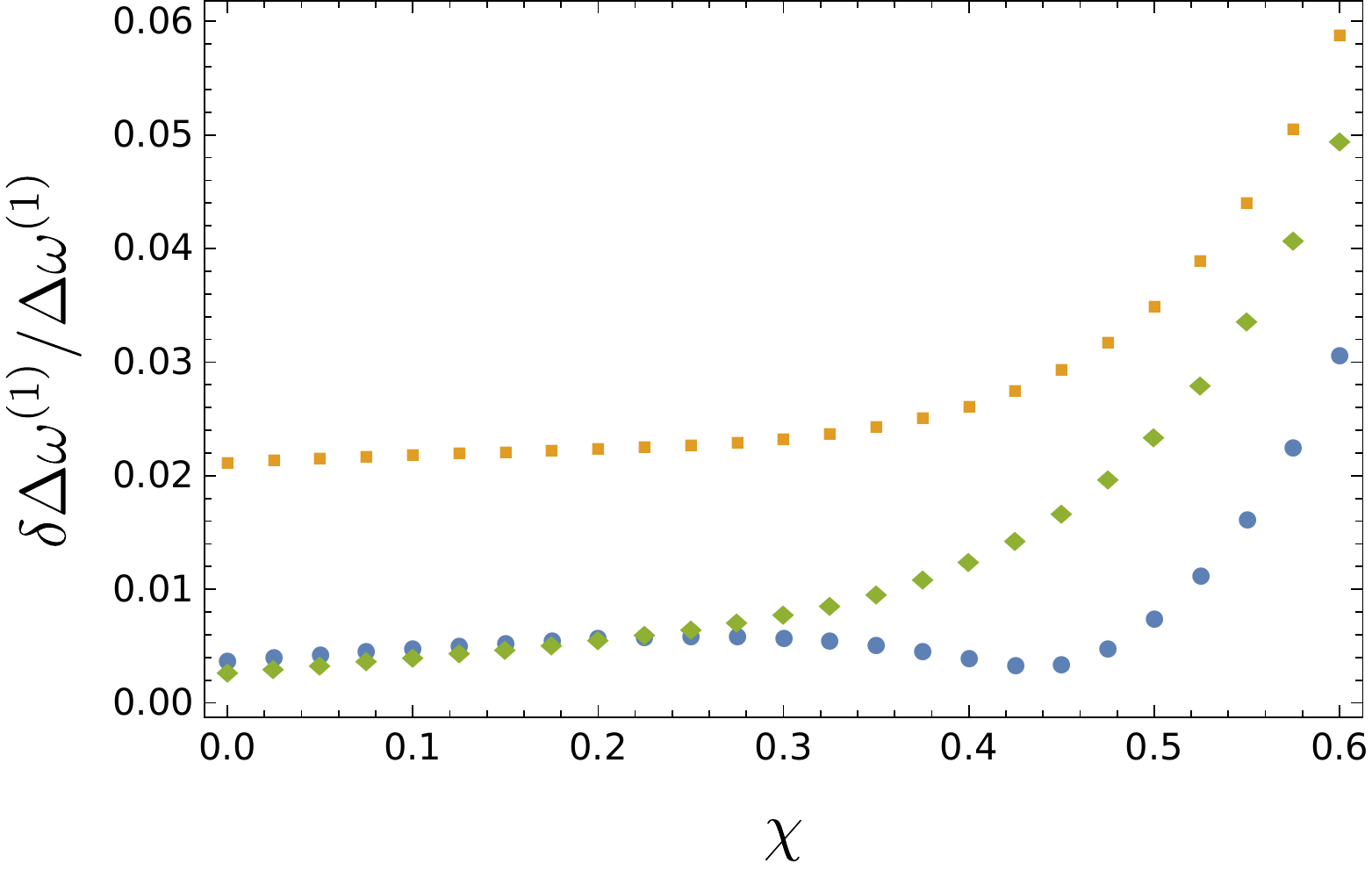}
		\includegraphics[width=0.49\textwidth]{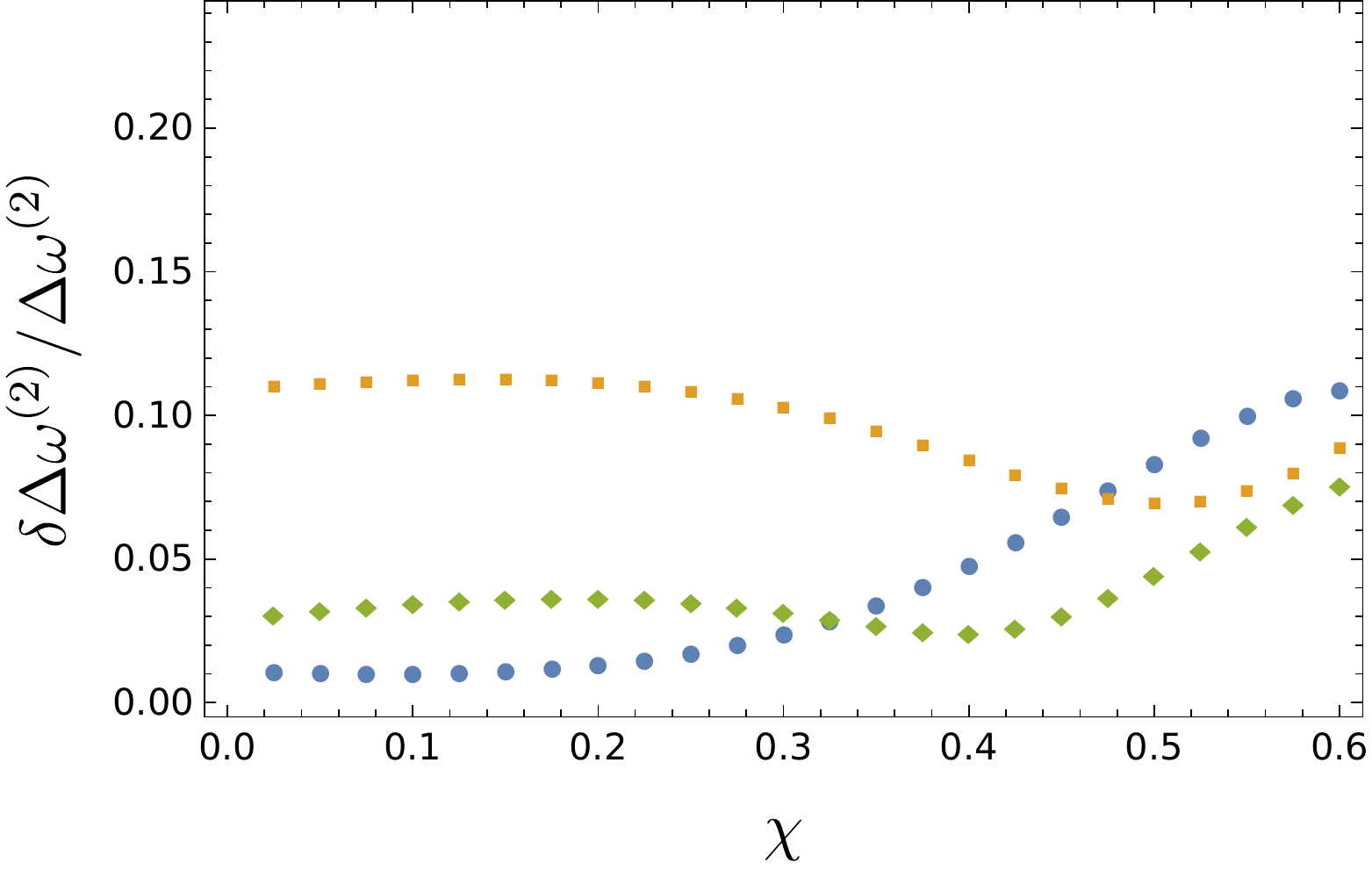}
		\includegraphics[width=0.49\textwidth]{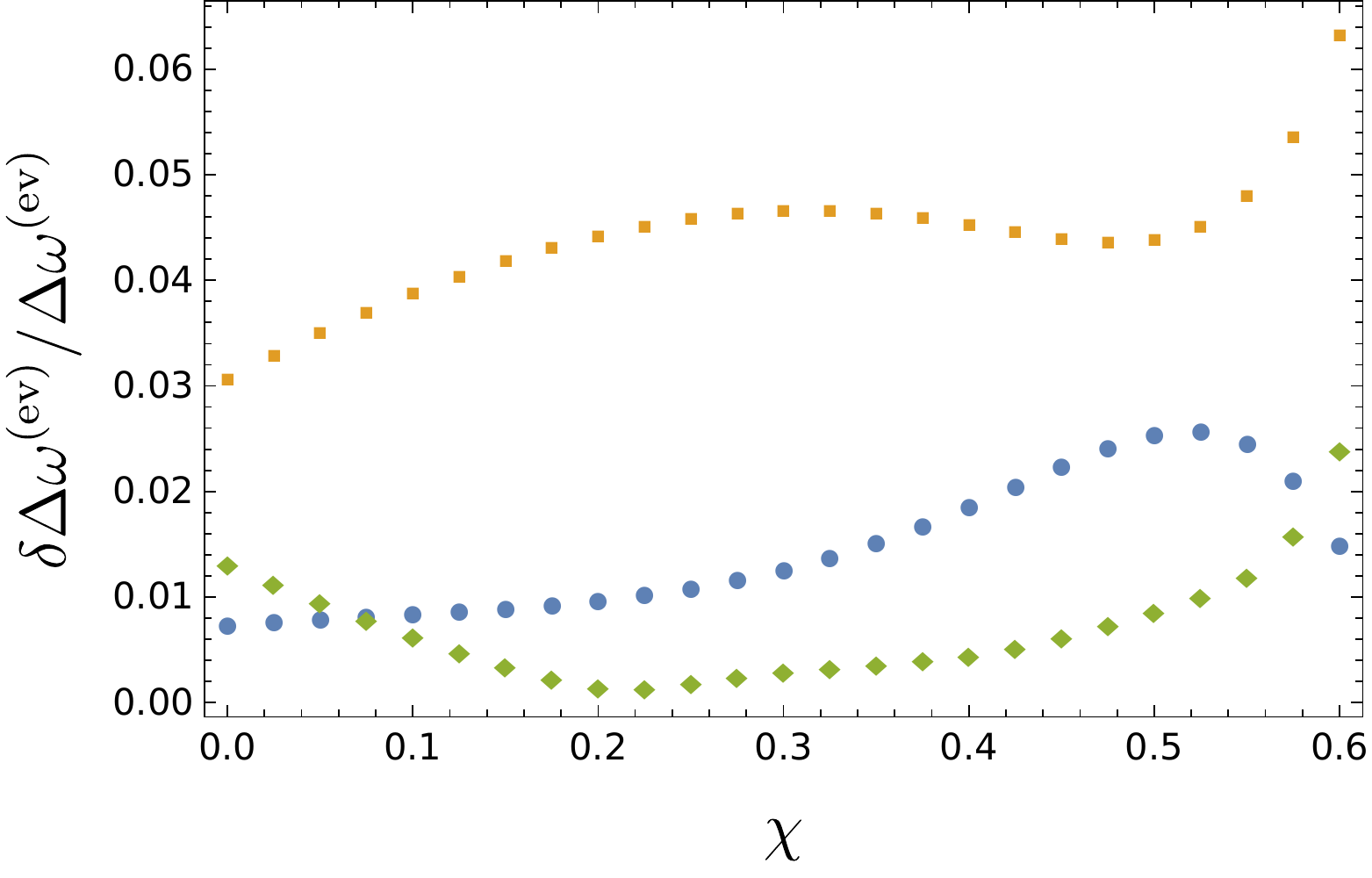}
		\caption{Absolute relative error between the approximations \eqref{eqn:PT55}, \eqref{eqn:RM55}, \eqref{eqn:WKB55} and numerical result for the Kerr quasinormal frequency $M\omega^{(0)}_{5,5,0}$, and of the correction coefficients $\Delta \omega^{(i)}_{5,5,0}$ as a function of the black hole spin.}
		\label{fig:analyticapprox}
	\end{center}
\end{figure}
\bea
\frac{\beta}{d^2}&=& -V_{-\infty},\\
\frac{\gamma}{d^2}&=&\frac{1}{4}\left(-\frac{\beta}{d^2}-2V_0-2\sqrt{V_0^2+V_0 \frac{\beta}{d^2}}\right), \\
d^2 &=& \frac{1}{128 V''_0}\left(\frac{d^2}{\gamma}\right)^3\left(\frac{\beta^2}{d^4}-16\frac{\gamma^2}{d^4}\right)^{2}
\eea
where $V_{-\infty}=V_{l,m}(-\infty)$, $V_{0}=V_{l,m}(y_0)$, $V''_{0}=V''_{l,m}(y_0)$, and $y_0$ is the position of the maximum, \textit{i.e.}, where $V_{lm}'(y_0)=0$.
We note that on setting $\beta = 0$, one has $\gamma = -V_0 d^2$, $d^2=-2 \frac{V0}{V_0''}$, and therefore
\be
\omega_n = -\sqrt{V_0+\frac{V_0''}{8 V_0}}-i\sqrt{\frac{-V_0''}{2 V_0}}(n+\frac{1}{2})\, ,
\ee
which corresponds to \eqref{eqn:PT}. In section \ref{subsec:analytic}, we have compared the different approximation methods --- solved perturbatively for low spin --- with the numerical result for several particular effective potentials as described in that section. Fig.~\ref{fig:analyticapprox} illustrates this comparison from which we conclude that overall the Rosen-Morse approximation is most effective. Note however that for large $l$, where these methods should perform best, one should ideally make a large $l$ expansion to perturbatively solve the spheroidal wave equation rather than solving it for small dimensionless spin as we have done here.

\subsection{Numeric integration}
Let us explain here how we solve the wave equation numerically in order to obtain a the quasinormal frequencies. The problem we consider is 

\begin{equation}\label{radialeqrotapp}
\frac{d^2\varphi}{dy^2}+\left(\omega^2-V(y;\omega)\right)\varphi=0\, ,
\end{equation}
together with the boundary conditions

\begin{equation}
\varphi\propto\begin{cases}
e^{i\omega y}\quad &\text{when} \quad y\rightarrow\infty\\
e^{-i\hat\omega y}\quad &\text{when} \quad y\rightarrow-\infty\, ,
\end{cases} 
\end{equation}
where $\hat\omega=\sqrt{\omega^2-V(-\infty;\omega)}$ since we are allowing for the possibility of $V(-\infty;\omega)\neq 0$. The potential is required to tend to its asymptotic values at $y=\pm \infty$ faster than $1/y$, since otherwise the amplitude of $\varphi$ would scale as $y^{\alpha}$. Now, in order to find the quasinormal modes, a naive strategy would consist in solving numerically the equation above starting with the appropriate boundary condition at some negative $y=y_-$, and then obtaining the solution for a sufficiently large $y_+$, so that we can identify
\begin{equation}
\varphi(y_2)=Ae^{i\omega y_+}+B e^{-i\omega y_+}\, .
\end{equation}
 Then, we would search for the value of $\omega$ such that $B=0$. The problem is that quasinormal frequencies are complex, with a negative imaginary part. Hence, the outgoing mode at infinity is exponentially larger than the ingoing mode, so that identifying the value of $B$ numerically becomes very hard as we increase $y_2$. 
 
 In order to avoid this issue, it is more convenient to work with the phase by defining 
 \begin{equation}
 \varphi=e^{i\int dy \alpha(y)}\, .
 \end{equation}
 In terms of $\alpha$, the equation \req{radialeqrotapp} becomes
 \begin{equation}
 i\alpha'-\alpha^2+\omega^2-V(y;\omega)=0\, .
 \end{equation}
 Then our strategy will be to obtain two solutions of this equation, $\alpha_{\pm}$, such that both will satisfy a gluing condition when $\omega$ is a quasinormal frequency. These solutions correspond to those that have only outgoing modes at $+\infty$ and at $-\infty$, respectively:
 \begin{align}
 \alpha_{+}&\rightarrow \omega \quad\text{when}\quad y\rightarrow\infty\, ,\\
  \alpha_{-}&\rightarrow -\hat\omega \quad\text{when}\quad y\rightarrow-\infty\, .
 \end{align}
 Then, if $\omega$ is a quasinormal frequency these solutions are actually the same and they must satisfy 
 \begin{equation}
 \alpha_{+}(y_0)= \alpha_{-}(y_0)\, ,
 \end{equation}
 at some (actually any) point $y_0$. In practice, instead of working with $\alpha_{\pm}$, it is useful to introduce other two functions
 \begin{equation}
 \alpha_{+}=\omega+\beta_{+}\, ,\quad  \alpha_{-}=-\hat\omega+\beta_{-}\, .
 \end{equation}
 These satisfy 
 \begin{align}
 i\beta_{+}'-\beta_{+}^2-2\beta_{+}\omega-V(y;\omega)&=0\, ,\quad \beta_{+}(y\rightarrow\infty)\rightarrow 0\, ,\\
 i\beta_{-}'-\beta_{-}^2+2\beta_{-}\hat\omega+V(-\infty;\omega)-V(y;\omega)&=0\, ,\quad \beta_{-}(y\rightarrow-\infty)\rightarrow 0\, ,
\end{align}
 and in turn, the QNF condition reads
 \begin{equation}
\beta_{+}(y_0)-\beta_{-}(y_0)+\omega+\hat\omega=0\, .
 \end{equation}
 The last point that remains to be addressed is how to generate boundary conditions for the the functions $\beta_{\pm}$. The boundary conditions have to be determined accurately in order to obtain the correct result for the quasinormal frequencies. For instance, it is not valid to set $\beta_{+}(y_+)=0$ for some large $y_{+}$. We can generate boundary conditions by assuming a $1/\omega$ expansion of $\beta_{\pm}$ in the following form (let us exemplify this in the case of $\beta_{+}$). We rewrite the equation for $\beta_+$ as follows
 \begin{equation}
\beta_{+}=\frac{i\beta_{+}'-\beta_{+}^2-V(y;\omega)}{2\omega}\, .
 \end{equation}
 Then we iterate this expression and we generate a sequence $\beta_{+}^{(n)}$, where we start with $\beta_{+}^{(0)}=0$. In the $n$-th iteration we only need to keep terms up to order $\omega^{-n}$ (also note that here we do not expand $V(y;\omega)$ in $1/\omega$). Thus, for instance we get
 \begin{equation}
\beta_{+}^{(1)}=-\frac{V(y;\omega)}{2\omega}\, ,\quad \beta_{+}^{(2)}=-\frac{V(y;\omega)}{2\omega}-\frac{iV'(y;\omega)}{4\omega^2}\, , \quad\ldots
 \end{equation}
 Then, if we evaluate these expressions for some large enough $y=y_{+}$ we generate an initial condition for $\beta_{+}$. In order to get good precision we may need to go to high-enough $n$ (in our calculations $n\sim 10$). We note that this method works well at least for the first iterations $\beta_{+}^{(n)}$, since $\omega>>V(y_+;\omega)$ for large enough $y_+$, however the convergence for $n\rightarrow\infty$ is not guaranteed. We can improve the convergence by constructing Pad\'e approximants in $1/\omega$ starting from the $\beta_{+}^{(n)}$. This already provides enough precision to determine the quasinormal modes accurately. 
 An analogous procedure can be done in the case of $\beta_{-}$ (in this case we expand in $1/\hat\omega$). However, since in the case of black holes the potential decays exponentially fast for $y\rightarrow-\infty$, we find that it is usually enough to use the approximation $\beta_{-}^{(2)}$ for the initial condition.

\section{Spheroidal harmonics}\label{app:spheroidal}
We summarize the properties relevant for this work. For more on spheroidal harmonics see \cite{Flammer1957,meixner1954} or \cite{Berti:2005gp} for the spin-weighted harmonics relevant in the gravitational case. We define spheroidal harmonics as the eigenfunctions of \eqref{eqn:spheroidal}
\be
(1-x^2) S''_{l,m}-2xS'_{l,m}+\left(A_{l,m}(c)+c^2x^2-\frac{m^2}{1-x^2}\right)S_{l,m}=0,
\ee
such that $S_{l,m}(x;c)e^{i m \phi}$ is regular on the sphere with $x=\cos{\theta}$. $A_{l,m}(c)$ is the corresponding eigenvalue.\footnote{We supress the argument when there is no potential confusion.} For $c=0$, \eqref{eqn:spheroidal} reduces to Legendre's differential equation, so that the solutions are given by

\be
S_{l,m}(x,0)= N_{l,m} P^m_l(x), \quad A_{l,m}(0)=l(l+1),
\ee
with
\be
N_{l,m}= \sqrt{\frac{(2l+1)(l-m)!}{2(l+m)!}}
\ee
given our normalization
\be
\int^{1}_{-1} S_{l,m}(x) S_{l',m}(x) = \delta_{l l'}.
\ee
For $c \neq 0$, we will compute $S_{l,m}$ perturbatively around this zeroth order solution, as already indicated in the main text \eqref{eqn:spheroidalexpand}

\be
S_{l,m}(x;c) = \sum^{\infty}_{n=|m|}c^{2(n-|m|)}a_{n,m}P_n^{m}(x)\, .
\ee
In this expression, it has already been made manifest that the perturbing term $c^2x^2$ preserves parity, so even and odd Legendre polynomials do not couple.\footnote{This would not be the case for spin-weighted spheroidal harmonics.} The method to compute the coefficients $a_{n,m}$ efficiently is to derive a three-term recurrence relation 
\be
\alpha_k a_{k+1,m}+(\beta_k-A_{l,m}) a_{k,m}+ \gamma_k a_{k-1,m}=0,
\ee
which is subsequently solved by the method of continued fractions. The details have been described many times, for instance, in \cite{falloon2003theory} so we do not repeat them. We stress, however, that Mathematica 11 does not perform this expansion correctly. We recommend instead using the implementation in the BHperturbation toolkit \cite{BHPToolkit}, which moreover also includes the spin-weighted spheroidal harmonics relevant for the gravitational case. 

\section{Scalar Laplacian}\label{app:laplacian}
We write $\psi=e^{-i\omega t}e^{im\phi}\psi_{m}$ and then we have $\nabla^2\psi=e^{-i\omega t}e^{im\phi}\mathcal{D}^2\psi_m$. The operator $\mathcal{D}^2$ can be decomposed as the sum of its zeroth-order part plus the corrections:
\begin{equation}
\mathcal{D}^2=\mathcal{D}_{(0)}^2+\mathcal{D}_{(1)}^2\, .
\end{equation}
The operator $\mathcal{D}^2_{(0)}$ is given by
\begin{align}
\mathcal{D}^2_{(0)}\psi&=\frac{1}{\Sigma}\partial_{\rho}\left(\Delta\partial_{\rho}\psi\right)+\frac{1}{\Sigma}\partial_{x}\left((1-x^2)\partial_{x}\psi\right)\\
&-\frac{\psi}{\Delta\Sigma}\bigg[\frac{(\Sigma-2M\rho)m^2}{(1-x^2)}+4Ma\rho m\omega-\left(2M\rho a^2+\frac{\Sigma(\rho^2+a^2)}{(1-x^2)}\right)\omega^2\bigg]\, .
\end{align}
On the other hand, we have

\begin{align}
\mathcal{D}^2_{(1)}\psi&=-\frac{H_2\Delta}{\Sigma} \partial^{2}_{\rho}\psi-\frac{(1-x^2)H_2}{\Sigma}\partial^{2}_x\psi+\frac{P}{\Sigma}\partial_{\rho}\psi+\frac{Q}{\Sigma}\partial_{x}\psi+\frac{S}{\Sigma}\psi\, ,
\end{align}
where the quantities $P$, $Q$ and $S$ read
{\allowdisplaybreaks
\begin{align}
\notag
P=&-\frac{H_3{}^{(1,0)}}{\Sigma^2}4 a^2 M^2 \rho ^2 \left(x^2-1\right)+2 H_2 (M-\rho )\\\notag
& +\frac{\Delta H_4{}^{(1,0)}-\Sigma H_1{}^{(1,0)}}{2 \Sigma^2}  \left(a^4 x^2+a^2 \rho  \left(-2 M x^2+2 M+\rho +\rho  x^2\right)+\rho ^4\right)\\\notag
&+\frac{4 a^2 (H_4-2H_3)}{\Delta\Sigma^3} M^2 \rho  \left(x^2-1\right) \left(a^4 x^2-a^2 \rho  \left(M x^2+\rho \right)+\rho ^3 (3 M-2 \rho )\right)\\
&+\frac{H_1 M}{\Delta\Sigma^2} \left(-a^6 x^2+a^4 \rho ^2 \left(1-2 x^2\right)+a^2 \rho ^3 \left(4 M \left(x^2-1\right)-\rho  \left(x^2-2\right)\right)+\rho ^6\right)\, ,\\\notag
&&\\\notag
Q=&\frac{H_3{}^{(0,1)} }{\Delta\Sigma^2}4 a^2 M^2 \rho ^2 \left(x^2-1\right)^2-\frac{2 a^2 H_1}{\Delta\Sigma^2} M \rho  x \left(x^2-1\right) \left(a^2+\rho ^2\right)+2 H_2 x\\\notag
&-\frac{\left(x^2-1\right) H_4{}^{(0,1)}}{2\Delta\Sigma^2} \Big[a^6 x^4+a^4 \rho  x^2 \left(\rho  \left(x^2+2\right)-2 M x^2\right)\\\notag
&+a^2 \rho ^2 \left(4 M^2 \left(x^2-1\right)-4 M \rho  x^2+\rho ^2 \left(2 x^2+1\right)\right)+\rho ^5 (\rho -2 M)\Big]\\\notag
&+\frac{4 a^2 (H_4-2H_3)}{\Delta \Sigma^3} M^2 \rho ^2 x \left(x^2-1\right) \left(a^2 \left(x^2-2\right)-\rho ^2\right)\\
&+\frac{\left(x^2-1\right) H_1{}^{(0,1)}}{2\Delta\Sigma} \left(a^4 x^2+a^2 \rho  \left(-2 M x^2+2 M+\rho +\rho  x^2\right)+\rho ^4\right)\, ,\\\notag
&&\\\notag
S=&\frac{4 a H_3}{\Delta^2 \Sigma^2} M \rho  \Big(2 a m^2 M \rho  \left(a^2 x^2+\rho  (\rho -2 M)\right)-8 a^2 m M^2 \rho ^2 \left(x^2-1\right) \omega \\\notag
&-m \omega  \Delta \Sigma^2+2 a M \rho  \left(x^2-1\right) \omega ^2 \left(a^4 x^2+a^2 \rho  \left(-2 M x^2+2 M+\rho +\rho  x^2\right)+\rho ^4\right)\Big)\\\notag
&+\frac{H_1}{\Delta^2\Sigma} \left(a^4 x^2 \omega +a^2 \rho  \omega  \left(-2 M x^2+2 M+\rho +\rho  x^2\right)-2 a m M \rho +\rho ^4 \omega \right)^2\\\notag
&-\frac{H_4}{\left(x^2-1\right) \Delta^2 \Sigma^2} \Big(a^4 x^2+a^2 \rho  \left(-2 M x^2+2 M+\rho +\rho  x^2\right)+\rho ^4\Big) \Big(a^2 m x^2\\
&-2 a M \rho  \left(x^2-1\right) \omega +m \rho  (\rho -2 M)\Big)^2\, .
\end{align}
}

\section{Fits to the quasinormal frequencies}\label{app:fit}
In this appendix we provide polynomial fits in $\chi$ to the quasinormal frequencies. 
For each value of $l$ and $m$, we write the quasinormal frequencies as in Eqs.~\req{qnfrotation} and \req{qnfrotationquartic}, \textit{i.e.},
\begin{equation}
M\omega= M\omega^{(0)}+\frac{\ell^4}{M^4}\left(\alpha_{1}^{2}\Delta\omega^{(1)}+\alpha_{2}^{2}\Delta\omega^{(2)}+\lambda_{\rm ev}\Delta\omega^{(\rm ev)}\right)+\frac{\ell^6}{M^6}\left(\epsilon_1\Delta\omega^{(\epsilon_1)}+\epsilon_2\Delta\omega^{(\epsilon_2)}\right)\, .
\end{equation}
At the same time, we are going to fit the zeroth-order frequencies $\omega^{(0)}$ and the correction coefficients $\Delta\omega^{(i)}$ to a polynomial in $\chi$:\footnote{Note that we define the $\omega^{(0)}_{k}$ so that they are dimensionless and independent of the mass.}
\begin{equation}
M\omega^{(0)}=\sum_{k=0}^{k_{\rm max}}\omega^{(0)}_{k}\chi^k\, ,\quad \Delta\omega^{(i)}=\sum_{k=0}^{k_{\rm max}}\Delta\omega^{(i)}_{k}\chi^k\, ,
\end{equation}
where the order $k_{\rm max}$ can be made higher if we want to get a more accurate fit. For the fits we present we have taken $k_{\rm max}=6$. Let us note that these fits are not equivalent to a Taylor expansion, and in general they will be different --- though the first terms can be similar to those of the Taylor expansion. In particular, a Taylor expansion will provide a higher accuracy for small values of $\chi$, but the convergence is slow --- according to our estimates we need up to $14$ terms in order to get a good precision for $\chi=0.7$.
In any case, we will make a few assumptions for these coefficients based on the behaviour that their Taylor expansions would have. In the case of $m=0$, we will only include even powers of $\chi$ in the fit in analogy with the Taylor expansion, while in the case of the $\alpha_2$ corrections we will set $\Delta\omega^{(2)}_{0}=0$, since these corrections must vanish in the absence of rotation.

Taking this into account, in the following tables we present the values of the best-fit coefficients for all the fundamental quasinormal frequencies with $l\le 2$ and, in the case of the $\mathcal{O}(\ell^4)$ corrections, also for the ones with $l=m=3$. In each case we indicate what is the interval of $\chi$ for which the numerical data that has been fitted. This is the interval in which the fit is fully reliable, nevertheless for slightly higher values of $\chi$ the fit should still yield an approximate result.

\clearpage

\bgroup
\def\arraystretch{1.25}
\setlength{\tabcolsep}{4pt}
\begin{table*}[h]
	\centering
	\begin{tabular}{|c||c|c|c|c|c|c|c|c|}
		\hline
		$k$&$\text{Re }\omega^{(0)}_{k}$&$\text{Im }\omega^{(0)}_{k}$&$\text{Re }\Delta\omega^{(1)}_{k}$&$\text{Im }\Delta\omega^{(1)}_{k}$&$\text{Re }\Delta\omega^{(2)}_{k}$&$\text{Im }\Delta\omega^{(2)}_{k}$&$\text{Re }\Delta\omega^{(\rm ev)}_{k}$&$\text{Im }\Delta\omega^{(\rm ev)}_{k}$\\
		\hline\hline
		0& 0.11045 & -0.1049 & 0.05165 & 0.00135 & 0. & 0. & 0.02615 & 0.01529 \\ \hline
2& 0.00779 & 0.0094 & 0.00133 & 0.02564 & 0.03043 & -0.00007 & -0.01736 & 0.00258 \\ \hline
4& 0.00039 & 0.00477 & -0.0125 & 0.01019 & 0.02432 & 0.00561 & 0.02418 & -0.00799 \\ \hline
6& -0.00295 & 0.00419 & 0.00481 & -0.00137 & 0.03501 & 0.02325 & -0.09844 & -0.00071 \\ \hline
		\end{tabular}
	\caption{Best-fit coefficients for the fundamental quasinormal frequencies with $(l,m)=(0,0)$. Numerical data fitted in the interval $0\le\chi\le 0.6$.}
	\label{tablefit:l0m0}
\end{table*}
\egroup

\bgroup
\def\arraystretch{1.25}
\setlength{\tabcolsep}{4pt}
\begin{table*}[h]
	\centering
	\begin{tabular}{|c||c|c|c|c|c|c|c|c|}
		\hline
		$k$&$\text{Re }\omega^{(0)}_{k}$&$\text{Im }\omega^{(0)}_{k}$&$\text{Re }\Delta\omega^{(1)}_{k}$&$\text{Im }\Delta\omega^{(1)}_{k}$&$\text{Re }\Delta\omega^{(2)}_{k}$&$\text{Im }\Delta\omega^{(2)}_{k}$&$\text{Re }\Delta\omega^{(\rm ev)}_{k}$&$\text{Im }\Delta\omega^{(\rm ev)}_{k}$\\
		\hline\hline
		0& 0.29294 & -0.09766 & 0.07099 & 0.00712 & 0. & 0. & 0.01055 & 0.01335  \\ \hline
1&  0.07708 & 0.00029 & 0.08126 & -0.0004 & -0.02817 & -0.04086 & 0.02772 & 0.01571 \\ \hline
2&  0.03857 & 0.00821 & 0.037 & -0.00283 & -0.0225 & -0.02337 & 0.02021 & 0.00837 \\ \hline
3&  0.00813 & -0.00184 & 0.00595 & 0.11589 & 0.00599 & -0.01063 & 0.01006 & 0.00438 \\ \hline
4& 0.064 & 0.03475 & 0.02698 & -0.45681 & -0.03122 & -0.10408 & 0.02379 & 0.0045 \\ \hline
5&  -0.08355 & -0.05101 & 0.01048 & 0.78395 & 0.05706 & 0.17702 & -0.02536 & -0.00407 \\ \hline
6& 0.07838 & 0.046 & -0.09413 & -0.5883 & -0.0749 & -0.22699 & 0.04849 & 0.00642 \\ \hline
		\end{tabular}
	\caption{Best-fit coefficients for the fundamental quasinormal frequencies with $(l,m)=(1,1)$. Numerical data fitted in the interval $0\le\chi\le 0.65$.}
	\label{tablefit:l1m1}
\end{table*}
\egroup

\bgroup
\def\arraystretch{1.25}
\setlength{\tabcolsep}{4pt}
\begin{table*}[h]
	\centering
	\begin{tabular}{|c||c|c|c|c|c|c|c|c|}
		\hline
		$k$&$\text{Re }\omega^{(0)}_{k}$&$\text{Im }\omega^{(0)}_{k}$&$\text{Re }\Delta\omega^{(1)}_{k}$&$\text{Im }\Delta\omega^{(1)}_{k}$&$\text{Re }\Delta\omega^{(2)}_{k}$&$\text{Im }\Delta\omega^{(2)}_{k}$&$\text{Re }\Delta\omega^{(\rm ev)}_{k}$&$\text{Im }\Delta\omega^{(\rm ev)}_{k}$\\
		\hline\hline
		0& 0.29294 & -0.09766 & 0.07099 & 0.00712 & 0. & 0. & 0.01058 & 0.01339 \\ \hline
2& 0.01905 & 0.00806 & 0.03156 & 0.01034 & -0.00442 & 0.00974 & 0.00919 & -0.02552  \\ \hline
4& 0.00359 & 0.0037 & -0.00059 & 0.00487 & 0.00587 & 0.00412 & 0.02652 & 0.01386 \\ \hline
6&0.00052 & 0.0032 & -0.00894 & 0.01132 & 0.03407 & 0.0049 & -0.04393 & -0.05869 \\ \hline
		\end{tabular}
	\caption{Best-fit coefficients for the fundamental quasinormal frequencies with $(l,m)=(1,0)$. Numerical data fitted in the interval $0\le\chi\le 0.65$.}
	\label{tablefit:l1m0}
\end{table*}
\egroup

\bgroup
\def\arraystretch{1.25}
\setlength{\tabcolsep}{4pt}
\begin{table*}[h]
	\centering
	\begin{tabular}{|c||c|c|c|c|c|c|c|c|}
		\hline
		$k$&$\text{Re }\omega^{(0)}_{k}$&$\text{Im }\omega^{(0)}_{k}$&$\text{Re }\Delta\omega^{(1)}_{k}$&$\text{Im }\Delta\omega^{(1)}_{k}$&$\text{Re }\Delta\omega^{(2)}_{k}$&$\text{Im }\Delta\omega^{(2)}_{k}$&$\text{Re }\Delta\omega^{(\rm ev)}_{k}$&$\text{Im }\Delta\omega^{(\rm ev)}_{k}$\\
		\hline\hline
		0&   0.29294 & -0.09766 & 0.07099 & 0.00713 & 0. & 0. & 0.01055 & 0.01335 \\ \hline
1&   -0.07716 & -0.00033 & -0.08129 & 0.00099 & 0.0282 & 0.04093 & -0.02778 & -0.01567 \\ \hline
2&   0.03699 & 0.00727 & 0.0314 & 0.01285 & -0.02274 & -0.02278 & 0.02133 & 0.00748 \\ \hline
3&   -0.0212 & -0.00559 & -0.00078 & -0.00799 & 0.00242 & 0.0332 & -0.02246 & 0.00175\\ \hline
4&  0.01301 & 0.00418 & -0.07871 & 0.02091 & -0.02282 & -0.05111 & 0.04581 & -0.02033 \\ \hline
5&  -0.00751 & -0.00199 & 0.12324 & -0.04641 & 0.01833 & 0.06798 & -0.06336 & 0.03523 \\ \hline
6&  0.00244 & 0.0003 & -0.07752 & 0.01462 & -0.00941 & -0.05411 & 0.03725 & -0.02113  \\ \hline
		\end{tabular}
	\caption{Best-fit coefficients for the fundamental quasinormal frequencies with $(l,m)=(1,-1)$. Numerical data fitted in the interval $0\le\chi\le 0.65$.}
	\label{tablefit:l1m-1}
\end{table*}
\egroup

\bgroup
\def\arraystretch{1.24}
\setlength{\tabcolsep}{4pt}
\begin{table*}[h]
	\centering
	\begin{tabular}{|c||c|c|c|c|c|c|c|c|}
		\hline
		$k$&$\text{Re }\omega^{(0)}_{k}$&$\text{Im }\omega^{(0)}_{k}$&$\text{Re }\Delta\omega^{(1)}_{k}$&$\text{Im }\Delta\omega^{(1)}_{k}$&$\text{Re }\Delta\omega^{(2)}_{k}$&$\text{Im }\Delta\omega^{(2)}_{k}$&$\text{Re }\Delta\omega^{(\rm ev)}_{k}$&$\text{Im }\Delta\omega^{(\rm ev)}_{k}$\\
		\hline\hline
		0& 0.48365 & -0.09676 & 0.10448 & 0.00798 & 0. & 0. & 0.01277 & 0.01001\\ \hline
1&  0.15013 & 0.00008 & 0.15036 & 0.00487 & -0.05364 & -0.04549 & 0.04024 & 0.02119 \\ \hline
2&   0.08157 & 0.00837 & 0.09239 & -0.0239 & 0.00893 & -0.05012 & 0.00319 & 0.01263 \\ \hline
3&   -0.01213 & -0.00345 & -0.04352 & 0.22076 & -0.38405 & 0.03937 & 0.20292 & 0.03732\\ \hline
4&   0.23056 & 0.04265 & 0.14706 & -0.80785 & 1.12651 & -0.33955 & -0.54716 & -0.09\\ \hline
5&  -0.31813 & -0.06224 & 0.00446 & 1.31401 & -1.59675 & 0.57161 & 0.79409 & 0.14973 \\ \hline
6&  0.25178 & 0.05362 & -0.27766 & -0.92678 & 0.67854 & -0.51428 & -0.35505 & -0.08484  \\ \hline
		\end{tabular}
	\caption{Best-fit coefficients for the fundamental quasinormal frequencies with $(l,m)=(2,2)$. Numerical data fitted in the interval $0\le\chi\le 0.7$.}
	\label{tablefit:l2m2}
\end{table*}
\egroup

\clearpage

\bgroup
\def\arraystretch{1.24}
\setlength{\tabcolsep}{4pt}
\begin{table*}[h]
	\centering
	\begin{tabular}{|c||c|c|c|c|c|c|c|c|}
		\hline
		$k$&$\text{Re }\omega^{(0)}_{k}$&$\text{Im }\omega^{(0)}_{k}$&$\text{Re }\Delta\omega^{(1)}_{k}$&$\text{Im }\Delta\omega^{(1)}_{k}$&$\text{Re }\Delta\omega^{(2)}_{k}$&$\text{Im }\Delta\omega^{(2)}_{k}$&$\text{Re }\Delta\omega^{(\rm ev)}_{k}$&$\text{Im }\Delta\omega^{(\rm ev)}_{k}$\\
		\hline\hline
		0& 0.48364 & -0.09676 & 0.10449 & 0.00799 & 0. & 0. & 0.01275 & 0.01\\ \hline
1&   0.07516 & 0.00002 & 0.07368 & 0.002 & -0.0277 & -0.02259 & 0.0223 & 0.01213 \\ \hline
2&  0.04378 & 0.00833 & 0.09571 & 0.00488 & 0.01994 & -0.01112 & -0.02474 & -0.03284 \\ \hline
3&  0.0066 & -0.0037 & -0.22682 & 0.04119 & -0.36303 & 0.04743 & 0.40041 & 0.2142\\ \hline
4&  0.05725 & 0.03181 & 0.84182 & -0.16931 & 1.19518 & -0.21456 & -1.22029 & -0.70041\\ \hline
5& -0.06926 & -0.0448 & -1.2083 & 0.29016 & -1.85175 & 0.3513 & 1.81083 & 1.02423\\ \hline
6&   0.06078 & 0.03673 & 0.65095 & -0.21579 & 1.0752 & -0.25814 & -0.98215 & -0.57558   \\ \hline
		\end{tabular}
	\caption{Best-fit coefficients for the fundamental quasinormal frequencies with $(l,m)=(2,1)$. Numerical data fitted in the interval $0\le\chi\le 0.7$.}
	\label{tablefit:l2m1}
\end{table*}
\egroup

\bgroup
\def\arraystretch{1.24}
\setlength{\tabcolsep}{4pt}
\begin{table*}[h]
	\centering
	\begin{tabular}{|c||c|c|c|c|c|c|c|c|}
		\hline
		$k$&$\text{Re }\omega^{(0)}_{k}$&$\text{Im }\omega^{(0)}_{k}$&$\text{Re }\Delta\omega^{(1)}_{k}$&$\text{Im }\Delta\omega^{(1)}_{k}$&$\text{Re }\Delta\omega^{(2)}_{k}$&$\text{Im }\Delta\omega^{(2)}_{k}$&$\text{Re }\Delta\omega^{(\rm ev)}_{k}$&$\text{Im }\Delta\omega^{(\rm ev)}_{k}$\\
		\hline\hline
			0&  0.48364 & -0.09676 & 0.10448 & 0.00798 & 0. & 0. & 0.01278 & 0.01006 \\ \hline
2&  0.03151 & 0.00746 & 0.05188 & 0.01259 & -0.01977 & 0.00869 & 0.0307 & -0.01066  \\ \hline
4& 0.00642 & 0.0034 & -0.00151 & 0.00638 & -0.00649 & 0.00333 & 0.00676 & 0.02744 \\ \hline
6& 0.00258 & 0.00326 & -0.03061 & 0.00368 & 0.06307 & 0.00719 & -0.00278 & -0.04022 \\ \hline
		\end{tabular}
	\caption{Best-fit coefficients for the fundamental quasinormal frequencies with $(l,m)=(2,0)$. Numerical data fitted in the interval $0\le\chi\le 0.7$.}
	\label{tablefit:l2m0}
\end{table*}
\egroup

\bgroup
\def\arraystretch{1.24}
\setlength{\tabcolsep}{4pt}
\begin{table*}[h]
	\centering
	\begin{tabular}{|c||c|c|c|c|c|c|c|c|}
		\hline
		$k$&$\text{Re }\omega^{(0)}_{k}$&$\text{Im }\omega^{(0)}_{k}$&$\text{Re }\Delta\omega^{(1)}_{k}$&$\text{Im }\Delta\omega^{(1)}_{k}$&$\text{Re }\Delta\omega^{(2)}_{k}$&$\text{Im }\Delta\omega^{(2)}_{k}$&$\text{Re }\Delta\omega^{(\rm ev)}_{k}$&$\text{Im }\Delta\omega^{(\rm ev)}_{k}$\\
		\hline\hline
		0& 0.48364 & -0.09676 & 0.10447 & 0.00799 & 0. & 0. & 0.01278 & 0.01001\\ \hline
1&  -0.07525 & -0.00007 & -0.07521 & -0.00185 & 0.02552 & 0.02288 & -0.01961 & -0.01047 \\ \hline
2& 0.04219 & 0.00728 & 0.05057 & 0.01269 & -0.02349 & -0.00379 & 0.03451 & -0.00235 \\ \hline
3&  -0.01926 & -0.00337 & 0.02529 & -0.01194 & -0.01582 & 0.00749 & -0.04148 & 0.00933\\ \hline
4& 0.01269 & 0.00257 & -0.19476 & 0.04779 & 0.08576 & 0.00193 & 0.1019 & -0.01949\\ \hline
5& -0.00709 & 0.00033 & 0.31157 & -0.07185 & -0.14109 & -0.00497 & -0.14067 & 0.02909\\ \hline
6&  0.00237 & -0.00036 & -0.2025 & 0.03429 & 0.09928 & 0.0054 & 0.07601 & -0.01611  \\ \hline
		\end{tabular}
	\caption{Best-fit coefficients for the fundamental quasinormal frequencies with $(l,m)=(2,-1)$. Numerical data fitted in the interval $0\le\chi\le 0.7$.}
	\label{tablefit:l2m-1}
\end{table*}
\egroup

\clearpage

\bgroup
\def\arraystretch{1.24}
\setlength{\tabcolsep}{4pt}
\begin{table*}[h]
	\centering
	\begin{tabular}{|c||c|c|c|c|c|c|c|c|}
		\hline
		$k$&$\text{Re }\omega^{(0)}_{k}$&$\text{Im }\omega^{(0)}_{k}$&$\text{Re }\Delta\omega^{(1)}_{k}$&$\text{Im }\Delta\omega^{(1)}_{k}$&$\text{Re }\Delta\omega^{(2)}_{k}$&$\text{Im }\Delta\omega^{(2)}_{k}$&$\text{Re }\Delta\omega^{(\rm ev)}_{k}$&$\text{Im }\Delta\omega^{(\rm ev)}_{k}$\\
		\hline\hline
		0&  0.48364 & -0.09676 & 0.10447 & 0.00799 & 0. & 0. & 0.01278 & 0.01001\\ \hline
1& -0.15049 & -0.00014 & -0.15038 & -0.00367 & 0.05159 & 0.0458 & -0.03896 & -0.02099\\ \hline
2&   0.07413 & 0.00709 & 0.06167 & 0.00901 & -0.04846 & -0.04272 & 0.03313 & 0.01627 \\ \hline
3&  -0.04423 & -0.0061 & 0.08825 & -0.02782 & 0.0652 & 0.03665 & -0.04166 & -0.00732\\ \hline
4&  0.02805 & 0.00402 & -0.39612 & 0.08147 & -0.18006 & -0.04624 & 0.09165 & -0.00016\\ \hline
5& -0.01544 & -0.00154 & 0.63235 & -0.14455 & 0.25774 & 0.04068 & -0.12671 & 0.006 \\ \hline
6&  0.00481 & 0.00012 & -0.36687 & 0.08613 & -0.13258 & -0.02436 & 0.07158 & -0.00408  \\ \hline
		\end{tabular}
	\caption{Best-fit coefficients for the fundamental quasinormal frequencies with $(l,m)=(2,-2)$. Numerical data fitted in the interval $0\le\chi\le 0.7$.}
	\label{tablefit:l2m-2}
\end{table*}
\egroup

\bgroup
\def\arraystretch{1.24}
\setlength{\tabcolsep}{4pt}
\begin{table*}[h]
	\centering
	\begin{tabular}{|c||c|c|c|c|c|c|c|c|}
		\hline
		$k$&$\text{Re }\omega^{(0)}_{k}$&$\text{Im }\omega^{(0)}_{k}$&$\text{Re }\Delta\omega^{(1)}_{k}$&$\text{Im }\Delta\omega^{(1)}_{k}$&$\text{Re }\Delta\omega^{(2)}_{k}$&$\text{Im }\Delta\omega^{(2)}_{k}$&$\text{Re }\Delta\omega^{(\rm ev)}_{k}$&$\text{Im }\Delta\omega^{(\rm ev)}_{k}$\\
		\hline\hline
		0&  0.67537 & -0.0965 & 0.14087 & 0.00817 & 0. & 0. & 0.01767 & 0.00854\\ \hline
1&0.22402 & 0.00008 & 0.21982 & 0.00608 & -0.07668 & -0.04739 & 0.05231 & 0.02172 \\ \hline
2&  0.11294 & 0.00703 & 0.12123 & 0.00073 & -0.04613 & -0.05069 & 0.0316 & 0.0191 \\ \hline
3&  0.06363 & 0.00757 & 0.06436 & 0.02026 & -0.15188 & -0.05168 & 0.07222 & 0.02693\\ \hline
4&  0.07388 & 0.00186 & -0.09679 & -0.14218 & 0.57269 & 0.00605 & -0.23133 & -0.06056\\ \hline
5& -0.03812 & 0.01125 & 0.23416 & 0.27696 & -1.43677 & -0.11812 & 0.5903 & 0.15652 \\ \hline
6&  0.11367 & 0.00252 & -0.35831 & -0.33504 & 1.14031 & -0.00669 & -0.4772 & -0.12771  \\ \hline
		\end{tabular}
	\caption{Best-fit coefficients for the fundamental quasinormal frequencies with $(l,m)=(3,3)$. Numerical data fitted in the interval $0\le\chi\le 0.5$.}
	\label{tablefit:l3m3}
\end{table*}
\egroup

\bgroup
\def\arraystretch{1.24}
\setlength{\tabcolsep}{4pt}
\begin{table*}[h]
	\centering
	\begin{tabular}{|c||c|c|c|c|c|c|c|c|}
		\hline
		$k$&$\text{Re }\Delta\omega^{(\epsilon_1)}_{k}$&$\text{Im }\Delta\omega^{(\epsilon_1)}_{k}$&$\text{Re }\Delta\omega^{(\epsilon_2)}_{k}$&$\text{Im }\Delta\omega^{(\epsilon_2)}_{k}$\\
		\hline\hline
			0& 0.05563 & 0.05074 & 0. & 0. \\ \hline
2&   -0.01986 & 0.00681 & 0.08529 & 0.06786  \\ \hline
4&  -0.01605 & 0.00357 & 0.06874 & 0.06671  \\ \hline
6&0.21976 & 0.03214 & -0.03405 & 0.11615  \\ \hline
		\end{tabular}
	\caption{Best-fit coefficients for the fundamental quasinormal frequencies with $(l,m)=(0,0)$. Numerical data fitted in the interval $0\le\chi\le 0.55$.}
	\label{tablefitQ:l0m0}
\end{table*}
\egroup

\clearpage

\bgroup
\def\arraystretch{1.24}
\setlength{\tabcolsep}{4pt}
\begin{table*}[h]
	\centering
	\begin{tabular}{|c||c|c|c|c|c|c|c|c|}
		\hline
		$k$&$\text{Re }\Delta\omega^{(\epsilon_1)}_{k}$&$\text{Im }\Delta\omega^{(\epsilon_1)}_{k}$&$\text{Re }\Delta\omega^{(\epsilon_2)}_{k}$&$\text{Im }\Delta\omega^{(\epsilon_2)}_{k}$\\
		\hline\hline
0& 0.02268 & 0.04957 & 0. & 0. \\ \hline
1&  0.06603 & 0.03729 & 0.01918 & -0.06619 \\ \hline
2&   0.066 & -0.06576 & -0.0575 & -0.06562  \\ \hline
3&   -0.59936 & 0.48246 & 0.54102 & 0.26582  \\ \hline
4&  2.09244 & -2.27312 & -1.83777 & -1.24738  \\ \hline
5&   -3.55373 & 4.08949 & 3.34512 & 2.40635   \\ \hline
6& 2.01729 & -3.37237 & -2.3341 & -2.05855 \\ \hline
		\end{tabular}
	\caption{Best-fit coefficients for the fundamental quasinormal frequencies with $(l,m)=(1,1)$. Numerical data fitted in the interval $0\le\chi\le 0.6$.}
	\label{tablefitQ:l1m1}
\end{table*}
\egroup

\bgroup
\def\arraystretch{1.24}
\setlength{\tabcolsep}{4pt}
\begin{table*}[h]
	\centering
	\begin{tabular}{|c||c|c|c|c|c|c|c|c|}
		\hline
		$k$&$\text{Re }\Delta\omega^{(\epsilon_1)}_{k}$&$\text{Im }\Delta\omega^{(\epsilon_1)}_{k}$&$\text{Re }\Delta\omega^{(\epsilon_2)}_{k}$&$\text{Im }\Delta\omega^{(\epsilon_2)}_{k}$\\
		\hline\hline
			0& 0.02254 & 0.04943 & 0. & 0. \\ \hline
2&  0.11835 & -0.01758 & -0.04988 & 0.06689 \\ \hline
4&  -0.06671 & -0.12347 & 0.10883 & 0.11394 \\ \hline
6& 0.3679 & 0.44587 & -0.03133 & -0.21393  \\ \hline
		\end{tabular}
	\caption{Best-fit coefficients for the fundamental quasinormal frequencies with $(l,m)=(1,0)$. Numerical data fitted in the interval $0\le\chi\le 0.6$.}
	\label{tablefitQ:l1m0}
\end{table*}
\egroup

\bgroup
\def\arraystretch{1.24}
\setlength{\tabcolsep}{4pt}
\begin{table*}[h]
	\centering
	\begin{tabular}{|c||c|c|c|c|c|c|c|c|}
		\hline
		$k$&$\text{Re }\Delta\omega^{(\epsilon_1)}_{k}$&$\text{Im }\Delta\omega^{(\epsilon_1)}_{k}$&$\text{Re }\Delta\omega^{(\epsilon_2)}_{k}$&$\text{Im }\Delta\omega^{(\epsilon_2)}_{k}$\\
		\hline\hline
0& 0.02265 & 0.04959 & 0. & 0. \\ \hline
1& -0.06894 & -0.03486 & -0.01715 & 0.06744 \\ \hline
2&   -0.01511 & -0.00246 & -0.00338 & -0.04172  \\ \hline
3&  0.07621 & 0.03351 & -0.07162 & 0.06305  \\ \hline
4& -0.30424 & -0.03019 & 0.0152 & -0.24391  \\ \hline
5&   0.45248 & -0.01903 & -0.03135 & 0.40034  \\ \hline
6&  -0.29976 & -0.04325 & -0.00458 & -0.33342 \\ \hline
		\end{tabular}
	\caption{Best-fit coefficients for the fundamental quasinormal frequencies with $(l,m)=(1,-1)$. Numerical data fitted in the interval $0\le\chi\le 0.6$.}
	\label{tablefitQ:l1m-1}
\end{table*}
\egroup

\clearpage

\bgroup
\def\arraystretch{1.24}
\setlength{\tabcolsep}{4pt}
\begin{table*}[h]
	\centering
	\begin{tabular}{|c||c|c|c|c|c|c|c|c|}
		\hline
		$k$&$\text{Re }\Delta\omega^{(\epsilon_1)}_{k}$&$\text{Im }\Delta\omega^{(\epsilon_1)}_{k}$&$\text{Re }\Delta\omega^{(\epsilon_2)}_{k}$&$\text{Im }\Delta\omega^{(\epsilon_2)}_{k}$\\
		\hline\hline
0& 0.02196 & 0.04401 & 0. & 0. \\ \hline
1& 0.09087 & 0.06379 & 0.0146 & -0.05694 \\ \hline
2& -0.03684 & -0.0665 & 0.00196 & -0.14295 \\ \hline
3&  0.09022 & 0.58874 & 0.43019 & 0.43705   \\ \hline
4&  -1.06044 & -2.74112 & -1.466 & -2.13431 \\ \hline
5&   2.08651 & 5.02003 & 3.11198 & 3.99247 \\ \hline
6&  -2.39044 & -4.27682 & -2.73093 & -3.34858\\ \hline
		\end{tabular}
	\caption{Best-fit coefficients for the fundamental quasinormal frequencies with $(l,m)=(2,2)$. Numerical data fitted in the interval $0\le\chi\le 0.6$.}
	\label{tablefitQ:l2m2}
\end{table*}
\egroup

\bgroup
\def\arraystretch{1.24}
\setlength{\tabcolsep}{4pt}
\begin{table*}[h]
	\centering
	\begin{tabular}{|c||c|c|c|c|c|c|c|c|}
		\hline
		$k$&$\text{Re }\Delta\omega^{(\epsilon_1)}_{k}$&$\text{Im }\Delta\omega^{(\epsilon_1)}_{k}$&$\text{Re }\Delta\omega^{(\epsilon_2)}_{k}$&$\text{Im }\Delta\omega^{(\epsilon_2)}_{k}$\\
		\hline\hline
0& 0.02198 & 0.04403 & 0. & 0. \\ \hline
1& 0.04241 & 0.02976 & 0.00653 & -0.02828 \\ \hline
2&0.1701 & 0.01084 & -0.0433 & -0.01456 \\ \hline
3&  0.09022 & 0.58874 & 0.43019 & 0.43705   \\ \hline
4&-0.45995 & -0.15267 & -0.07654 & 0.28357 \\ \hline
5&  -3.53665 & -0.51708 & -0.4018 & 1.93122 \\ \hline
6&  2.4638 & 0.11233 & 0.28765 & -1.39317\\ \hline
		\end{tabular}
	\caption{Best-fit coefficients for the fundamental quasinormal frequencies with $(l,m)=(2,1)$. Numerical data fitted in the interval $0\le\chi\le 0.6$.}
	\label{tablefitQ:l2m1}
\end{table*}
\egroup

\bgroup
\def\arraystretch{1.24}
\setlength{\tabcolsep}{4pt}
\begin{table*}[h]
	\centering
	\begin{tabular}{|c||c|c|c|c|c|c|c|c|}
		\hline
		$k$&$\text{Re }\Delta\omega^{(\epsilon_1)}_{k}$&$\text{Im }\Delta\omega^{(\epsilon_1)}_{k}$&$\text{Re }\Delta\omega^{(\epsilon_2)}_{k}$&$\text{Im }\Delta\omega^{(\epsilon_2)}_{k}$\\
		\hline\hline
			0& 0.02195 & 0.04399 & 0. & 0. \\ \hline
2& 0.15007 & -0.0028 & -0.07404 & 0.04972\\ \hline
4& 0.00256 & -0.02278 & 0.05823 & 0.04803 \\ \hline
6& 0.05537 & 0.10027 & 0.19755 & 0.01568  \\ \hline
		\end{tabular}
	\caption{Best-fit coefficients for the fundamental quasinormal frequencies with $(l,m)=(2,0)$. Numerical data fitted in the interval $0\le\chi\le 0.6$.}
	\label{tablefitQ:l2m0}
\end{table*}
\egroup

\clearpage

\bgroup
\def\arraystretch{1.24}
\setlength{\tabcolsep}{4pt}
\begin{table*}[h]
	\centering
	\begin{tabular}{|c||c|c|c|c|c|c|c|c|}
		\hline
		$k$&$\text{Re }\Delta\omega^{(\epsilon_1)}_{k}$&$\text{Im }\Delta\omega^{(\epsilon_1)}_{k}$&$\text{Re }\Delta\omega^{(\epsilon_2)}_{k}$&$\text{Im }\Delta\omega^{(\epsilon_2)}_{k}$\\
		\hline\hline
0& 0.02196 & 0.04403 & 0. & 0.  \\ \hline
1& -0.04479 & -0.03058 & -0.0067 & 0.02933 \\ \hline
2& 0.09883 & 0.00015 & -0.0472 & 0.01785 \\ \hline
3& 0.00037 & 0.01481 & 0.02956 & -0.03795  \\ \hline
4&-0.21416 & 0.096 & 0.02624 & 0.11391 \\ \hline
5&0.42318 & -0.19835 & -0.00313 & -0.17872\\ \hline
6& -0.27925 & 0.17271 & 0.04023 & 0.17706\\ \hline
		\end{tabular}
	\caption{Best-fit coefficients for the fundamental quasinormal frequencies with $(l,m)=(2,-1)$. Numerical data fitted in the interval $0\le\chi\le 0.6$.}
	\label{tablefitQ:l2m-1}
\end{table*}
\egroup

\bgroup
\def\arraystretch{1.24}
\setlength{\tabcolsep}{4pt}
\begin{table*}[h]
	\centering
	\begin{tabular}{|c||c|c|c|c|c|c|c|c|}
		\hline
		$k$&$\text{Re }\Delta\omega^{(\epsilon_1)}_{k}$&$\text{Im }\Delta\omega^{(\epsilon_1)}_{k}$&$\text{Re }\Delta\omega^{(\epsilon_2)}_{k}$&$\text{Im }\Delta\omega^{(\epsilon_2)}_{k}$\\
		\hline\hline
0&  0.02195 & 0.04403 & 0. & 0. \\ \hline
1& -0.08973 & -0.06095 & -0.01301 & 0.05876 \\ \hline
2& -0.03295 & 0.00667 & 0.02763 & -0.08893\\ \hline
3&0.20992 & 0.02785 & -0.00447 & 0.03235  \\ \hline
4&-0.55375 & -0.03307 & -0.30014 & -0.00253\\ \hline
5&0.89057 & -0.04055 & 0.6136 & -0.07965\\ \hline
6&-0.57514 & 0.04579 & -0.43004 & 0.06671\\ \hline
		\end{tabular}
	\caption{Best-fit coefficients for the fundamental quasinormal frequencies with $(l,m)=(2,-2)$. Numerical data fitted in the interval $0\le\chi\le 0.6$.}
	\label{tablefitQ:l2m-2}
	\end{table*}
\egroup

\bibliography{Gravities}

\newcommand{\noop}[1]{}
\providecommand{\href}[2]{#2}\begingroup\raggedright\begin{thebibliography}{10}

\bibitem{LIGOScientific:2018mvr}
{\bf LIGO Scientific, Virgo} Collaboration, B.~Abbott {\em et.~al.}, {\it
  {GWTC-1: A Gravitational-Wave Transient Catalog of Compact Binary Mergers
  Observed by LIGO and Virgo during the First and Second Observing Runs}},
  {\em Phys. Rev. X} {\bf 9} (2019), no.~3 031040
  [\href{http://arXiv.org/abs/1811.12907}{{\tt 1811.12907}}].

\bibitem{TheLIGOScientific:2016src}
{\bf LIGO Scientific, Virgo} Collaboration, B.~P. Abbott {\em et.~al.}, {\it
  {Tests of general relativity with GW150914}},  {\em Phys. Rev. Lett.} {\bf
  116} (2016), no.~22 221101 [\href{http://arXiv.org/abs/1602.03841}{{\tt
  1602.03841}}]. [Erratum: Phys. Rev. Lett.121,no.12,129902(2018)].

\bibitem{Yunes:2016jcc}
N.~Yunes, K.~Yagi and F.~Pretorius, {\it {Theoretical Physics Implications of
  the Binary Black-Hole Mergers GW150914 and GW151226}},  {\em Phys. Rev.} {\bf
  D94} (2016), no.~8 084002 [\href{http://arXiv.org/abs/1603.08955}{{\tt
  1603.08955}}].

\bibitem{Berti:2018cxi}
E.~Berti, K.~Yagi and N.~Yunes, {\it {Extreme Gravity Tests with Gravitational
  Waves from Compact Binary Coalescences: (I) Inspiral-Merger}},  {\em Gen.
  Rel. Grav.} {\bf 50} (2018), no.~4 46
  [\href{http://arXiv.org/abs/1801.03208}{{\tt 1801.03208}}].

\bibitem{Berti:2018vdi}
E.~Berti, K.~Yagi, H.~Yang and N.~Yunes, {\it {Extreme Gravity Tests with
  Gravitational Waves from Compact Binary Coalescences: (II) Ringdown}},  {\em
  Gen. Rel. Grav.} {\bf 50} (2018), no.~5 49
  [\href{http://arXiv.org/abs/1801.03587}{{\tt 1801.03587}}].

\bibitem{Barack:2018yly}
L.~Barack {\em et.~al.}, {\it {Black holes, gravitational waves and fundamental
  physics: a roadmap}},  {\em Class. Quant. Grav.} {\bf 36} (2019), no.~14
  143001 [\href{http://arXiv.org/abs/1806.05195}{{\tt 1806.05195}}].

\bibitem{Abbott:2018lct}
{\bf LIGO Scientific, Virgo} Collaboration, B.~Abbott {\em et.~al.}, {\it
  {Tests of General Relativity with GW170817}},  {\em Phys. Rev. Lett.} {\bf
  123} (2019), no.~1 011102 [\href{http://arXiv.org/abs/1811.00364}{{\tt
  1811.00364}}].

\bibitem{LIGOScientific:2019fpa}
{\bf LIGO Scientific, Virgo} Collaboration, B.~Abbott {\em et.~al.}, {\it
  {Tests of General Relativity with the Binary Black Hole Signals from the
  LIGO-Virgo Catalog GWTC-1}},  {\em Phys. Rev. D} {\bf 100} (2019), no.~10
  104036 [\href{http://arXiv.org/abs/1903.04467}{{\tt 1903.04467}}].

\bibitem{Okounkova:2019zjf}
M.~Okounkova, L.~C. Stein, J.~Moxon, M.~A. Scheel and S.~A. Teukolsky, {\it
  {Numerical relativity simulation of GW150914 beyond general relativity}},
  \href{http://arXiv.org/abs/1911.02588}{{\tt 1911.02588}}.

\bibitem{Sennett:2019bpc}
N.~Sennett, R.~Brito, A.~Buonanno, V.~Gorbenko and L.~Senatore, {\it
  {Gravitational-Wave Constraints on an Effective--Field-Theory Extension of
  General Relativity}},  \href{http://arXiv.org/abs/1912.09917}{{\tt
  1912.09917}}.

\bibitem{Carson:2020cqb}
Z.~Carson and K.~Yagi, {\it {Probing string-inspired gravity with the
  inspiral-merger-ringdown consistency tests of gravitational waves}},
  \href{http://arXiv.org/abs/2002.08559}{{\tt 2002.08559}}.

\bibitem{Carson:2020ter}
Z.~Carson and K.~Yagi, {\it {Probing Einstein-dilaton Gauss-Bonnet Gravity with
  the inspiral and ringdown of gravitational waves}},  {\em Phys. Rev. D} {\bf
  101} (2020), no.~10 104030 [\href{http://arXiv.org/abs/2003.00286}{{\tt
  2003.00286}}].

\bibitem{Carson:2020iik}
Z.~Carson and K.~Yagi, {\it {Probing beyond-Kerr spacetimes with
  inspiral-ringdown corrections to gravitational waves}},
  \href{http://arXiv.org/abs/2003.02374}{{\tt 2003.02374}}.

\bibitem{Okounkova:2020rqw}
M.~Okounkova, {\it {Numerical relativity simulation of GW150914 in Einstein
  dilaton Gauss-Bonnet gravity}},  \href{http://arXiv.org/abs/2001.03571}{{\tt
  2001.03571}}.

\bibitem{Endlich:2017tqa}
S.~Endlich, V.~Gorbenko, J.~Huang and L.~Senatore, {\it {An effective formalism
  for testing extensions to General Relativity with gravitational waves}},
  {\em JHEP} {\bf 09} (2017) 122 [\href{http://arXiv.org/abs/1704.01590}{{\tt
  1704.01590}}].

\bibitem{Cano:2019ore}
P.~A. Cano and A.~Ruipérez, {\it {Leading higher-derivative corrections to
  Kerr geometry}},  {\em JHEP} {\bf 05} (2019) 189
  [\href{http://arXiv.org/abs/1901.01315}{{\tt 1901.01315}}]. [Erratum: JHEP
  03, 187 (2020)].

\bibitem{Berti:2009kk}
E.~Berti, V.~Cardoso and A.~O. Starinets, {\it {Quasinormal modes of black
  holes and black branes}},  {\em Class. Quant. Grav.} {\bf 26} (2009) 163001
  [\href{http://arXiv.org/abs/0905.2975}{{\tt 0905.2975}}].

\bibitem{Maselli:2019mjd}
A.~Maselli, P.~Pani, L.~Gualtieri and E.~Berti, {\it {Parametrized ringdown
  spin expansion coefficients: a data-analysis framework for black-hole
  spectroscopy with multiple events}},  {\em Phys. Rev.} {\bf D101} (2020),
  no.~2 024043 [\href{http://arXiv.org/abs/1910.12893}{{\tt 1910.12893}}].

\bibitem{Cardoso:2009pk}
V.~Cardoso and L.~Gualtieri, {\it {Perturbations of Schwarzschild black holes
  in Dynamical Chern-Simons modified gravity}},  {\em Phys. Rev.} {\bf D80}
  (2009) 064008 [\href{http://arXiv.org/abs/0907.5008}{{\tt 0907.5008}}].
  [Erratum: Phys. Rev.D81,089903(2010)].

\bibitem{Blazquez-Salcedo:2016enn}
J.~L. Blázquez-Salcedo, C.~F.~B. Macedo, V.~Cardoso, V.~Ferrari, L.~Gualtieri,
  F.~S. Khoo, J.~Kunz and P.~Pani, {\it {Perturbed black holes in
  Einstein-dilaton-Gauss-Bonnet gravity: Stability, ringdown, and
  gravitational-wave emission}},  {\em Phys. Rev.} {\bf D94} (2016), no.~10
  104024 [\href{http://arXiv.org/abs/1609.01286}{{\tt 1609.01286}}].

\bibitem{Blazquez-Salcedo:2017txk}
J.~L. Blázquez-Salcedo, F.~S. Khoo and J.~Kunz, {\it {Quasinormal modes of
  Einstein-Gauss-Bonnet-dilaton black holes}},  {\em Phys. Rev. D} {\bf 96}
  (2017), no.~6 064008 [\href{http://arXiv.org/abs/1706.03262}{{\tt
  1706.03262}}].

\bibitem{Tattersall:2017erk}
O.~J. Tattersall, P.~G. Ferreira and M.~Lagos, {\it {General theories of linear
  gravitational perturbations to a Schwarzschild Black Hole}},  {\em Phys. Rev.
  D} {\bf 97} (2018), no.~4 044021 [\href{http://arXiv.org/abs/1711.01992}{{\tt
  1711.01992}}].

\bibitem{Tattersall:2018nve}
O.~J. Tattersall and P.~G. Ferreira, {\it {Quasinormal modes of black holes in
  Horndeski gravity}},  {\em Phys. Rev. D} {\bf 97} (2018), no.~10 104047
  [\href{http://arXiv.org/abs/1804.08950}{{\tt 1804.08950}}].

\bibitem{Cardoso:2018ptl}
V.~Cardoso, M.~Kimura, A.~Maselli and L.~Senatore, {\it {Black holes in an
  Effective Field Theory extension of GR}},  {\em Phys. Rev. Lett.} {\bf 121}
  (2018), no.~25 251105 [\href{http://arXiv.org/abs/1808.08962}{{\tt
  1808.08962}}].

\bibitem{Konoplya:2020bxa}
R.~Konoplya and A.~Zinhailo, {\it {Quasinormal modes, stability and shadows of
  a black hole in the novel 4D Einstein-Gauss-Bonnet gravity}},
  \href{http://arXiv.org/abs/2003.01188}{{\tt 2003.01188}}.

\bibitem{Cardoso:2019mqo}
V.~Cardoso, M.~Kimura, A.~Maselli, E.~Berti, C.~F. Macedo and R.~McManus, {\it
  {Parametrized black hole quasinormal ringdown: Decoupled equations for
  nonrotating black holes}},  {\em Phys. Rev. D} {\bf 99} (2019), no.~10 104077
  [\href{http://arXiv.org/abs/1901.01265}{{\tt 1901.01265}}].

\bibitem{McManus:2019ulj}
R.~McManus, E.~Berti, C.~F. Macedo, M.~Kimura, A.~Maselli and V.~Cardoso, {\it
  {Parametrized black hole quasinormal ringdown. II. Coupled equations and
  quadratic corrections for nonrotating black holes}},  {\em Phys. Rev. D} {\bf
  100} (2019), no.~4 044061 [\href{http://arXiv.org/abs/1906.05155}{{\tt
  1906.05155}}].

\bibitem{Kerr:1963ud}
R.~P. Kerr, {\it {Gravitational field of a spinning mass as an example of
  algebraically special metrics}},  {\em Phys. Rev. Lett.} {\bf 11} (1963)
  237--238.

\bibitem{PhysRevLett.29.1114}
S.~A. Teukolsky, {\it Rotating black holes: Separable wave equations for
  gravitational and electromagnetic perturbations},  {\em Phys. Rev. Lett.}
  {\bf 29} (Oct, 1972) 1114--1118.

\bibitem{Teukolsky:1973ha}
S.~A. Teukolsky, {\it {Perturbations of a rotating black hole. 1. Fundamental
  equations for gravitational electromagnetic and neutrino field
  perturbations}},  {\em Astrophys. J.} {\bf 185} (1973) 635--647.

\bibitem{PhysRevD.30.295}
V.~Ferrari and B.~Mashhoon, {\it New approach to the quasinormal modes of a
  black hole},  {\em Phys. Rev. D} {\bf 30} (Jul, 1984) 295--304.

\bibitem{Schutz:1985km}
B.~F. Schutz and C.~M. Will, {\it {Black hole normal modes: a semianalytic
  approach}},  {\em Astrophys. J.} {\bf 291} (1985) L33--L36.

\bibitem{Leaver:1985ax}
E.~Leaver, {\it {An Analytic representation for the quasi normal modes of Kerr
  black holes}},  {\em Proc. Roy. Soc. Lond. A} {\bf A402} (1985) 285--298.

\bibitem{doi:10.1063/1.527130}
E.~W. Leaver, {\it Solutions to a generalized spheroidal wave equation:
  Teukolsky’s equations in general relativity, and the two-center problem in
  molecular quantum mechanics},  {\em Journal of Mathematical Physics} {\bf 27}
  (1986), no.~5 1238--1265
  [\href{http://arXiv.org/abs/https://doi.org/10.1063/1.527130}{{\tt
  https://doi.org/10.1063/1.527130}}].

\bibitem{Leaver:1986gd}
E.~W. Leaver, {\it {Spectral decomposition of the perturbation response of the
  Schwarzschild geometry}},  {\em Phys. Rev. D} {\bf 34} (1986) 384--408.

\bibitem{Zimmerman:2014aha}
A.~Zimmerman, H.~Yang, Z.~Mark, Y.~Chen and L.~Lehner, {\it {Quasinormal Modes
  Beyond Kerr}},  {\em Astrophys. Space Sci. Proc.} {\bf 40} (2015) 217--223
  [\href{http://arXiv.org/abs/1406.4206}{{\tt 1406.4206}}].

\bibitem{Mark:2014aja}
Z.~Mark, H.~Yang, A.~Zimmerman and Y.~Chen, {\it {Quasinormal modes of weakly
  charged Kerr-Newman spacetimes}},  {\em Phys. Rev.} {\bf D91} (2015), no.~4
  044025 [\href{http://arXiv.org/abs/1409.5800}{{\tt 1409.5800}}].

\bibitem{Camanho:2014apa}
X.~O. Camanho, J.~D. Edelstein, J.~Maldacena and A.~Zhiboedov, {\it {Causality
  Constraints on Corrections to the Graviton Three-Point Coupling}},  {\em
  JHEP} {\bf 02} (2016) 020 [\href{http://arXiv.org/abs/1407.5597}{{\tt
  1407.5597}}].

\bibitem{Kanti:1995vq}
P.~Kanti, N.~E. Mavromatos, J.~Rizos, K.~Tamvakis and E.~Winstanley, {\it
  {Dilatonic black holes in higher curvature string gravity}},  {\em Phys.
  Rev.} {\bf D54} (1996) 5049--5058
  [\href{http://arXiv.org/abs/hep-th/9511071}{{\tt hep-th/9511071}}].

\bibitem{Alexeev:1996vs}
S.~O. Alexeev and M.~V. Pomazanov, {\it {Black hole solutions with dilatonic
  hair in higher curvature gravity}},  {\em Phys. Rev.} {\bf D55} (1997)
  2110--2118 [\href{http://arXiv.org/abs/hep-th/9605106}{{\tt
  hep-th/9605106}}].

\bibitem{Torii:1996yi}
T.~Torii, H.~Yajima and K.-i. Maeda, {\it {Dilatonic black holes with
  Gauss-Bonnet term}},  {\em Phys. Rev.} {\bf D55} (1997) 739--753
  [\href{http://arXiv.org/abs/gr-qc/9606034}{{\tt gr-qc/9606034}}].

\bibitem{Alexander:2009tp}
S.~Alexander and N.~Yunes, {\it {Chern-Simons Modified General Relativity}},
  {\em Phys. Rept.} {\bf 480} (2009) 1--55
  [\href{http://arXiv.org/abs/0907.2562}{{\tt 0907.2562}}].

\bibitem{Yunes:2009hc}
N.~Yunes and F.~Pretorius, {\it {Dynamical Chern-Simons Modified Gravity. I.
  Spinning Black Holes in the Slow-Rotation Approximation}},  {\em Phys. Rev.}
  {\bf D79} (2009) 084043 [\href{http://arXiv.org/abs/0902.4669}{{\tt
  0902.4669}}].

\bibitem{Pani:2011gy}
P.~Pani, C.~F.~B. Macedo, L.~C.~B. Crispino and V.~Cardoso, {\it {Slowly
  rotating black holes in alternative theories of gravity}},  {\em Phys. Rev.}
  {\bf D84} (2011) 087501 [\href{http://arXiv.org/abs/1109.3996}{{\tt
  1109.3996}}].

\bibitem{Kleihaus:2011tg}
B.~Kleihaus, J.~Kunz and E.~Radu, {\it {Rotating Black Holes in Dilatonic
  Einstein-Gauss-Bonnet Theory}},  {\em Phys. Rev. Lett.} {\bf 106} (2011)
  151104 [\href{http://arXiv.org/abs/1101.2868}{{\tt 1101.2868}}].

\bibitem{Yagi:2012ya}
K.~Yagi, N.~Yunes and T.~Tanaka, {\it {Slowly Rotating Black Holes in Dynamical
  Chern-Simons Gravity: Deformation Quadratic in the Spin}},  {\em Phys. Rev.}
  {\bf D86} (2012) 044037 [\href{http://arXiv.org/abs/1206.6130}{{\tt
  1206.6130}}]. [Erratum: Phys. Rev.D89,049902(2014)].

\bibitem{Ayzenberg:2014aka}
D.~Ayzenberg and N.~Yunes, {\it {Slowly-Rotating Black Holes in
  Einstein-Dilaton-Gauss-Bonnet Gravity: Quadratic Order in Spin Solutions}},
  {\em Phys. Rev.} {\bf D90} (2014) 044066
  [\href{http://arXiv.org/abs/1405.2133}{{\tt 1405.2133}}]. [Erratum: Phys.
  Rev.D91,no.6,069905(2015)].

\bibitem{Maselli:2015tta}
A.~Maselli, P.~Pani, L.~Gualtieri and V.~Ferrari, {\it {Rotating black holes in
  Einstein-Dilaton-Gauss-Bonnet gravity with finite coupling}},  {\em Phys.
  Rev.} {\bf D92} (2015), no.~8 083014
  [\href{http://arXiv.org/abs/1507.00680}{{\tt 1507.00680}}].

\bibitem{Kleihaus:2015aje}
B.~Kleihaus, J.~Kunz, S.~Mojica and E.~Radu, {\it {Spinning black holes in
  Einstein--Gauss-Bonnet--dilaton theory: Nonperturbative solutions}},  {\em
  Phys. Rev. D} {\bf 93} (2016), no.~4 044047
  [\href{http://arXiv.org/abs/1511.05513}{{\tt 1511.05513}}].

\bibitem{Delsate:2018ome}
T.~Delsate, C.~Herdeiro and E.~Radu, {\it {Non-perturbative spinning black
  holes in dynamical Chern–Simons gravity}},  {\em Phys. Lett.} {\bf B787}
  (2018) 8--15 [\href{http://arXiv.org/abs/1806.06700}{{\tt 1806.06700}}].

\bibitem{Reall:2019sah}
H.~S. Reall and J.~E. Santos, {\it {Higher derivative corrections to Kerr black
  hole thermodynamics}},  {\em JHEP} {\bf 04} (2019) 021
  [\href{http://arXiv.org/abs/1901.11535}{{\tt 1901.11535}}].

\bibitem{Burger:2019wkq}
D.~J. Burger, W.~T. Emond and N.~Moynihan, {\it {Rotating Black Holes in Cubic
  Gravity}},  {\em Phys. Rev. D} {\bf 101} (2020), no.~8 084009
  [\href{http://arXiv.org/abs/1910.11618}{{\tt 1910.11618}}].

\bibitem{Adair:2020vso}
C.~Adair, P.~Bueno, P.~A. Cano, R.~A. Hennigar and R.~B. Mann, {\it {Slowly
  rotating black holes in Einsteinian cubic gravity}},
  \href{http://arXiv.org/abs/2004.09598}{{\tt 2004.09598}}.

\bibitem{meixner1954}
J.~Meixner and F.~Sch\"afke, {\it Mathieusche funktionen und
  sph\"aroidfunktionen (mathieu functions and spheroidal functions)},  1954.

\bibitem{WebBerti}
E.~Berti, {\it Webpage with mathematica notebooks and numerical quasinormal
  mode tables: https://pages.jh.edu/$\sim$eberti2/ringdown/}, .

\bibitem{Yang:2012he}
H.~Yang, D.~A. Nichols, F.~Zhang, A.~Zimmerman, Z.~Zhang and Y.~Chen, {\it
  {Quasinormal-mode spectrum of Kerr black holes and its geometric
  interpretation}},  {\em Phys. Rev.} {\bf D86} (2012) 104006
  [\href{http://arXiv.org/abs/1207.4253}{{\tt 1207.4253}}].

\bibitem{PhysRevD.82.104003}
S.~R. Dolan, {\it Quasinormal mode spectrum of a kerr black hole in the eikonal
  limit},  {\em Phys. Rev. D} {\bf 82} (Nov, 2010) 104003.

\bibitem{Cardoso:2008bp}
V.~Cardoso, A.~S. Miranda, E.~Berti, H.~Witek and V.~T. Zanchin, {\it {Geodesic
  stability, Lyapunov exponents and quasinormal modes}},  {\em Phys. Rev.} {\bf
  D79} (2009) 064016 [\href{http://arXiv.org/abs/0812.1806}{{\tt 0812.1806}}].

\bibitem{Fishbach:2017dwv}
M.~Fishbach, D.~E. Holz and B.~Farr, {\it {Are LIGO's Black Holes Made From
  Smaller Black Holes?}},  {\em Astrophys. J.} {\bf 840} (2017), no.~2 L24
  [\href{http://arXiv.org/abs/1703.06869}{{\tt 1703.06869}}].

\bibitem{Punturo:2010zz}
M.~Punturo {\em et.~al.}, {\it {The Einstein Telescope: A third-generation
  gravitational wave observatory}},  {\em Class. Quant. Grav.} {\bf 27} (2010)
  194002.

\bibitem{Healy:2017vuz}
J.~Healy, C.~O. Lousto, I.~Ruchlin and Y.~Zlochower, {\it {Evolutions of
  unequal mass, highly spinning black hole binaries}},  {\em Phys. Rev.} {\bf
  D97} (2018), no.~10 104026 [\href{http://arXiv.org/abs/1711.09041}{{\tt
  1711.09041}}].

\bibitem{Berti:2005ys}
E.~Berti, V.~Cardoso and C.~M. Will, {\it {On gravitational-wave spectroscopy
  of massive black holes with the space interferometer LISA}},  {\em Phys.
  Rev.} {\bf D73} (2006) 064030 [\href{http://arXiv.org/abs/gr-qc/0512160}{{\tt
  gr-qc/0512160}}].

\bibitem{salvatier2016probabilistic}
J.~Salvatier, T.~V. Wiecki and C.~Fonnesbeck, {\it Probabilistic programming in
  python using pymc3},  {\em PeerJ Computer Science} {\bf 2} (2016) e55.

\bibitem{Berti:2016lat}
E.~Berti, A.~Sesana, E.~Barausse, V.~Cardoso and K.~Belczynski, {\it
  {Spectroscopy of Kerr black holes with Earth- and space-based
  interferometers}},  {\em Phys. Rev. Lett.} {\bf 117} (2016), no.~10 101102
  [\href{http://arXiv.org/abs/1605.09286}{{\tt 1605.09286}}].

\bibitem{LIGOScientific:2020stg}
{\bf LIGO Scientific, Virgo} Collaboration, R.~Abbott {\em et.~al.}, {\it
  {GW190412: Observation of a Binary-Black-Hole Coalescence with Asymmetric
  Masses}},  \href{http://arXiv.org/abs/2004.08342}{{\tt 2004.08342}}.

\bibitem{Cabero:2019zyt}
M.~Cabero, J.~Westerweck, C.~D. Capano, S.~Kumar, A.~B. Nielsen and
  B.~Krishnan, {\it {Black hole spectroscopy in the next decade}},  {\em Phys.
  Rev. D} {\bf 101} (2020), no.~6 064044
  [\href{http://arXiv.org/abs/1911.01361}{{\tt 1911.01361}}].

\bibitem{rosen1932vibrations}
N.~Rosen and P.~M. Morse, {\it On the vibrations of polyatomic molecules},
  {\em Physical Review} {\bf 42} (1932), no.~2 210.

\bibitem{Flammer1957}
C.~Flammer, {\it Spheroidal wave functions},  1957.

\bibitem{Berti:2005gp}
E.~Berti, V.~Cardoso and M.~Casals, {\it {Eigenvalues and eigenfunctions of
  spin-weighted spheroidal harmonics in four and higher dimensions}},  {\em
  Phys. Rev. D} {\bf 73} (2006) 024013
  [\href{http://arXiv.org/abs/gr-qc/0511111}{{\tt gr-qc/0511111}}]. [Erratum:
  Phys.Rev.D 73, 109902 (2006)].

\bibitem{falloon2003theory}
P.~E. Falloon, P.~Abbott and J.~Wang, {\it Theory and computation of spheroidal
  wavefunctions},  {\em Journal of Physics A: Mathematical and General} {\bf
  36} (2003), no.~20 5477.

\bibitem{BHPToolkit}
``{Black Hole Perturbation Toolkit}.''
  (\href{http://bhptoolkit.org/}{bhptoolkit.org}).

\end{thebibliography}\endgroup
\bibliographystyle{JHEP-2}
\label{biblio}

\end{document}